\documentclass[12pt]{article}
\pdfoutput=1 % if your are submitting a pdflatex (i.e. if you have
\usepackage{jheppub} % for details on the use of the package, please
\usepackage{tikz}

\usepackage[T1]{fontenc} % if needed
\usepackage{amsmath,physics}
\usepackage{amsthm}
\usepackage{comment}
\usepackage[percent]{overpic}
\usepackage[toc,page]{appendix}
\usepackage{etoolbox}
\usepackage{breakcites}
\usepackage{placeins}

\let\Oldsection\section
\renewcommand{\section}{\FloatBarrier\Oldsection}

\let\Oldsubsection\subsection
\renewcommand{\subsection}{\FloatBarrier\Oldsubsection}

\let\Oldsubsubsection\subsubsection
\renewcommand{\subsubsection}{\FloatBarrier\Oldsubsubsection}

\BeforeBeginEnvironment{appendices}{\clearpage}
\AfterEndEnvironment{appendices}{\clearpage}
\AfterEndEnvironment{appendices}{\pretocmd{\section}{\clearpage}{}{}}{}
\AfterEndEnvironment{appendices}{\pretocmd{\subsubsection}{\clearpage}{}{}}{}

\usepackage{graphicx,float,caption,subcaption,xcolor}
\usepackage{hyperref}
\hypersetup{
    colorlinks=true,
    linkcolor=blue,
    filecolor=magenta,      
    urlcolor=cyan,
    pdftitle={Discrete Bulk Reconstruction},
    pdfpagemode=FullScreen,
    }

\graphicspath{{./img/}} 
\usepackage{multirow}

\usepackage{bbm}%for that nice \mathbbm{1}

\newcommand{\Hil}{\mathcal{H}}
\def\ket#1{{|{#1}\rangle}}
\newtheorem{theorem}{Theorem}
\numberwithin{theorem}{section}
\newtheorem{lemma}[theorem]{Lemma}
\newtheorem{problem}[theorem]{Problem}

\newtheorem{corollary}[theorem]{Corollary}
\newtheorem{proposition}[theorem]{Proposition}

\title{Discrete Bulk Reconstruction}

\author[a]{Scott Aaronson}\note{Supported by a Vannevar Bush Fellowship
from the US Department of Defense, the Berkeley NSF-QLCI CIQC Center, a Simons Investigator Award, and the Simons “It from Qubit” collaboration.}
\author[a]{and Jason Pollack}\note{Supported by the Simons Foundation through \emph{It from Qubit: Simons Collaboration on Quantum Fields, Gravity, and Information}.}

\affiliation[a]{Quantum Information Center, Department of Computer Science, The University of Texas at Austin,
2317 Speedway, Austin, TX 78712, USA}

\emailAdd{scott@scottaaronson.com}
\emailAdd{jasonpollack@gmail.com}

\abstract{According to the \emph{AdS/CFT correspondence}, the geometries of certain spacetimes are fully determined by quantum states that live on their boundaries---indeed, by the von Neumann entropies of portions of those boundary states. \ This work investigates to what extent the geometries can be reconstructed from the entropies \emph{in polynomial time}. \ Bouland, Fefferman, and Vazirani (2019) argued that the AdS/CFT map can be exponentially complex if one wants to reconstruct regions such as the interiors of black holes. \ Our main result provides a sort of converse: we show that, in the special case of a single 1D boundary, if the input data consists of a list of entropies of \emph{contiguous} boundary regions, and if the entropies satisfy a single inequality called Strong Subadditivity, then we can construct a graph model for the bulk in linear time. \ Moreover, the bulk graph is planar, it has $O(N^2)$ vertices (the information-theoretic minimum), and it's ``universal,'' with only the edge weights depending on the specific entropies in question. \ From a combinatorial perspective, our problem boils down to an ``inverse'' of the famous min-cut problem: rather than being given a graph and asked to find a min-cut, here we're given the values of min-cuts separating various sets of vertices, and need to find a weighted undirected graph consistent with those values. \ Our solution to this problem relies on the notion of a ``bulkless'' graph, which might be of independent interest for AdS/CFT. \ We also make initial progress on the case of multiple 1D boundaries---where the boundaries could be connected via wormholes---including an upper bound of $O(N^4)$ vertices whenever an embeddable bulk graph exists (thus putting the problem into the complexity class $\mathsf{NP}$).

}

\begin{document} 
\maketitle
\flushbottom

%%%%%%%%%%%%%%%%%%%%%%%%%%%%%%%%%%%
\section{Introduction}

The \emph{anti-de Sitter / conformal field theory} or AdS/CFT correspondence is one of the most important developments in theoretical physics of the past quarter century. \ AdS/CFT posits a duality between two superficially different theories:

\begin{enumerate}
\item[(i)] a theory of quantum gravity in a $D$-dimensional ``bulk'' space with a negative cosmological constant (called anti-de Sitter or AdS space), and
\item[(ii)] a quantum field theory, with no gravity, that lives on\footnote{Strictly speaking, since the two theories share the same Hilbert space, the field theory does not literally live on the boundary of the bulk space, but rather on a \emph{copy} of it: that is, on a space isometric to the boundary. We will often elide such distinctions when speaking informally.} the $(D-1)$-dimensional boundary of the $D$-dimensional AdS space.
\end{enumerate}

The relation between the two theories is often called ``holographic''; it necessarily fails to respect spatial locality. \ Though it originally emerged from string theory, AdS/CFT is now often discussed in the context of the black hole information problem and so on with no explicit reference to strings.

The universe described by AdS/CFT is not our universe: among other differences, it has a negative cosmological constant whereas ours appears to have a positive one. \ Nevertheless, AdS/CFT stands as the most explicit known example where a classical spacetime emerges from more fundamental, quantum degrees of freedom, as envisioned by the ``It from Qubit'' approach to physics.

What does it mean for the bulk and boundary theories to be ``equivalent''? \ It means that there's a mapping, or ``dictionary,'' mapping all states and observables in one theory to corresponding states and observables in the other, justifying the view that the two theories are just different ways to look at the same underlying Hilbert space. \ This dictionary is not yet completely known\footnote{Some string theorists would disagree}---indeed, the AdS side lacks an independent fully-rigorous definition---but enough calculations have been done in both theories and have yielded the same answers to convince most experts that the dictionary exists.

\begin{figure}[!htb]
\begin{center}\includegraphics{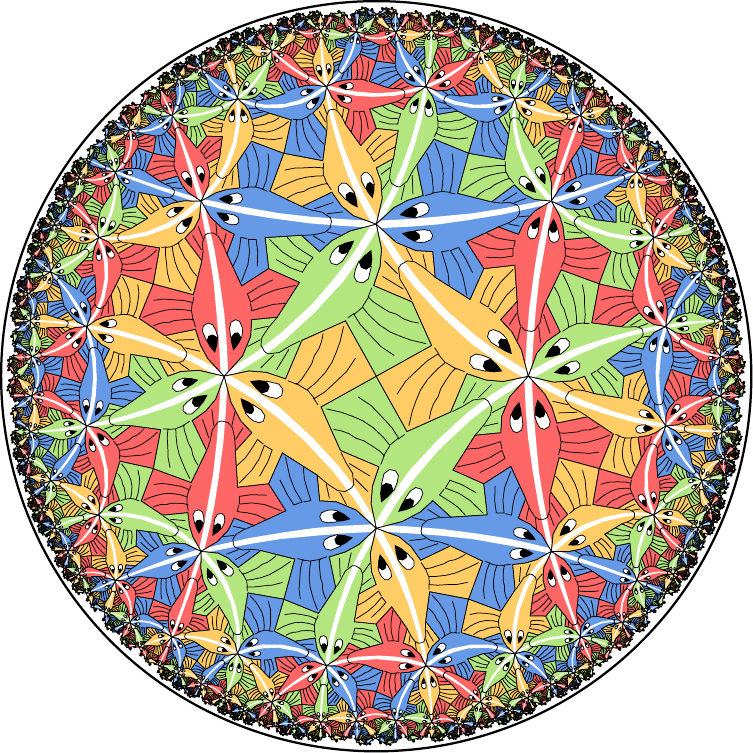}
\caption{Representation of the hyperbolic plane on the Poincar\'e disk. White lines denote boundary-anchored geodesics. \ The piling up of fish at the boundary indicates that lengths become infinite as they approach it. \ (Figure from \url{https://www.d.umn.edu/~ddunham/dunbrid07.pdf}, modeled on \emph{Circle Limit III}, M.C. Escher, 1959.) 
\label{fig:escher}}
\end{center}
\end{figure}

A key entry in the holographic dictionary is the \emph{Ryu-Takayanagi (RT) formula} \cite{Ryu:2006bv}, which equates the von Neumann entropy of a subregion of the boundary state with the minimal area among all bulk surfaces that end on that boundary subregion (see Figure \ref{fig:escher}). \ Hence, if we are given a CFT state, we could imagine systematically \emph{reconstructing} the bulk geometry by looking at the entropies of various boundary subregions to determine areas. \ In practice, we will not be able to reconstruct beyond event horizons in the bulk spacetime (or, more generally, extremal surface barriers \cite{Engelhardt:2013tra}), but we can nevertheless often learn the metric in a large portion of the spacetime \cite{Bao:2019bib,Bao:2020abm}.

The RT formula is extremely interesting from a quantum-information perspective because it implies that, in holographic states, entropies behave like areas, and thus must satisfy all the same constraints that areas do. \ Entropies in \emph{all} quantum states satisfy certain inequalities, such as \emph{subadditivity} and \emph{strong subadditivity}:

\begin{equation}
S(\rho_{AB}) \le S(\rho_A) + S(\rho_B), \hspace{0.5 cm}
S(\rho_{ABC})+S(\rho_B) \le S(\rho_{AB}) + S(\rho_{BC}).
\end{equation}

\noindent But entropies in \emph{holographic} states must satisfy additional inequalities, such as \emph{monogamy of mutual information} \cite{Hayden:2011ag}:

\begin{equation} 
S(\rho_A) + S(\rho_B) + S(\rho_C) + S(\rho_{ABC}) \le S(\rho_{AB}) + S(\rho_{BC})+ S(\rho_{AC}),
\end{equation}

\noindent as well as an infinite family of further inequalities \cite{Bao:2015bfa}. \ Note that the difficult-sounding problem of which boundary quantum states describe classical bulk geometries thereby gets reduced to a much more concrete problem, of checking that a list of entropy inequalities is satisfied.

It is natural to wonder just how simple AdS/CFT can be made. \ Can we see the main ideas in a model scenario, without using the machinery of quantum field theory? \ Work over the past decade has aimed to answer this: for instance, \cite{Almheiri:2014lwa,Harlow:2016vwg} interpreted AdS/CFT as literally an example of a quantum error-correcting code, while \cite{Pastawski:2015qua,Hayden:2016cfa} proposed tensor networks which exhibit a form of the RT formula. \ Meanwhile, \cite{Bao:2015bfa} proposed modeling the bulk space by a weighted graph with finitely many vertices. \ In this case, minimal surfaces turn into \emph{min-cuts}: that is, sets of edges of minimum total weight that separate the graph into disconnected components.

In this work, we seek to push the simplification process as far as possible. \ We dispense entirely with the bulk spatial manifold and the boundary quantum state, and just proceed directly from a list of entropies, for unions of ``atomic'' boundary regions, to a graph model of the bulk---i.e., a weighted graph whose min-cuts correspond to the given entropies. \ We ask both when this graph exists, and how easily it can be constructed when it does.

The mapping from entropies to a graph will not be unique. \ The mapping from the graph to a full bulk geometry will \emph{also} not be unique. \ Nevertheless, the graph can be thought of as a particular coarse-graining of a geometry, or as the common data shared by an equivalence class of possible geometries.

\subsection{Our Results}

In the important case $D=2$, where the spatial boundary consists of one or more circles, we make significant progress. \ When the spatial boundary is a \emph{single} circle, we show that a planar bulk graph is fully determined by the entropies of the quadratically many \emph{contiguous} boundary regions, and always exists provided those entropies satisfy Strong Subadditivity. \  Furthermore, this graph has $O(N^2)$ vertices and is \emph{universal}, meaning that we can reproduce \emph{any} valid vector of contiguous entropies merely adjusting the graph's edge weights. \ The mapping from contiguous entropies to edge weights is just a linear transformation, and is computable in $O(N^2)$ time---linear in the amount of input data.

The fact that we're only trying to explain the entropies of \emph{contiguous} boundary regions is crucial here. \ Note that, \emph{assuming} the bulk graph is planar, it's already known how to pass from the vector of contiguous entropies to the entropy of any desired non-contiguous region: the work of \cite{Bao:2016rbj} reduced that problem to an instance of minimum-weight bipartite perfect matching, which is solvable in polynomial time. \ On the other hand, there are also valid holographic states with non-contiguous entropies that violate the Bao-Chatwin-Davies prescription. \ We can interpret these states as describing bulk graphs that fail to be planar because they contain \emph{wormholes} connecting faraway boundary regions.

Note also that, with $D\ge 3$ bulk dimensions, there are already $\exp(N)$ \emph{contiguous} boundary regions, and the bulk reconstruction problem is much more complicated for that reason, among others.

When there are multiple boundary regions, the resulting bulk geometry may have wormholes connecting the boundaries, and the problem again becomes more complicated. \ Although our main graph construction is unable to handle the multi-boundary case, we prove some weaker results and no-go theorems.

\subsection{The Ryu-Takayanagi Formula and Min-Cut}

For computer scientists, a striking feature of AdS/CFT is the role played by \emph{min-cuts}, a central concept in combinatorics and graph theory. \ Given a finite weighted undirected graph $G$ with real edge weights $w(e)\ge 0$, as well as two disjoint sets of vertices $R,R'$, a min-cut is a set $C$ of edges of minimum total weight, $W=\sum_{e\in C} w(e)$, whose removal disconnects $R$ from $R'$. \ The famous Max-Flow/Min-Cut Theorem \cite{ford1956maximal} says that $W$, the weight of the min-cut, equals the maximum amount of ``flow'' that can be routed from $R$ to $R'$ via the edges of $G$, where the flow along any edge cannot exceed its weight, and the total flow entering any vertex must equal the flow leaving it. \ Polynomial-time algorithms to compute the min-cut, or equivalently max-flow, between $R$ and $R'$ are staples of the undergraduate CS curriculum, e.g., \cite{CLRS}. \ Nearly linear-time algorithms for these problems were even recently achieved \cite{chen2022maximum}.

The connection between min-cuts and AdS/CFT comes via the RT formula discussed previously. \ Recall what the RT formula says: that, up to tiny corrections, the area of a \emph{minimal surface} in the bulk separating a boundary region $R$ from its complement is proportional to $S(\rho_R)$, the von Neumann entropy of the reduced quantum state of CFT on region $R$, with the coefficient of proportionality given by $1/(4 G_\mathrm{N})$, with $G_\mathrm{N}$ the gravitational constant. \ When we replace manifolds by graphs, the geometric concept of a minimal surface becomes \emph{precisely} the combinatorial concept of a min-cut. \ Likewise, the problem of constructing a bulk geometry to represent the boundary entropies, becomes the problem of constructing a graph with prescribed min-cut values between various distinguished sets of vertices.

Our concern, then, is with the ``inverse'' of the classic min-cut problem: rather than being given a graph and asked to find a min-cut, we will be given a list of min-cut values and asked to find a graph consistent with them.

\subsection{Previous Work}

Beyond the ``It from Qubit'' program in general, this paper was specifically inspired by research into the entropies of subregions of holographic states---a line of work initiated by \cite{Bao:2015bfa} and continued by, for example, \cite{Rota:2017ubr,Czech:2019lps,He:2019ttu,Bao:2020mqq,Fadel:2021urx}. 

For the most part, authors have focused either on generalizing to classes of states beyond holographic states, or better mathematical understanding of the structure of the space of holographic states, or ways of obtaining the Ryu-Takayanagi formula using tools other than minimal surfaces. \ Our work does not fall into any of these categories. \ Instead, we work directly with minimal areas, and are concerned with reproducing them as the min-cuts of a graph. 

As we review in Section \ref{sub:doubly_exponential}, \cite{Bao:2015bfa} already considered the problem of going back and forth between geometries and graphs, and gave a general method to obtain a graph given a geometry and a set of subregions. \ Unfortunately, their method generically produces a graph with a number of vertices \emph{doubly exponential} in the number of regions. \ \cite{Bao:2016rbj} later explicitly studied the complexity of obtaining the minimal surfaces of regions in a given geometry.

Our central point of departure from these earlier works is that we never assume a manifold or a metric. \ Instead, we take as input only a finite list of the entropies themselves, and then try to pass \emph{directly} to a graph model of the bulk---now taking care to minimize the graph's size, as well as the computational complexity of building the graph. \ As we'll see, in an important special case---namely, when we're given as input the entropies of contiguous regions along a 1D boundary---we'll be able to construct a graph with $O(N^2)$ vertices in linear time, a vast improvement over the doubly-exponential generic construction of \cite{Bao:2015bfa}.

Our work can be seen as a sort of converse to the widely-discussed paper of \cite{Bouland:2019pvu}, on the computational complexity of the holographic dictionary. \ Those authors argued that, when the bulk geometry contains black holes or wormholes, the problem of reconstructing the geometry from boundary data can be exponentially hard even for a quantum computer. \ Indeed, they showed this task to be at least as hard as distinguishing various candidate pseudorandom $n$-qubit states from Haar-random states. \ By contrast, this work aims to show that, when the bulk \emph{lacks} event horizons or nontrivial topology (or, more generally, the extremal surfaces which create the ``python's lunch'' phenomenon \cite{Brown:2019rox,Engelhardt:2021qjs}), its geometry can be completely reconstructed from boundary data in classical polynomial time. \ We make significant progress toward this conjecture by proving it in the special case of a single 1D boundary.

% \subsection{Summary of Results}

% \draftnoteJP{Summary figure: bulkless --> planar --> diamondwork. Possibly wormhole version as well?}

%%%%%%%%%%%%%%%%%%%%%%%%%%%%%%%%%%%
\section{Preliminaries}\label{sec:basic}

\subsection{Statement of the Problem}

In the previous section, we introduced the problem of reconstructing a bulk graph from the entropies of boundary regions. \ In general, we might only have access to the entropies of a limited set of subregions. \ Given a factorization of a finite-dimensional Hilbert space as an $N$-fold tensor-product,
\begin{equation}
    \Hil = \otimes_{i=1}^N \Hil_i,
\end{equation}
we can group the entropies of the various reduced density matrices constructible from a state $\ket{\Psi}\in\Hil$ into an \emph{entropy vector} of length $2^{N-1}-1$,
\begin{equation}
    \vec{S}\left(\ket{\Psi}\right) \equiv
    \left( S_1, S_2, \ldots, S_N, S_{1,2}, S_{1,3}, \ldots, S_{1,N},
    \ldots, S_{N-1,N}, S_{1,2,3}, \ldots \right),    
\end{equation}
where $S_{a_1,\ldots,a_k}$ is the von Neumann entropy of the reduced state $\rho_{a_1,\ldots,a_k}$ on $a_1,\ldots,a_k$, and the indices are over all sets of elements in $\{1,\ldots,N\}$, except that because of the purity of $\ket{\Psi}$ there is a redundancy between the entropy of a set and that of its complement (and $S_\emptyset = S_{1,\ldots,N} = 0$ is omitted).

One possible approach would therefore be to start explicitly with a quantum state $\ket{\Psi}\in \Hil$ and work with its entropy vector relative to some factorization. \ However, specifying the state in, say, the field value basis would require a continuous function's worth of degrees of freedom, which would defeat our goal of working with purely discrete objects. \ Indeed, the Hilbert space of a conformal field theory is infinite-dimensional and not isomorphic to the tensor product of a finite number of smaller factors. \ In principle, these problems could be cured by explicitly fixing an ultraviolet and infrared cutoff for the field theory, so that the Hilbert space became explicitly finite-dimensional, but in practice this procedure is difficult to carry out and sensitive to details of how the cutoff is implemented.

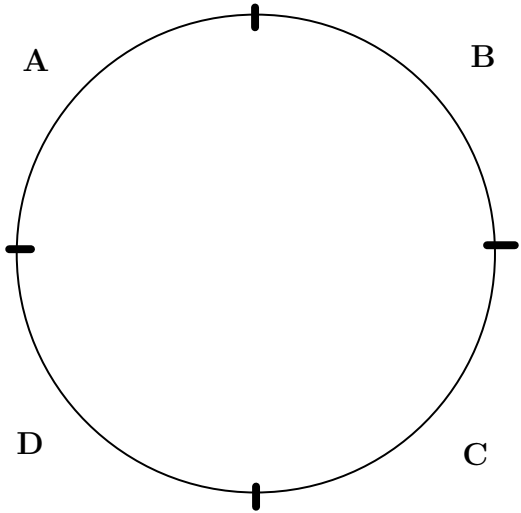
\begin{figure}[!htb]
\centering

\tikzset{every picture/.style={line width=0.75pt}} %set default line width to 0.75pt        

\begin{tikzpicture}[x=0.75pt,y=0.75pt,yscale=-1,xscale=1]

%Shape: Circle [id:dp6337493530440246] 
\draw   (208.59,145.61) .. controls (208.59,79) and (262.58,25) .. (329.19,25) .. controls (395.8,25) and (449.8,79) .. (449.8,145.61) .. controls (449.8,212.22) and (395.8,266.21) .. (329.19,266.21) .. controls (262.58,266.21) and (208.59,212.22) .. (208.59,145.61) -- cycle ;
%Shape: Free Drawing [id:dp029613617400735093] 
\draw  [line width=3] [line join = round][line cap = round] (445.8,141.41) .. controls (450.47,141.41) and (455.13,141.41) .. (459.8,141.41) ;
%Shape: Free Drawing [id:dp8811102290156974] 
\draw  [line width=3] [line join = round][line cap = round] (329.3,273.41) .. controls (329.3,270.01) and (329.3,266.61) .. (329.3,263.21) ;
%Shape: Free Drawing [id:dp8301339134071475] 
\draw  [line width=3] [line join = round][line cap = round] (204.8,143.41) .. controls (208.47,143.41) and (212.13,143.41) .. (215.8,143.41) ;
%Shape: Free Drawing [id:dp3747691733324743] 
\draw  [line width=3] [line join = round][line cap = round] (328.8,21.41) .. controls (328.8,24.75) and (328.8,28.08) .. (328.8,31.41) ;

% Text Node
\draw (210,41) node [anchor=north west][inner sep=0.75pt]   [align=left] {\textbf{A}};
% Text Node
\draw (436,39) node [anchor=north west][inner sep=0.75pt]   [align=left] {\textbf{B}};
% Text Node
\draw (432,240) node [anchor=north west][inner sep=0.75pt]   [align=left] {\textbf{C}};
% Text Node
\draw (207,234) node [anchor=north west][inner sep=0.75pt]   [align=left] {\textbf{D}};

\end{tikzpicture}

\caption{Division of the boundary into atomic regions.\label{fig:atomic}}
\end{figure}

Our approach is simpler: we drop the demand that the entropies be derived from some particular quantum state $\ket{\Psi}$. \ Instead, we take our input to be an abstract list of entropies, either $2^{N-1}-1$ entries arranged into an entropy vector,
\begin{equation}\label{eq:entropy_vector}
   \vec{S}^{(N)} \equiv
    \left( S_1, S_2, \ldots, S_N, S_{1,2}, S_{1,3}, \ldots, S_{1,N},
    \ldots, S_{N-1,N}, S_{1,2,3}, \ldots \right), 
\end{equation}
or some subset of this list. \ For all $i\in [N]$, we will say that $S_i$ is an entropy of the $i^{th}$ \emph{atomic region}, and have in mind that the union of the $N$ atomic regions comprises the entire boundary. \ For example, as shown in Figure \ref{fig:atomic}, we could think of the $N$ atomic regions as subintervals of a spatial circle. \ In this case, it would be natural to think of the entropy vector $\vec{S}^{(N)}$ as derived from some collection of qubits (or qudits) living on the circle, but we will not actually demand this. \ (We will return in Section \ref{sec:discussion} to the question of when we can construct a concrete quantum state with a specified entropy vector.)

To avoid excessive notation, we often label atomic regions by Latin letters $A,B,C,\ldots$. \ We write, for example, $S(AC)$ for the entropy of $AC\equiv A \cup C$, the union of the atomic regions $A$ and $C$.

The discrete bulk reconstruction problem can now be formulated as follows: 

\begin{problem}[discrete bulk reconstruction problem (DBRP)]
We're given as input a list of atomic boundary regions labeled $1,\ldots,N$, a list of subsets of the regions $R_1,\ldots,R_k \subseteq [N]$, and a real-valued entropy $S(R_i)\ge 0$ for each $R_i$. \ The problem is to construct a weighted undirected graph $G$, with $N$ distinguished boundary vertices $1,\ldots,N$ that we identify with the atomic boundary regions, such that for each $R_i$, the weight of the minimum cut separating $R_i$ from the rest of the boundary vertices (i.e., from $[N]- R_i$) is equal to $S(R_i)$. \ (Or to output that no such graph exists.)
\end{problem}

When such a graph does exist, we would like to know if it is essentially unique, how many vertices it has, whether it is planar, and so on.

\subsection{Computability with Doubly Exponential Vertices}\label{sub:doubly_exponential}

A priori, one might worry that the DBRP could be uncomputable (equivalent to the halting problem), with no upper bound $f(N)$ on the number of bulk vertices needed as a function of the number $N$ of boundary vertices. \ Fortunately, the work of \cite{Bao:2015bfa} showed that this is not the case, and that $f(N) = 2^{2^N}$ vertices always suffice. \ Let us prove this for completeness.

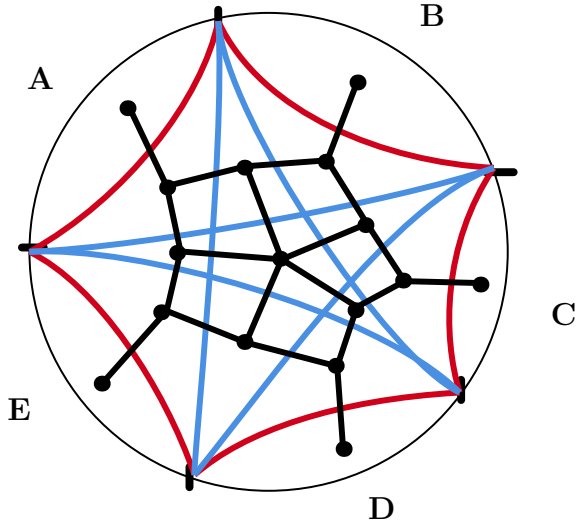
\begin{figure}[!htb]
\centering

\tikzset{every picture/.style={line width=0.75pt}} %set default line width to 0.75pt        

\begin{tikzpicture}[x=0.75pt,y=0.75pt,yscale=-1,xscale=1]

%Shape: Free Drawing [id:dp8940787419257947] 
\draw  [line width=3] [line join = round][line cap = round] (438.8,105.41) .. controls (443.47,105.41) and (448.13,105.41) .. (452.8,105.41) ;
%Shape: Free Drawing [id:dp5582059795594412] 
\draw  [line width=3] [line join = round][line cap = round] (426.3,220.41) .. controls (426.3,217.01) and (426.3,213.61) .. (426.3,210.21) ;
%Shape: Free Drawing [id:dp08850854480355763] 
\draw  [line width=3] [line join = round][line cap = round] (204.8,143.41) .. controls (208.47,143.41) and (212.13,143.41) .. (215.8,143.41) ;
%Shape: Free Drawing [id:dp4145595964186515] 
\draw  [line width=3] [line join = round][line cap = round] (303.8,23.41) .. controls (303.8,26.75) and (303.8,30.08) .. (303.8,33.41) ;
%Shape: Circle [id:dp3729033397238102] 
\draw   (208.59,145.61) .. controls (208.59,79) and (262.58,25) .. (329.19,25) .. controls (395.8,25) and (449.8,79) .. (449.8,145.61) .. controls (449.8,212.22) and (395.8,266.21) .. (329.19,266.21) .. controls (262.58,266.21) and (208.59,212.22) .. (208.59,145.61) -- cycle ;
%Shape: Free Drawing [id:dp9596451131997783] 
\draw  [line width=3] [line join = round][line cap = round] (289.3,264.41) .. controls (289.3,261.01) and (289.3,257.61) .. (289.3,254.21) ;
%Curve Lines [id:da23450595253333195] 
\draw [color={rgb, 255:red, 208; green, 2; blue, 27 }  ,draw opacity=1 ][line width=2.25]    (208.59,145.61) .. controls (257.8,118.21) and (295.8,68.21) .. (303.8,29.21) ;
%Curve Lines [id:da26643241054178857] 
\draw [color={rgb, 255:red, 208; green, 2; blue, 27 }  ,draw opacity=1 ][line width=2.25]    (442.8,103.21) .. controls (387.8,100.21) and (325.8,76.21) .. (303.8,29.21) ;
%Curve Lines [id:da5447105537310994] 
\draw [color={rgb, 255:red, 208; green, 2; blue, 27 }  ,draw opacity=1 ][line width=2.25]    (425.59,216.61) .. controls (416.8,188.21) and (416.8,142.21) .. (442.8,103.21) ;
%Curve Lines [id:da6080239500799434] 
\draw [color={rgb, 255:red, 208; green, 2; blue, 27 }  ,draw opacity=1 ][line width=2.25]    (291.8,258.21) .. controls (316.8,236.21) and (362.8,220.21) .. (425.59,216.61) ;
%Curve Lines [id:da1865393785929017] 
\draw [color={rgb, 255:red, 208; green, 2; blue, 27 }  ,draw opacity=1 ][line width=2.25]    (208.59,145.61) .. controls (242.8,162.21) and (276.8,214.21) .. (291.8,258.21) ;
%Curve Lines [id:da6738931058941031] 
\draw [color={rgb, 255:red, 74; green, 144; blue, 226 }  ,draw opacity=1 ][line width=2.25]    (208.59,145.61) .. controls (268.8,143.21) and (393.8,120.21) .. (442.8,103.21) ;
%Curve Lines [id:da6433353982348533] 
\draw [color={rgb, 255:red, 74; green, 144; blue, 226 }  ,draw opacity=1 ][line width=2.25]    (303.8,29.21) .. controls (307.8,78.21) and (389.8,200.21) .. (425.59,216.61) ;
%Curve Lines [id:da8113791894184315] 
\draw [color={rgb, 255:red, 74; green, 144; blue, 226 }  ,draw opacity=1 ][line width=2.25]    (291.8,258.21) .. controls (331.8,209.21) and (391.8,121.21) .. (442.8,103.21) ;
%Curve Lines [id:da6506138625473936] 
\draw [color={rgb, 255:red, 74; green, 144; blue, 226 }  ,draw opacity=1 ][line width=2.25]    (208.59,145.61) .. controls (260.8,142.21) and (382.8,174.21) .. (425.59,216.61) ;
%Curve Lines [id:da2213896607865562] 
\draw [color={rgb, 255:red, 74; green, 144; blue, 226 }  ,draw opacity=1 ][line width=2.25]    (303.8,29.21) .. controls (307.8,92.21) and (290.8,255.21) .. (291.8,258.21) ;
%Shape: Free Drawing [id:dp016146667925268865] 
\draw  [line width=6] [line join = round][line cap = round] (278.01,113.11) .. controls (278.2,113.11) and (278.4,113.11) .. (278.59,113.11) ;
%Shape: Free Drawing [id:dp7571553169692553] 
\draw  [line width=6] [line join = round][line cap = round] (258.01,73.11) .. controls (258.2,73.11) and (258.4,73.11) .. (258.59,73.11) ;
%Shape: Free Drawing [id:dp7324144852397254] 
\draw  [line width=6] [line join = round][line cap = round] (374.01,60.11) .. controls (374.2,60.11) and (374.4,60.11) .. (374.59,60.11) ;
%Shape: Free Drawing [id:dp11914568777587853] 
\draw  [line width=6] [line join = round][line cap = round] (436.01,162.11) .. controls (436.2,162.11) and (436.4,162.11) .. (436.59,162.11) ;
%Shape: Free Drawing [id:dp7336019431141567] 
\draw  [line width=6] [line join = round][line cap = round] (367.01,245.11) .. controls (367.2,245.11) and (367.4,245.11) .. (367.59,245.11) ;
%Shape: Free Drawing [id:dp8867406360307959] 
\draw  [line width=6] [line join = round][line cap = round] (245.01,212.11) .. controls (245.2,212.11) and (245.4,212.11) .. (245.59,212.11) ;
%Shape: Free Drawing [id:dp5989234038878695] 
\draw  [line width=6] [line join = round][line cap = round] (358.01,100.11) .. controls (358.2,100.11) and (358.4,100.11) .. (358.59,100.11) ;
%Shape: Free Drawing [id:dp2188557499031385] 
\draw  [line width=6] [line join = round][line cap = round] (275.01,176.11) .. controls (275.2,176.11) and (275.4,176.11) .. (275.59,176.11) ;
%Shape: Free Drawing [id:dp8849520937116535] 
\draw  [line width=6] [line join = round][line cap = round] (363.01,203.11) .. controls (363.2,203.11) and (363.4,203.11) .. (363.59,203.11) ;
%Shape: Free Drawing [id:dp2858288715905508] 
\draw  [line width=6] [line join = round][line cap = round] (397.01,160.11) .. controls (397.2,160.11) and (397.4,160.11) .. (397.59,160.11) ;
%Shape: Free Drawing [id:dp047126422568335435] 
\draw  [line width=6] [line join = round][line cap = round] (283.01,146.11) .. controls (283.2,146.11) and (283.4,146.11) .. (283.59,146.11) ;
%Shape: Free Drawing [id:dp11933168936431171] 
\draw  [line width=6] [line join = round][line cap = round] (335.01,149.11) .. controls (335.2,149.11) and (335.4,149.11) .. (335.59,149.11) ;
%Shape: Free Drawing [id:dp8524844847382536] 
\draw  [line width=6] [line join = round][line cap = round] (317.01,191.11) .. controls (317.2,191.11) and (317.4,191.11) .. (317.59,191.11) ;
%Shape: Free Drawing [id:dp5989326108958655] 
\draw  [line width=6] [line join = round][line cap = round] (317.01,103.11) .. controls (317.2,103.11) and (317.4,103.11) .. (317.59,103.11) ;
%Shape: Free Drawing [id:dp5352673905467935] 
\draw  [line width=6] [line join = round][line cap = round] (378.01,132.11) .. controls (378.2,132.11) and (378.4,132.11) .. (378.59,132.11) ;
%Shape: Free Drawing [id:dp6561908061309656] 
\draw  [line width=6] [line join = round][line cap = round] (373.01,175.11) .. controls (373.2,175.11) and (373.4,175.11) .. (373.59,175.11) ;
%Straight Lines [id:da27948288928252185] 
\draw [line width=2.25]    (277.8,175.41) -- (245.8,211.41) ;
%Straight Lines [id:da5874678244874887] 
\draw [line width=2.25]    (317.8,191.41) -- (277.8,175.41) ;
%Straight Lines [id:da9587069832255923] 
\draw [line width=2.25]    (284.8,145.41) -- (277.8,175.41) ;
%Straight Lines [id:da1759287701564567] 
\draw [line width=2.25]    (363.8,203.41) -- (366.8,245.41) ;
%Straight Lines [id:da9817301194872639] 
\draw [line width=2.25]    (437.8,161.41) -- (397.8,160.41) ;
%Straight Lines [id:da4072525469005148] 
\draw [line width=2.25]    (374.8,57.41) -- (358.8,99.41) ;
%Straight Lines [id:da03936094537919943] 
\draw [line width=2.25]    (317.8,191.41) -- (363.8,203.41) ;
%Straight Lines [id:da897354416197411] 
\draw [line width=2.25]    (363.8,203.41) -- (373.8,173.41) ;
%Straight Lines [id:da972593656929253] 
\draw [line width=2.25]    (397.8,160.41) -- (373.8,173.41) ;
%Straight Lines [id:da5135043831583774] 
\draw [line width=2.25]    (378.8,131.41) -- (397.8,160.41) ;
%Straight Lines [id:da3474011948746982] 
\draw [line width=2.25]    (358.8,99.41) -- (378.8,131.41) ;
%Straight Lines [id:da4864433159370436] 
\draw [line width=2.25]    (258.8,72.41) -- (277.8,112.41) ;
%Straight Lines [id:da9458357966603479] 
\draw [line width=2.25]    (277.8,112.41) -- (284.8,145.41) ;
%Straight Lines [id:da29791415976561697] 
\draw [line width=2.25]    (317.8,102.41) -- (277.8,112.41) ;
%Straight Lines [id:da7958504313728423] 
\draw [line width=2.25]    (317.8,102.41) -- (358.8,99.41) ;
%Straight Lines [id:da19759893061031897] 
\draw [line width=2.25]    (317.8,102.41) -- (335.8,149.41) ;
%Straight Lines [id:da05386945139184318] 
\draw [line width=2.25]    (378.8,131.41) -- (335.8,149.41) ;
%Straight Lines [id:da12487036172320654] 
\draw [line width=2.25]    (373.8,173.41) -- (335.8,149.41) ;
%Straight Lines [id:da36846852257266227] 
\draw [line width=2.25]    (335.8,149.41) -- (317.8,191.41) ;
%Straight Lines [id:da7402096726536034] 
\draw [line width=2.25]    (335.8,149.41) -- (284.8,145.41) ;

% Text Node
\draw (206,51) node [anchor=north west][inner sep=0.75pt]   [align=left] {\textbf{A}};
% Text Node
\draw (404,19) node [anchor=north west][inner sep=0.75pt]   [align=left] {\textbf{B}};
% Text Node
\draw (470,170) node [anchor=north west][inner sep=0.75pt]   [align=left] {\textbf{C}};
% Text Node
\draw (378,268) node [anchor=north west][inner sep=0.75pt]   [align=left] {\textbf{D}};
% Text Node
\draw (196,218) node [anchor=north west][inner sep=0.75pt]   [align=left] {\textbf{E}};

\end{tikzpicture}

\caption{A minimal graph with one vertex for every intersection of RT regions. (For a non-planar graph there will in general be many more non-empty intersections, not shown here.)\label{fig:RT_intersections}}
\end{figure}

\begin{proposition}[\cite{Bao:2015bfa}]
Whenever a graph exists that solves the DBRP on a given set of input data, there exists a (possibly identical) graph with at most $2^{2^N}$ vertices that also solves it.\label{thm:double_exponential}
\end{proposition}
\begin{proof}
By definition, any graph $G$ solving the DBRP has $N$ boundary vertices $\{v_1,\ldots,v_N\}$. \ For every subset $S\subset\{v_1,\ldots\,v_N\}$ of the boundary vertices, find the min-cut separating $S$ from $\{v_1,\ldots\,v_N\}- S$. \ The min-cut divides the graph into two subgraphs; call the one containing $S$ the ``RT region of $S$.'' \ There are $2^N$ distinct subsets of boundary vertices, and hence at most $2^N$ RT regions. \ Then construct all $2^{2^N}$ intersections of the RT regions (some of which may be the empty set). \ If there exists an intersection $H\subseteq G$ with more than one vertex, then we can construct a strictly smaller graph $G^\prime$ that also solves the DBRP with the same input data, by preserving all edges which cross the boundary of $H$, and connecting them all to a single internal vertex. \ See Figure \ref{fig:RT_intersections}. \ Repeat this procedure until no intersections with more than one vertex remain. \ Then the total number of vertices in the graph is at most $2^{2^N}$.
\end{proof}

Proposition \ref{thm:double_exponential} has the following important consequence.

\begin{corollary}
The DBRP is Turing-computable.
\end{corollary}
\begin{proof}
To solve the DBRP for given input data, we first construct the graph of size $2^{2^N}$ from Proposition \ref{thm:double_exponential}, with one vertex for each intersection. \ We then write down a list of $\exp(2^{2^N})$ inequalities on the edge weights to express that each of the $2^N$ min-cuts have at least the correct values. \ Next, we loop over all $\exp(2^{2^N})^{2^N}$ tuples of cuts to be forced to have at \emph{most} the correct values. \ Finally, for each such tuple, we search for a solution to the resulting system of inequalities using linear programming, halting whenever a solution is found. \ This takes $\exp(\exp(\exp (N)))$ time overall.
\end{proof}

Of course, we would like to be more efficient whenever possible.

\subsection{Planarity}

When trying to construct a graph representing a 2D bulk, one extremely natural requirement is that the graph be planar. \ A planar graph seems interpretable as a discrete analogue of a 2-manifold with trivial topology (so in particular, no wormholes). \ One of the main surprises of this paper is that, given any entropies for contiguous boundary regions, we can always construct a \emph{planar} bulk graph.

Having said that, we'll see that planarity can break down when there are entropies of \emph{non}-contiguous boundary regions to account for---an especially natural situation when there are multiple 1D boundaries. \ In this case, we can effectively force there to be wormholes in the bulk.

To show this, we'll use the famous Kuratowski's Theorem \cite{kuratowski1930probleme}, which says that a graph $G$ is planar if and only if it does not contain either the complete graph $K_5$ or the complete bipartite graph $K_{3,3}$ as minors (i.e., if neither can be obtained from $G$ via contractions and deletions). \ Thus, it suffices to construct boundary regions whose entropies can only be explained via $K_5$ or $K_{3,3}$.

\subsection{Entropic Inequalities}
\label{entropineq}
Among vectors of $2^N - 1$ nonnegative real numbers, not all are valid lists of entropies for the subsystems of some $N$-partite pure quantum state. \ The set that's valid is called the \textit{entropy cone}: ``cone'' because it can be shown to be closed under nonnegative linear combinations (for a review, see e.g., \cite{Walter:2014nxw}). Two inequalities which are valid for any quantum state are subadditivity (SA),
\begin{equation}
  S(\rho_{AB}) \leq S(\rho_A) + S(\rho_B),  \label{eq:SA}
\end{equation} 

\noindent and strong subadditivity (SSA),

\begin{equation}
S(\rho_{ABC}) + S(\rho_B) \leq S(\rho_{AB}) + S(\rho_{BC}),\label{eq:SSA}
\end{equation}

\noindent where $A$, $B$, and $C$ can be any $3$ disjoint subsystems. \ (Strong subadditivity implies subadditivity, as can be seen by taking the $B$ system to be trivial, but it will sometimes be convenient to refer to it separately.) \ For larger $N$, the complete list of inequalities satisfied by the entropies of arbitrary $N$-partite quantum states is not yet known, nor is it even known to be finite.

Within the entropy cone is the so-called \textit{holographic entropy cone}: the set of entropy vectors that correspond to $N$-partite quantum states with holographic duals. \ Formally, \cite{Bao:2015bfa} defined the $N^{th}$ holographic entropy cone as simply the set of entropy vectors $v\in \mathbb{R}^{2^N-1}$ for which there exists a weighted, undirected graph $G$, with $N$ boundary vertices and any finite number of bulk vertices, such that for all subsets $R\subseteq [N]$ of the boundary vertices, the value of a min-cut in $G$ separating $R$ from $[N]-R$ is exactly $S(R)$. \ They justified this definition by arguing that a holographic bulk dual, in the physics sense, can always be converted to such a graph and vice versa.

It's known that, for every $N$, the holographic entropy cone is defined by a finite list of inequalities \cite{Bao:2015bfa}, and those inequalities can in principle be computed in finite time given $N$. \ It's known further that, for a pure state divided into $N\le 3$ regions, all true entropy inequalities follow from SSA, meaning that the entropy cone and the holographic entropy cone coincide. \ For $N\ge 4$, however, there are additional inequalities that cause the holographic entropy cone to be a strict subset of the full entropy cone \cite{Walter:2020zvt}.

One such inequality is called \textit{monogamy of mutual information}, or MMI:

\begin{equation}
    S(\rho_{AB}) + S(\rho_{BC}) + S(\rho_{AC}) \ge S(\rho_A) + S(\rho_B) + S(\rho_C) + S(\rho_{ABC}). \label{eq:MMI}
\end{equation}

\begin{figure}[!htb]
\centering

\tikzset{every picture/.style={line width=0.75pt}} %set default line width to 0.75pt        

\begin{tikzpicture}[x=0.75pt,y=0.75pt,yscale=-1,xscale=1]

%Straight Lines [id:da00103526458897929] 
\draw [line width=3]    (155,70) -- (474.8,70.81) ;
%Shape: Free Drawing [id:dp8254052659534765] 
\draw  [line width=3] [line join = round][line cap = round] (205.8,65.41) .. controls (205.8,68.75) and (205.8,72.08) .. (205.8,75.41) ;
%Shape: Free Drawing [id:dp08738260147509758] 
\draw  [line width=3] [line join = round][line cap = round] (288.8,64.41) .. controls (288.8,67.75) and (288.8,71.08) .. (288.8,74.41) ;
%Shape: Free Drawing [id:dp6081637567648217] 
\draw  [line width=3] [line join = round][line cap = round] (370.8,65.41) .. controls (370.8,68.75) and (370.8,72.08) .. (370.8,75.41) ;
%Curve Lines [id:da3474357789332254] 
\draw [color={rgb, 255:red, 74; green, 144; blue, 226 }  ,draw opacity=1 ][line width=2.25]    (205.8,68.61) .. controls (269.8,-9.39) and (309.8,-9.39) .. (372.8,70.61) ;
%Curve Lines [id:da938139218034209] 
\draw [color={rgb, 255:red, 208; green, 2; blue, 27 }  ,draw opacity=1 ][line width=2.25]    (205.8,68.61) .. controls (225.24,47.63) and (237.88,37.39) .. (249.87,38.01) .. controls (261.32,38.61) and (272.17,49.1) .. (287.8,69.61) ;
%Curve Lines [id:da11984276550687101] 
\draw [color={rgb, 255:red, 208; green, 2; blue, 27 }  ,draw opacity=1 ][line width=2.25]    (287.8,69.61) .. controls (325.8,28.61) and (337.8,28.61) .. (369.8,70.61) ;

% Text Node
\draw (243,79) node [anchor=north west][inner sep=0.75pt]   [align=left] {\textbf{A}};
% Text Node
\draw (322,79) node [anchor=north west][inner sep=0.75pt]   [align=left] {\textbf{B}};
% Text Node
\draw (173,108.4) node [anchor=north west][inner sep=0.75pt]  [font=\Large]  {$\mathrm{S( AB)} \ \leq \ \mathrm{S( A) \ +\ S( B)}$};

\end{tikzpicture}

\caption{Cutting and pasting proof of Subadditivity. \ To form a curve (though not necessarily a minimal one) that separates the region $AB$ from its complement, we can simply take the union of a curve for $A$ and a curve for $B$, as in the triangle inequality.\label{fig:SA_proof}}
\end{figure}
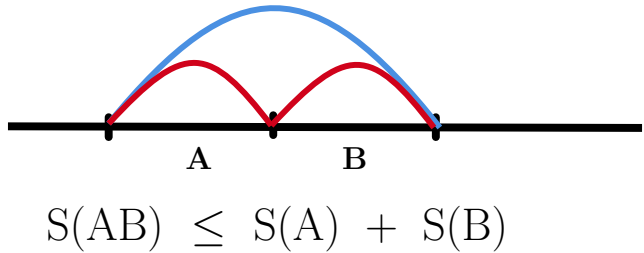

\begin{figure}[!htb]
\centering

\tikzset{every picture/.style={line width=0.75pt}} %set default line width to 0.75pt        

\begin{tikzpicture}[x=0.75pt,y=0.75pt,yscale=-1,xscale=1]

%Shape: Free Drawing [id:dp3807204962277846] 
\draw  [line width=3] [line join = round][line cap = round] (456.8,71.41) .. controls (456.8,74.75) and (456.8,78.08) .. (456.8,81.41) ;
%Straight Lines [id:da17408357542813602] 
\draw [line width=3]    (161,76) -- (480.8,76.81) ;
%Shape: Free Drawing [id:dp36366118193749375] 
\draw  [line width=3] [line join = round][line cap = round] (211.8,71.41) .. controls (211.8,74.75) and (211.8,78.08) .. (211.8,81.41) ;
%Shape: Free Drawing [id:dp6584956464853982] 
\draw  [line width=3] [line join = round][line cap = round] (294.8,70.41) .. controls (294.8,73.75) and (294.8,77.08) .. (294.8,80.41) ;
%Shape: Free Drawing [id:dp24356493904260623] 
\draw  [line width=3] [line join = round][line cap = round] (376.8,71.41) .. controls (376.8,74.75) and (376.8,78.08) .. (376.8,81.41) ;
%Curve Lines [id:da5300162253413816] 
\draw [color={rgb, 255:red, 74; green, 144; blue, 226 }  ,draw opacity=1 ][line width=2.25]    (209.8,75.61) .. controls (260.8,19.21) and (293.8,3.21) .. (335.8,33.21) ;
%Curve Lines [id:da41198165177587454] 
\draw [color={rgb, 255:red, 74; green, 144; blue, 226 }  ,draw opacity=1 ][line width=2.25]    (335.8,33.21) .. controls (383.8,2.21) and (429.8,32.21) .. (458.8,76.61) ;
%Curve Lines [id:da36952042602112933] 
\draw [color={rgb, 255:red, 208; green, 2; blue, 27 }  ,draw opacity=1 ][line width=2.25]    (291.8,75.61) .. controls (306.8,58.21) and (319.8,45.21) .. (335.8,33.21) ;
%Curve Lines [id:da8903726073899838] 
\draw [color={rgb, 255:red, 208; green, 2; blue, 27 }  ,draw opacity=1 ][line width=2.25]    (335.8,33.21) .. controls (338.8,35.21) and (376.8,69.21) .. (376.8,76.61) ;

% Text Node
\draw (249,85) node [anchor=north west][inner sep=0.75pt]   [align=left] {\textbf{A}};
% Text Node
\draw (327,85) node [anchor=north west][inner sep=0.75pt]   [align=left] {\textbf{B}};
% Text Node
\draw (408,85) node [anchor=north west][inner sep=0.75pt]   [align=left] {\textbf{C}};
% Text Node
\draw (149,115.4) node [anchor=north west][inner sep=0.75pt]  [font=\Large]  {$\mathrm{S( ABC)} \ +\ \mathrm{S( B)} \ \leq \ \mathrm{S( AB) \ +\ S( BC)}$};

\end{tikzpicture}

\caption{Cutting and pasting proof of Strong Subadditivity. \ The union of the blue and red curves has length $S(AB)+S(BC)$. But the blue curve ends on $ABC$, and thus has length at least $S(ABC)$, while the red curve ends on $B$, and thus has length at least $S(B).$  \label{fig:SSA_proof}}
\end{figure}
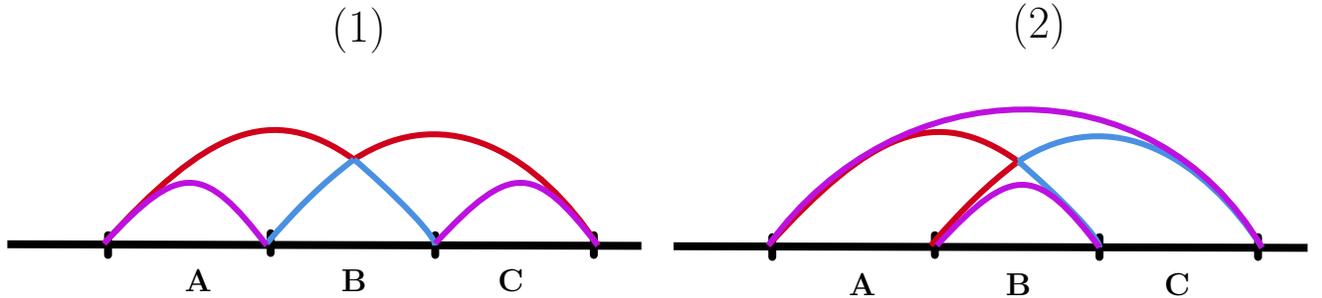

\begin{figure}[!htb]
\centering

\tikzset{every picture/.style={line width=0.75pt}} %set default line width to 0.75pt        

\begin{tikzpicture}[x=0.75pt,y=0.75pt,yscale=-1,xscale=1]

%Shape: Free Drawing [id:dp5440388512307233] 
\draw  [line width=3] [line join = round][line cap = round] (292.8,123.41) .. controls (292.8,126.75) and (292.8,130.08) .. (292.8,133.41) ;
%Straight Lines [id:da9859945009085542] 
\draw [line width=3]    (-3,128) -- (316.8,128.81) ;
%Shape: Free Drawing [id:dp34399130208036555] 
\draw  [line width=3] [line join = round][line cap = round] (47.8,123.41) .. controls (47.8,126.75) and (47.8,130.08) .. (47.8,133.41) ;
%Shape: Free Drawing [id:dp7596641009994283] 
\draw  [line width=3] [line join = round][line cap = round] (129.8,122.41) .. controls (129.8,125.75) and (129.8,129.08) .. (129.8,132.41) ;
%Shape: Free Drawing [id:dp05525798552053329] 
\draw  [line width=3] [line join = round][line cap = round] (212.8,123.41) .. controls (212.8,126.75) and (212.8,130.08) .. (212.8,133.41) ;
%Curve Lines [id:da6645787239495566] 
\draw [color={rgb, 255:red, 208; green, 2; blue, 27 }  ,draw opacity=1 ][line width=2.25]    (45.8,127.61) .. controls (96.8,71.21) and (129.8,55.21) .. (171.8,85.21) ;
%Curve Lines [id:da3723757245186776] 
\draw [color={rgb, 255:red, 208; green, 2; blue, 27 }  ,draw opacity=1 ][line width=2.25]    (171.8,85.21) .. controls (219.8,54.21) and (265.8,84.21) .. (294.8,128.61) ;
%Curve Lines [id:da28786703793601665] 
\draw [color={rgb, 255:red, 74; green, 144; blue, 226 }  ,draw opacity=1 ][line width=2.25]    (127.8,127.61) .. controls (142.8,110.21) and (155.8,97.21) .. (171.8,85.21) ;
%Curve Lines [id:da0570037478799168] 
\draw [color={rgb, 255:red, 74; green, 144; blue, 226 }  ,draw opacity=1 ][line width=2.25]    (171.8,85.21) .. controls (174.8,87.21) and (212.8,121.21) .. (212.8,128.61) ;
%Curve Lines [id:da9663422101404411] 
\draw [color={rgb, 255:red, 189; green, 16; blue, 224 }  ,draw opacity=1 ][line width=2.25]    (45.8,127.61) .. controls (65.24,106.63) and (77.88,96.39) .. (89.87,97.01) .. controls (101.32,97.61) and (112.17,108.1) .. (127.8,128.61) ;
%Curve Lines [id:da5456702376200306] 
\draw [color={rgb, 255:red, 189; green, 16; blue, 224 }  ,draw opacity=1 ][line width=2.25]    (212.8,127.61) .. controls (232.24,106.63) and (244.88,96.39) .. (256.87,97.01) .. controls (268.32,97.61) and (279.17,108.1) .. (294.8,128.61) ;
%Shape: Free Drawing [id:dp49356106719727433] 
\draw  [line width=3] [line join = round][line cap = round] (627.8,124.41) .. controls (627.8,127.75) and (627.8,131.08) .. (627.8,134.41) ;
%Straight Lines [id:da1118930466057031] 
\draw [line width=3]    (333,129) -- (652.8,129.81) ;
%Shape: Free Drawing [id:dp38385784660148703] 
\draw  [line width=3] [line join = round][line cap = round] (382.8,124.41) .. controls (382.8,127.75) and (382.8,131.08) .. (382.8,134.41) ;
%Shape: Free Drawing [id:dp9449169109714981] 
\draw  [line width=3] [line join = round][line cap = round] (464.8,123.41) .. controls (464.8,126.75) and (464.8,130.08) .. (464.8,133.41) ;
%Shape: Free Drawing [id:dp25263974957879975] 
\draw  [line width=3] [line join = round][line cap = round] (547.8,124.41) .. controls (547.8,127.75) and (547.8,131.08) .. (547.8,134.41) ;
%Curve Lines [id:da5921194734393129] 
\draw [color={rgb, 255:red, 208; green, 2; blue, 27 }  ,draw opacity=1 ][line width=2.25]    (380.8,128.61) .. controls (431.8,72.21) and (464.8,56.21) .. (506.8,86.21) ;
%Curve Lines [id:da38761850633323447] 
\draw [color={rgb, 255:red, 74; green, 144; blue, 226 }  ,draw opacity=1 ][line width=2.25]    (506.8,86.21) .. controls (554.8,55.21) and (600.8,85.21) .. (629.8,129.61) ;
%Curve Lines [id:da8678013284362807] 
\draw [color={rgb, 255:red, 208; green, 2; blue, 27 }  ,draw opacity=1 ][line width=2.25]    (462.8,128.61) .. controls (477.8,111.21) and (490.8,98.21) .. (506.8,86.21) ;
%Curve Lines [id:da6345462945231739] 
\draw [color={rgb, 255:red, 74; green, 144; blue, 226 }  ,draw opacity=1 ][line width=2.25]    (506.8,86.21) .. controls (509.8,88.21) and (547.8,122.21) .. (547.8,129.61) ;
%Curve Lines [id:da9756762411494675] 
\draw [color={rgb, 255:red, 189; green, 16; blue, 224 }  ,draw opacity=1 ][line width=2.25]    (465.8,128.61) .. controls (485.24,107.63) and (497.88,97.39) .. (509.87,98.01) .. controls (521.32,98.61) and (532.17,109.1) .. (547.8,129.61) ;
%Curve Lines [id:da3119478311124737] 
\draw [color={rgb, 255:red, 189; green, 16; blue, 224 }  ,draw opacity=1 ][line width=2.25]    (380.8,128.61) .. controls (445.8,38.43) and (574.8,35.43) .. (629.8,129.61) ;

% Text Node
\draw (85,139) node [anchor=north west][inner sep=0.75pt]   [align=left] {\textbf{A}};
% Text Node
\draw (164,139) node [anchor=north west][inner sep=0.75pt]   [align=left] {\textbf{B}};
% Text Node
\draw (243,139) node [anchor=north west][inner sep=0.75pt]   [align=left] {\textbf{C}};
% Text Node
\draw (10,171.4) node [anchor=north west][inner sep=0.75pt]  [font=\Large]  {$\mathrm{S( A) \ +\ S( B) \ +\ S( C) \ +\ S( ABC) \ \leq \ S( AB) \ +\ S( AC) \ +\ S( BC)}$};
% Text Node
\draw (420,141) node [anchor=north west][inner sep=0.75pt]   [align=left] {\textbf{A}};
% Text Node
\draw (499,141) node [anchor=north west][inner sep=0.75pt]   [align=left] {\textbf{B}};
% Text Node
\draw (579,141) node [anchor=north west][inner sep=0.75pt]   [align=left] {\textbf{C}};
% Text Node
\draw (159,7) node [anchor=north west][inner sep=0.75pt]  [font=\Large] [align=left] {(1)};
% Text Node
\draw (502,5) node [anchor=north west][inner sep=0.75pt]  [font=\Large] [align=left] {(2)};

\end{tikzpicture}

\caption{Cutting and pasting proof of MMI. The proof is divided into two cases according whether $S(AC)$ (in purple) is equal to $S(A) + S(C)$ (case (1)) or $S(ABC)+S(C)$ (case (2)); the two cases correspond to the disconnected and connected cases in Figure \ref{fig:disconnected_RT}, respectively. In both cases, the union of the blue and red curves has length $S(AB)+S(BC)$. In case (1), the length of the red curve is lower-bounded by $S(ABC)$ and the length of the blue curve is lower-bounded by $S(B)$. In case (2), the length of the red curve is lower-bounded by $S(A)$ and the length of the blue curve is lower-bounded by $S(B)$. (The holographic proof of MMI in \cite{Hayden:2011ag} combines these cases, but we have separated them here for clarity.) \label{fig:MMI_proof}}
\end{figure}

All known inequalities that define the holographic entropy cone have proofs that consist of ``cutting and pasting'' geodesics. \ As an example, we show cutting-and-pasting proofs of subadditivity, strong subadditivity, and MMI in Figures \ref{fig:SA_proof}-\ref{fig:MMI_proof}. \ We raise, as an open question, whether there are inequalities defining the holographic entropy cone that don't have cutting-and-pasting proofs of this kind.

\subsection{Entropies from Contiguous Data}\label{contig}

Given an ordering of $N$ atomic regions on a single boundary, we can easily check whether a set of regions is \emph{contiguous}, that is, whether their union consists of only a single connected component. \ For example, $AB$ and $BC$ are contiguous but (for $N>3$) $AC$ is not. \ Furthermore, the complement of a contiguous region is itself contiguous.

Hence there are $N(N-1)/2=\binom{N}{2}$ independent entropies of contiguous regions. (When $N$ is even, we only need half of the size-$N/2$ regions.) 

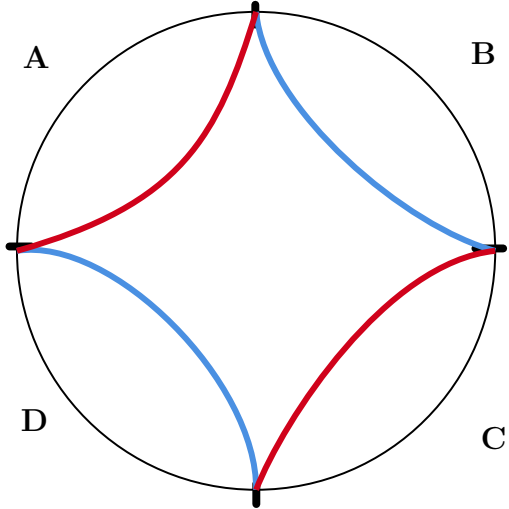
\begin{figure}[!htb]
\centering

\tikzset{every picture/.style={line width=0.75pt}} %set default line width to 0.75pt   

\begin{tikzpicture}[x=0.75pt,y=0.75pt,yscale=-1,xscale=1]

%Shape: Circle [id:dp2679266196532586] 
\draw   (208.59,145.61) .. controls (208.59,79) and (262.58,25) .. (329.19,25) .. controls (395.8,25) and (449.8,79) .. (449.8,145.61) .. controls (449.8,212.22) and (395.8,266.21) .. (329.19,266.21) .. controls (262.58,266.21) and (208.59,212.22) .. (208.59,145.61) -- cycle ;
%Shape: Free Drawing [id:dp514898679664584] 
\draw  [line width=3] [line join = round][line cap = round] (439.8,144.41) .. controls (444.47,144.41) and (449.13,144.41) .. (453.8,144.41) ;
%Shape: Free Drawing [id:dp1390339009547774] 
\draw  [line width=3] [line join = round][line cap = round] (329.3,273.41) .. controls (329.3,270.01) and (329.3,266.61) .. (329.3,263.21) ;
%Shape: Free Drawing [id:dp44398699045756085] 
\draw  [line width=3] [line join = round][line cap = round] (204.8,143.41) .. controls (208.47,143.41) and (212.13,143.41) .. (215.8,143.41) ;
%Shape: Free Drawing [id:dp6959769577174104] 
\draw  [line width=3] [line join = round][line cap = round] (328.8,21.41) .. controls (328.8,24.75) and (328.8,28.08) .. (328.8,31.41) ;
%Curve Lines [id:da06314722853616672] 
\draw [color={rgb, 255:red, 74; green, 144; blue, 226 }  ,draw opacity=1 ][line width=2.25]    (208.59,145.61) .. controls (256.8,138.21) and (329.8,212.21) .. (329.19,266.21) ;
%Curve Lines [id:da7325368330792033] 
\draw [color={rgb, 255:red, 74; green, 144; blue, 226 }  ,draw opacity=1 ][line width=2.25]    (329.19,25) .. controls (333.8,74.21) and (405.8,134.21) .. (449.8,145.61) ;
%Curve Lines [id:da6686774802745257] 
\draw [color={rgb, 255:red, 208; green, 2; blue, 27 }  ,draw opacity=1 ][line width=2.25]    (208.59,145.61) .. controls (291.8,122.21) and (309.8,88.21) .. (329.19,25) ;
%Curve Lines [id:da7064063900741919] 
\draw [color={rgb, 255:red, 208; green, 2; blue, 27 }  ,draw opacity=1 ][line width=2.25]    (329.19,266.21) .. controls (342.8,231.21) and (394.8,151.21) .. (449.8,145.61) ;

% Text Node
\draw (210,41) node [anchor=north west][inner sep=0.75pt]   [align=left] {\textbf{A}};
% Text Node
\draw (436,39) node [anchor=north west][inner sep=0.75pt]   [align=left] {\textbf{B}};
% Text Node
\draw (441,233) node [anchor=north west][inner sep=0.75pt]   [align=left] {\textbf{C}};
% Text Node
\draw (209,224) node [anchor=north west][inner sep=0.75pt]   [align=left] {\textbf{D}};

\end{tikzpicture}
\caption{Obtaining the RT surfaces of disconnected regions as unions of RT surfaces of connected regions. \ The red curve is the union of the RT surface of $A$ (with length $S(A)$) and the RT surface of $C$ (with length $S(C)$); the blue curve is the union of the RT surface of $ABC\equiv D$ (with length $S(D)$) and the RT surface of $B$ (with length $S(B)$). Both are extremal surfaces ending on $AC\equiv A\cup C$; we must therefore have $S(AC)\le \min\left(S(A)+S(C),S(B)+S(D)\right).$ For a single disconnected boundary, the validity of the RT formula implies that no further extremal surfaces exist, and the inequality is saturated. When the red curve is the RT surface, we say that the entanglement wedge of $AC$ is disconnected; when the blue curve is the RT surface, we say it is connected.  \label{fig:disconnected_RT}}
\end{figure}

One might wonder: what is so special about the \emph{contiguous} boundary regions, that could justify finding a bulk model to explain their entropies alone? \ The significance is that, as observed for example by \cite{Bao:2016rbj}:

\begin{proposition}[contiguous data suffices]
\label{contigprop}
If there is a single 1D boundary, and if the bulk geometry is topologically trivial (e.g., does not contain wormholes), then the entropies of the contiguous boundary regions determine the entropies of the non-contiguous boundary regions as well.
\end{proposition}

The proof of Proposition \ref{contigprop} is simply that, in the situation described, the RT surface of a non-contiguous region $R$ must be the union of RT surfaces of contiguous regions, or equivalently, geodesics that start and end on the boundary. \ (This is so because, if some component of $R$'s RT surface were \emph{not} a geodesic, then we could decrease the total area by replacing it with one.) \ But knowing the entropies of all the contiguous regions tells us the lengths of all boundary-anchored geodesics.

Thus, suppose $R$ is the union of $k$ contiguous boundary regions $R_1,\ldots,R_k$. \ Then to calculate the entropy $S(R)$, we ``merely'' need to solve a combinatorial optimization problem: namely, to take the minimum, over all sets of geodesics $g_1,\ldots,g_k$ whose union separates $R$ from its complement, of $l(g_1)+\cdots+l(g_k)$, where $l(g_i)$ is the length of $g_i$. \ To illustrate, in Figure \ref{fig:disconnected_RT}, we can calculate the entropy of the non-contiguous region $AC$ as
\begin{equation}
    S(AC) = \min\left\{S(A)+S(C),S(B)+S(D)\right\}\label{eq:noncontiguous},
\end{equation}
minimizing over the two possibilities for how to separate $AC$ from $BD$.

We remark that the validity of Proposition \ref{contigprop}, and (\ref{eq:noncontiguous}) in particular, is closely tied to holographic states being constrained to obey MMI (\ref{eq:MMI}). \ In particular, the two cases in Figure \ref{fig:MMI_proof} correspond to the two terms on the right-hand side of (\ref{eq:noncontiguous}). \ In a general quantum state we can make $S(AC)$ smaller than either term, for example by adding an EPR pair between $A$ and $C$, but this will violate MMI and render the state non-holographic.

In general, with $k$ contiguous boundary regions $R_1,\ldots,R_k$, there are $k!$ possibilities for geodesics separating $R = R_1 \cup \cdots \cup R_k$ from its complement. \ As observed by \cite{Bao:2016rbj}, the problem of minimizing over those $k!$ possibilities can be cast as an instance of the minimum-weight perfect matching problem, for a bipartite graph $H$ with $k$ vertices on each side. \ The left vertices of this $H$ correspond to the left endpoints of the $R_i$'s as we proceed around the circle clockwise, the right vertices correspond to their right endpoints, and the weight of an edge $(v,w)$ is just the length of the minimal geodesic connecting $v$ and $w$. \ This matching problem is well-known to be solvable in time polynomial in $k$.

Hence, under the assumptions of a single 1D boundary and trivial topology, we can represent the areas of \emph{exponentially many} RT surfaces---those of the non-contiguous regions---in terms of the areas of the $\binom{N}{2}$ RT surfaces of the contiguous regions. \ This fact is so useful that it motivates a definition: call an entropy vector $v\in \mathbb{R}_{\ge 0}^{2^{N-1}-1}$ a \emph{matching vector} if all non-contiguous entropies in it are obtained from the $\binom{N}{2}$ contiguous entropies via minimization over perfect matchings, as in the prescription above. \ A matching vector, despite its exponential length, is fully determined by polynomially many parameters.

Here we should pause to discuss a subtlety. \ As explained in Section \ref{entropineq}, the set of valid holographic entropy vectors $v\in \mathbb{R}_{\ge 0}^{2^{N-1}-1}$ forms a \emph{cone}; that is, it's closed under nonnegative linear combinations. \ The set of valid contiguous entropy vectors, $w\in \mathbb{R}_{\ge 0}^{\binom{N}{2}}$, \emph{also} forms a cone. \ By contrast, the set of matching vectors---that is, the subset of $\mathbb{R}_{\ge 0}^{2^{N-1}-1}$ that's obtainable from some contiguous entropy vector $w\in \mathbb{R}_{\ge 0}^{\binom{N}{2}}$ via the matching prescription---does \emph{not} form a cone. 

Here is an example that shows this: let $A,B,C,D$ be the $N=4$ atomic boundary
regions. \ In state $\rho$, we'll have
\begin{align}
S(A)&=S(C)=2,\nonumber \\
S(B)&=S(D)=1,\nonumber \\
S(AC) &= S(BD) = \min\{S(A)+S(C), S(B)+S(D)\} = 2,
\end{align}
with $S(AB)=S(CD)$ and $S(AD)=S(BC)$ taking any values consistent with
Strong Subadditivity. \ In state $\sigma$, we'll have
\begin{align}
S(A)&=S(C)=1,\nonumber \\
S(B)&=S(D)=2,\nonumber \\
S(AC) &= S(BD) = \min\{S(A)+S(C), S(B)+S(D)\} = 2.
\end{align}
Hence in state $\rho+\sigma$, assuming convexity we must have
\begin{align}
S(A)&=S(C)=3,\nonumber \\
S(B)&=S(D)=3,\nonumber \\
S(AC) &= S(BD) = 4.
\end{align}
But
$$\min\{S(A)+S(C), S(B)+S(D)\} = 6 > 4,$$ which gives us our contradiction.

%%%%%%%%%%%%%%%%%%%%%%%%%%%%%%%%%%%
\section{The Single 1D Boundary Case}\label{sec:single}

Having said what we could about the general DBRP in the previous section, in this section we radically specialize to the case where the $N$ atomic regions live on a single, $1$-dimensional boundary (homologous to a circle). \ Furthermore, we assume that the input data consists of entropies for \emph{contiguous} boundary regions only; Section \ref{contig} explained why this is a reasonable choice.

\subsection{Bulkless Graphs}\label{sub:bulkless}

Suppose we are given the $\binom{N}{2}$ independent contiguous entropies for $N$ atomic regions on a single boundary. \ By construction, any graph that solves the DBRP for this data must have at least $N$ vertices: the boundary vertices themselves. \ Assuming we don't care about planarity, how many additional vertices must be provided?

\begin{figure}[!htb]
\centering

\tikzset{every picture/.style={line width=0.75pt}} %set default line width to 0.75pt        

\begin{tikzpicture}[x=0.75pt,y=0.75pt,yscale=-1,xscale=1]

%Shape: Free Drawing [id:dp7994178947081909] 
\draw  [line width=3] [line join = round][line cap = round] (427.8,93.41) .. controls (432.47,93.41) and (437.13,93.41) .. (441.8,93.41) ;
%Shape: Free Drawing [id:dp1758084488012237] 
\draw  [line width=3] [line join = round][line cap = round] (443.3,192.41) .. controls (443.3,189.01) and (443.3,185.61) .. (443.3,182.21) ;
%Shape: Free Drawing [id:dp25495703559820604] 
\draw  [line width=3] [line join = round][line cap = round] (204.8,143.41) .. controls (208.47,143.41) and (212.13,143.41) .. (215.8,143.41) ;
%Shape: Free Drawing [id:dp5441003921715357] 
\draw  [line width=3] [line join = round][line cap = round] (251.8,51.41) .. controls (251.8,54.75) and (251.8,58.08) .. (251.8,61.41) ;
%Shape: Circle [id:dp7313390508841242] 
\draw  [line width=1.5]  (208.59,145.61) .. controls (208.59,79) and (262.58,25) .. (329.19,25) .. controls (395.8,25) and (449.8,79) .. (449.8,145.61) .. controls (449.8,212.22) and (395.8,266.21) .. (329.19,266.21) .. controls (262.58,266.21) and (208.59,212.22) .. (208.59,145.61) -- cycle ;
%Shape: Free Drawing [id:dp9013995075342591] 
\draw  [line width=3] [line join = round][line cap = round] (358.3,269.41) .. controls (358.3,266.01) and (358.3,262.61) .. (358.3,259.21) ;
%Shape: Free Drawing [id:dp787752605309751] 
\draw  [line width=3] [line join = round][line cap = round] (350.8,20.41) .. controls (350.8,23.75) and (350.8,27.08) .. (350.8,30.41) ;
%Shape: Free Drawing [id:dp0863495441224198] 
\draw  [line width=3] [line join = round][line cap = round] (256.8,237.41) .. controls (256.8,240.75) and (256.8,244.08) .. (256.8,247.41) ;
%Straight Lines [id:da015303718712253334] 
\draw [color={rgb, 255:red, 208; green, 2; blue, 27 }  ,draw opacity=1 ][line width=1.5]    (251.8,54.01) -- (208.59,145.61) ;
%Straight Lines [id:da6718285031758997] 
\draw [color={rgb, 255:red, 208; green, 2; blue, 27 }  ,draw opacity=1 ][line width=1.5]    (349.8,27.01) -- (251.8,54.01) ;
%Straight Lines [id:da9969768033679987] 
\draw [color={rgb, 255:red, 208; green, 2; blue, 27 }  ,draw opacity=1 ][line width=1.5]    (436.8,94.01) -- (349.8,27.01) ;
%Straight Lines [id:da8854164638697548] 
\draw [color={rgb, 255:red, 208; green, 2; blue, 27 }  ,draw opacity=1 ][line width=1.5]    (443.8,188.01) -- (436.8,94.01) ;
%Straight Lines [id:da6372093022482652] 
\draw [color={rgb, 255:red, 208; green, 2; blue, 27 }  ,draw opacity=1 ][line width=1.5]    (443.8,188.01) -- (357.8,263.01) ;
%Straight Lines [id:da8943233699115991] 
\draw [color={rgb, 255:red, 208; green, 2; blue, 27 }  ,draw opacity=1 ][line width=1.5]    (357.8,263.01) -- (257.8,242.01) ;
%Straight Lines [id:da15816041880534426] 
\draw [color={rgb, 255:red, 208; green, 2; blue, 27 }  ,draw opacity=1 ][line width=1.5]    (257.8,242.01) -- (208.59,145.61) ;
%Straight Lines [id:da2472038774555183] 
\draw [color={rgb, 255:red, 208; green, 2; blue, 27 }  ,draw opacity=1 ][line width=1.5]    (349.8,27.01) -- (213.8,144.01) ;
%Straight Lines [id:da5120072243204048] 
\draw [color={rgb, 255:red, 208; green, 2; blue, 27 }  ,draw opacity=1 ][line width=1.5]    (436.8,94.01) -- (208.59,145.61) ;
%Straight Lines [id:da5709143419922695] 
\draw [color={rgb, 255:red, 208; green, 2; blue, 27 }  ,draw opacity=1 ][line width=1.5]    (443.8,188.01) -- (213.8,144.01) ;
%Straight Lines [id:da3770021923023166] 
\draw [color={rgb, 255:red, 208; green, 2; blue, 27 }  ,draw opacity=1 ][line width=1.5]    (357.8,263.01) -- (208.59,145.61) ;
%Straight Lines [id:da5568177696967269] 
\draw [color={rgb, 255:red, 208; green, 2; blue, 27 }  ,draw opacity=1 ][line width=1.5]    (257.8,242.01) -- (251.8,54.01) ;
%Straight Lines [id:da8667551504967188] 
\draw [color={rgb, 255:red, 208; green, 2; blue, 27 }  ,draw opacity=1 ][line width=1.5]    (357.8,263.01) -- (251.8,54.01) ;
%Straight Lines [id:da38449811843194226] 
\draw [color={rgb, 255:red, 208; green, 2; blue, 27 }  ,draw opacity=1 ][line width=1.5]    (443.8,188.01) -- (251.8,54.01) ;
%Straight Lines [id:da16408284033415166] 
\draw [color={rgb, 255:red, 208; green, 2; blue, 27 }  ,draw opacity=1 ][line width=1.5]    (436.8,94.01) -- (251.8,54.01) ;
%Straight Lines [id:da3081203383964459] 
\draw [color={rgb, 255:red, 208; green, 2; blue, 27 }  ,draw opacity=1 ][line width=1.5]    (443.8,188.01) -- (349.8,27.01) ;
%Straight Lines [id:da9059660501967786] 
\draw [color={rgb, 255:red, 208; green, 2; blue, 27 }  ,draw opacity=1 ][line width=1.5]    (357.8,263.01) -- (349.8,27.01) ;
%Straight Lines [id:da44109050442923103] 
\draw [color={rgb, 255:red, 208; green, 2; blue, 27 }  ,draw opacity=1 ][line width=1.5]    (257.8,242.01) -- (349.8,27.01) ;
%Straight Lines [id:da031564016246758975] 
\draw [color={rgb, 255:red, 208; green, 2; blue, 27 }  ,draw opacity=1 ][line width=1.5]    (357.8,263.01) -- (436.8,94.01) ;
%Straight Lines [id:da06212939299241205] 
\draw [color={rgb, 255:red, 208; green, 2; blue, 27 }  ,draw opacity=1 ][line width=1.5]    (436.8,94.01) -- (257.8,242.01) ;
%Straight Lines [id:da08692014931979264] 
\draw [color={rgb, 255:red, 208; green, 2; blue, 27 }  ,draw opacity=1 ][line width=1.5]    (443.8,188.01) -- (257.8,242.01) ;
%Straight Lines [id:da011073905319051391] 
\draw [color={rgb, 255:red, 74; green, 144; blue, 226 }  ,draw opacity=1 ][line width=2.25]    (402.6,49.27) -- (223.8,202.02) ;

% Text Node
\draw (173,135) node [anchor=north west][inner sep=0.75pt]   [align=left] {\textbf{A}};
% Text Node
\draw (239,19) node [anchor=north west][inner sep=0.75pt]   [align=left] {\textbf{B}};
% Text Node
\draw (347,3) node [anchor=north west][inner sep=0.75pt]   [align=left] {\textbf{C}};
% Text Node
\draw (455,86) node [anchor=north west][inner sep=0.75pt]   [align=left] {\textbf{D}};
% Text Node
\draw (461,189) node [anchor=north west][inner sep=0.75pt]   [align=left] {\textbf{E}};
% Text Node
\draw (361,277) node [anchor=north west][inner sep=0.75pt]   [align=left] {\textbf{F}};
% Text Node
\draw (248,259) node [anchor=north west][inner sep=0.75pt]   [align=left] {\textbf{G}};

\end{tikzpicture}

\caption{The bulkless graph for $N=7$ vertices. \ We can think of the vertices as ordered around a circle and connected by all possible chords. \ Cuts are given by chords that separate the boundary vertices in a contiguous region $R$ from those in its complement $[N]-R$: for example, the blue chord is a cut between $ABC$ and $DEFG$. \label{fig:bulkless}}
\end{figure}
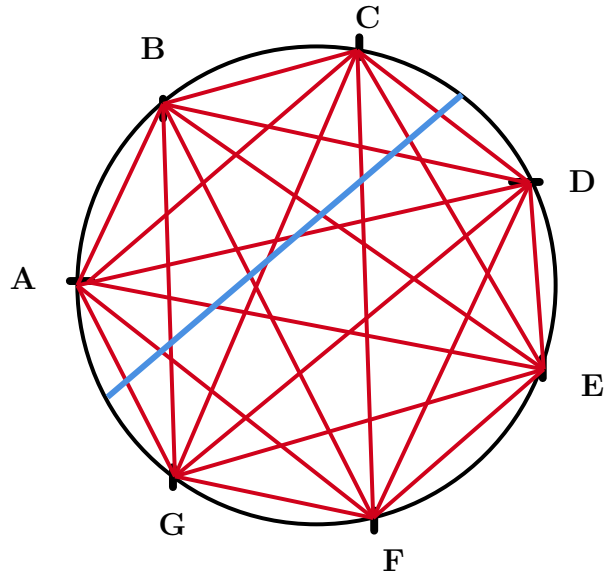

In this section, we show that the surprising answer is: zero additional vertices! \ More precisely, consider \emph{any} vector $v\in \mathbb{R}_{\ge 0}^{\binom{N}{2}}$ of contiguous entropies that obeys strong subadditivity (\ref{eq:SSA})---or, equivalently, that gives rise to a point in the quantum entropy cone. \ We show that $v$ admits an $N$-vertex graph, which we call a \emph{bulkless graph}, that solves the DBRP. The bulkless graph for $N=7$ is shown in Figure \ref{fig:bulkless}.

\begin{lemma}[bulkless graphs]
Let a \emph{bulkless graph} be a weighted undirected graph with only $N$ boundary vertices $A,B,C,\ldots$. \ Suppose we are given the $\binom{N}{2}$ entropies for the contiguous boundary regions, and suppose they obey strong subadditivity (SSA). \ Then there exists a bulkless graph (indeed, a unique such graph) that solves the DBRP for this data. \ The edge weights for this graph can be computed in $O(N^2)$ time (i.e., linear in the amount of input data). \label{lem:bulkless}
\end{lemma}
\begin{proof}
A bulkless graph is just a complete graph on $N$ vertices with appropriate edge weights. \ Our problem is how to assign nonnegative weights so that each min-cut of the graph gives the correct entropy. 

Label the weight of an edge by the two boundary vertices it connects, e.g.\ $w_{AB}$ is the weight of the edge connecting $A$ and $B$. \ For any contiguous set of vertices $L$, there is a unique min-cut separating $L$ from $[N]-L$: namely, the one that cuts all and only the edges $(i,j)$ with $i\in L$ and $j\in V-L$. \ This cut has total weight
\begin{equation}
S(L) =  \sum_{i\in L, j\in V-L} w_{ij}.\label{eq:bulkless_mincut}   
\end{equation}

When $L$ consists of more than one vertex, we must have, for any two disjoint contiguous sets $L_1, L_2$ such that $L=L_1 \cup L_2$, 
\begin{equation}
    S(L) = S(L_1) + S(L_2) - 2 \sum_{i\in L_1, j\in L_2} w_{ij}.
    \label{eq:SL}
\end{equation}
This is because the min-cut for $L$ is the sum of the weights of the edges separating $L_1$ from $[N]-L_1$ and $L_2$ from $[N]-L_2$, except that the weights of the edges connecting $L_1$ to $L_2$ (and $L_2$ to $L_1$) must be omitted.

This gives a system of linear equations that can be solved for the weights by induction in the size of $L$. \ For example, we have
\begin{align}
    S(AB) &= S(A) + S(B) - 2 w_{AB},\\
    S(BC) &= S(B) + S(C) - 2w_{BC},\label{eq:S_BC}
\end{align}
and so on. \ Rearranging, we see that 
\begin{align}
    w_{AB}&=\frac{ S(A) + S(B) - S(AB)}{2},\\
    w_{BC}&=\frac{S(B) + S(C) - S(BC)}{2},
\end{align}
and so on; all such weights are nonnegative because of Subadditivity (\ref{eq:SA}), which follows from Strong Subadditivity (\ref{eq:SSA}).

Next consider the size-$3$ contiguous regions, e.g.\ $ABC$:
\begin{align}
    S(ABC) &= S(AB) + S(C) - 2 w_{AC} - 2 w_{BC}\nonumber\\
    &= S(AB) + (S(BC) - S(B) - S(C) + 2 w_{BC}) + S(C) - 2 w_{AC} - 2 w_{BC} \nonumber\\
    &= S(AB) + S(BC) - S(B) - 2 w_{AC},
\end{align}
where in the first line we have taken $L_1=AB, L_2=C$ in (\ref{eq:SL}), and in the second line we have added and subtracted $S(BC)$ using (\ref{eq:S_BC}). \ The above can be considered as a more general form of (\ref{eq:SL}), where $L$ is divided into two \emph{overlapping} regions, $L=L_1\cup L_2:$
\begin{equation}
    S(L) = S(L_1) + S(L_2) - S(L_1 \cap L_2) - 2 \sum_{i\in L_1- L_2, j\in L_2 - L_1} w_{ij}.
    \label{eq:SL2}
\end{equation}
Formula (\ref{eq:SL2}) suffices to solve for the remaining edge weights in the graph by expanding the contiguous regions until they reach size $N/2$. \ For example, we have
\begin{align}
    w_{AC}&=\frac{S(AB) + S(BC) - S(B) - S(ABC) }{2},\\
    w_{AD}&=\frac{ S(ABC) + S(BCD) - S(BC) - S(ABCD) }{2},
\end{align}
and so forth.

All of these expressions for the weights involve only entropies of contiguous regions, which are included in the input data; and all of the weights are nonnegative by Strong Subadditivity (\ref{eq:SSA}), e.g.\ $S(ABC)+S(B)\leq S(AB)+S(BC)$.
So, since the input data obeys SSA, the $N$-vertex bulkless graph with weights given by (\ref{eq:SL2}) solves the DBRP for this data.

Finally, note that we calculated each of the $\binom{N}{2}$ edge weights as a linear combination of at most $4$ input entropies. \ This is trivial to do in $O(N^2)$ time.
\end{proof}

Of course, \emph{every} entropy vector obtained from a quantum state obeys SSA. \ So one might wonder: how, if at all, does the bulkless graph ``remember'' that the input vector $v\in \mathbb{R}^{\binom{N}{2}}_{\ge 0}$ came specifically from a \emph{holographic} state?

The answer is interesting and subtle. \ The bulkless graph indeed solves the DBRP for \emph{arbitrary} SSA-obeying contiguous data. \ Now, by relations such as (\ref{eq:noncontiguous}), the bulkless graph \emph{also} solves the DBRP when given as input a full $2^{N-1}-1$-dimensional entropy vector---but only \emph{assuming} the entropies of non-contiguous regions are given by the Bao-Chatwin-Davies formula from Section \ref{contig}. \ In other words, our construction of a bulkless graph doesn't require the contiguous boundary entropies to behave like areas. \ It doesn't even care about (e.g.) the ordering of $N$ atomic regions around the boundary, except insofar as the ordering affects SSA! \ On the other hand, as soon as we ask about the entropies of \emph{non-contiguous} boundary regions, our construction will generally return the correct answers only if it was applied to a genuine holographic state. \ More concretely, as we discussed in Section \ref{contig} above, the Bao-Chatwin-Davies formula will fail when the MMI inequality (\ref{eq:MMI}) is violated, as occurs in generic non-holographic states.

On the positive side, the bulkless graph has only $N$ vertices and $\binom{N}{2}$ edges, and can be constructed in linear time. \ On the negative side, it's extremely far from being planar or even ``geometric.'' \ However, as the next sections will show, if we use the bulkless graph as a starting point, we can easily construct planar graphs that have the same min-cut structure.

\subsection{The Chord Construction}

As Figure \ref{fig:bulkless} already suggests, a bulkless graph can be ``planarized''---made into a planar graph---by placing the $N$ vertices around a circle, drawing chords for the edges, and then creating a new vertex at each intersection of chords.

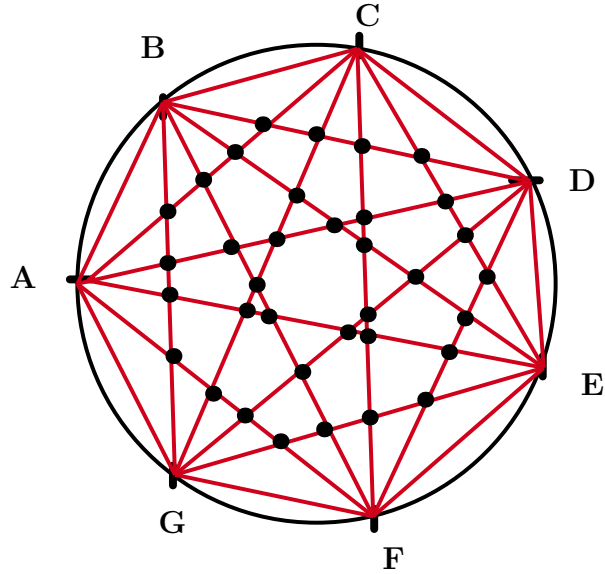
\begin{figure}[!htb]
\centering

\tikzset{every picture/.style={line width=0.75pt}} %set default line width to 0.75pt        

\begin{tikzpicture}[x=0.75pt,y=0.75pt,yscale=-1,xscale=1]

%Shape: Free Drawing [id:dp6882375617131433] 
\draw  [line width=3] [line join = round][line cap = round] (423.8,98.41) .. controls (428.47,98.41) and (433.13,98.41) .. (437.8,98.41) ;
%Shape: Free Drawing [id:dp7371228585473741] 
\draw  [line width=3] [line join = round][line cap = round] (439.3,197.41) .. controls (439.3,194.01) and (439.3,190.61) .. (439.3,187.21) ;
%Shape: Free Drawing [id:dp6275753584568688] 
\draw  [line width=3] [line join = round][line cap = round] (200.8,148.41) .. controls (204.47,148.41) and (208.13,148.41) .. (211.8,148.41) ;
%Shape: Free Drawing [id:dp3357035265880093] 
\draw  [line width=3] [line join = round][line cap = round] (247.8,56.41) .. controls (247.8,59.75) and (247.8,63.08) .. (247.8,66.41) ;
%Shape: Circle [id:dp7897898254300133] 
\draw  [line width=1.5]  (204.59,150.61) .. controls (204.59,84) and (258.58,30) .. (325.19,30) .. controls (391.8,30) and (445.8,84) .. (445.8,150.61) .. controls (445.8,217.22) and (391.8,271.21) .. (325.19,271.21) .. controls (258.58,271.21) and (204.59,217.22) .. (204.59,150.61) -- cycle ;
%Shape: Free Drawing [id:dp4018973080636865] 
\draw  [line width=3] [line join = round][line cap = round] (354.3,274.41) .. controls (354.3,271.01) and (354.3,267.61) .. (354.3,264.21) ;
%Shape: Free Drawing [id:dp1114676819495668] 
\draw  [line width=3] [line join = round][line cap = round] (346.8,25.41) .. controls (346.8,28.75) and (346.8,32.08) .. (346.8,35.41) ;
%Shape: Free Drawing [id:dp8306655505691294] 
\draw  [line width=3] [line join = round][line cap = round] (252.8,242.41) .. controls (252.8,245.75) and (252.8,249.08) .. (252.8,252.41) ;
%Straight Lines [id:da9446988375729461] 
\draw [color={rgb, 255:red, 208; green, 2; blue, 27 }  ,draw opacity=1 ][line width=1.5]    (247.8,59.01) -- (204.59,150.61) ;
%Straight Lines [id:da3999306275809371] 
\draw [color={rgb, 255:red, 208; green, 2; blue, 27 }  ,draw opacity=1 ][line width=1.5]    (345.8,32.01) -- (247.8,59.01) ;
%Straight Lines [id:da9082893725499146] 
\draw [color={rgb, 255:red, 208; green, 2; blue, 27 }  ,draw opacity=1 ][line width=1.5]    (432.8,99.01) -- (345.8,32.01) ;
%Straight Lines [id:da018406097927198495] 
\draw [color={rgb, 255:red, 208; green, 2; blue, 27 }  ,draw opacity=1 ][line width=1.5]    (439.8,193.01) -- (432.8,99.01) ;
%Straight Lines [id:da5875263783886098] 
\draw [color={rgb, 255:red, 208; green, 2; blue, 27 }  ,draw opacity=1 ][line width=1.5]    (439.8,193.01) -- (353.8,268.01) ;
%Straight Lines [id:da04061598737443428] 
\draw [color={rgb, 255:red, 208; green, 2; blue, 27 }  ,draw opacity=1 ][line width=1.5]    (353.8,268.01) -- (253.8,247.01) ;
%Straight Lines [id:da9385413435695618] 
\draw [color={rgb, 255:red, 208; green, 2; blue, 27 }  ,draw opacity=1 ][line width=1.5]    (253.8,247.01) -- (204.59,150.61) ;
%Straight Lines [id:da5712090023358489] 
\draw [color={rgb, 255:red, 208; green, 2; blue, 27 }  ,draw opacity=1 ][line width=1.5]    (345.8,32.01) -- (209.8,149.01) ;
%Straight Lines [id:da036247303347975546] 
\draw [color={rgb, 255:red, 208; green, 2; blue, 27 }  ,draw opacity=1 ][line width=1.5]    (432.8,99.01) -- (204.59,150.61) ;
%Straight Lines [id:da9420366782366998] 
\draw [color={rgb, 255:red, 208; green, 2; blue, 27 }  ,draw opacity=1 ][line width=1.5]    (439.8,193.01) -- (209.8,149.01) ;
%Straight Lines [id:da6115011446635972] 
\draw [color={rgb, 255:red, 208; green, 2; blue, 27 }  ,draw opacity=1 ][line width=1.5]    (353.8,268.01) -- (204.59,150.61) ;
%Straight Lines [id:da7840046459290848] 
\draw [color={rgb, 255:red, 208; green, 2; blue, 27 }  ,draw opacity=1 ][line width=1.5]    (253.8,247.01) -- (247.8,59.01) ;
%Straight Lines [id:da5076759857055748] 
\draw [color={rgb, 255:red, 208; green, 2; blue, 27 }  ,draw opacity=1 ][line width=1.5]    (353.8,268.01) -- (247.8,59.01) ;
%Straight Lines [id:da45053234420942134] 
\draw [color={rgb, 255:red, 208; green, 2; blue, 27 }  ,draw opacity=1 ][line width=1.5]    (439.8,193.01) -- (247.8,59.01) ;
%Straight Lines [id:da683769865738044] 
\draw [color={rgb, 255:red, 208; green, 2; blue, 27 }  ,draw opacity=1 ][line width=1.5]    (432.8,99.01) -- (247.8,59.01) ;
%Straight Lines [id:da26644332353376377] 
\draw [color={rgb, 255:red, 208; green, 2; blue, 27 }  ,draw opacity=1 ][line width=1.5]    (439.8,193.01) -- (345.8,32.01) ;
%Straight Lines [id:da652531659321643] 
\draw [color={rgb, 255:red, 208; green, 2; blue, 27 }  ,draw opacity=1 ][line width=1.5]    (353.8,268.01) -- (345.8,32.01) ;
%Straight Lines [id:da8798112878627087] 
\draw [color={rgb, 255:red, 208; green, 2; blue, 27 }  ,draw opacity=1 ][line width=1.5]    (253.8,247.01) -- (345.8,32.01) ;
%Straight Lines [id:da17295031735803978] 
\draw [color={rgb, 255:red, 208; green, 2; blue, 27 }  ,draw opacity=1 ][line width=1.5]    (353.8,268.01) -- (432.8,99.01) ;
%Straight Lines [id:da3822605221769011] 
\draw [color={rgb, 255:red, 208; green, 2; blue, 27 }  ,draw opacity=1 ][line width=1.5]    (432.8,99.01) -- (253.8,247.01) ;
%Shape: Free Drawing [id:dp017658094536647795] 
\draw  [line width=6] [line join = round][line cap = round] (250.01,114.11) .. controls (250.2,114.11) and (250.4,114.11) .. (250.59,114.11) ;
%Shape: Free Drawing [id:dp7687135004841872] 
\draw  [line width=6] [line join = round][line cap = round] (268.01,98.11) .. controls (268.2,98.11) and (268.4,98.11) .. (268.59,98.11) ;
%Shape: Free Drawing [id:dp3586297148156534] 
\draw  [line width=6] [line join = round][line cap = round] (284.01,84.11) .. controls (284.2,84.11) and (284.4,84.11) .. (284.59,84.11) ;
%Shape: Free Drawing [id:dp8082017479726367] 
\draw  [line width=6] [line join = round][line cap = round] (298.01,70.11) .. controls (298.2,70.11) and (298.4,70.11) .. (298.59,70.11) ;
%Shape: Free Drawing [id:dp7126303048394547] 
\draw  [line width=6] [line join = round][line cap = round] (325.01,75.11) .. controls (325.2,75.11) and (325.4,75.11) .. (325.59,75.11) ;
%Shape: Free Drawing [id:dp5276580390046692] 
\draw  [line width=6] [line join = round][line cap = round] (348.01,81.11) .. controls (348.2,81.11) and (348.4,81.11) .. (348.59,81.11) ;
%Shape: Free Drawing [id:dp946286904667341] 
\draw  [line width=6] [line join = round][line cap = round] (378.01,86.11) .. controls (378.2,86.11) and (378.4,86.11) .. (378.59,86.11) ;
%Shape: Free Drawing [id:dp6618361870398679] 
\draw  [line width=6] [line join = round][line cap = round] (390.01,109.11) .. controls (390.2,109.11) and (390.4,109.11) .. (390.59,109.11) ;
%Shape: Free Drawing [id:dp43720006715409077] 
\draw  [line width=6] [line join = round][line cap = round] (400.01,126.11) .. controls (400.2,126.11) and (400.4,126.11) .. (400.59,126.11) ;
%Shape: Free Drawing [id:dp6036279386963781] 
\draw  [line width=6] [line join = round][line cap = round] (411.01,147.11) .. controls (411.2,147.11) and (411.4,147.11) .. (411.59,147.11) ;
%Shape: Free Drawing [id:dp7258790965289723] 
\draw  [line width=6] [line join = round][line cap = round] (400.01,168.11) .. controls (400.2,168.11) and (400.4,168.11) .. (400.59,168.11) ;
%Shape: Free Drawing [id:dp23680349680275703] 
\draw  [line width=6] [line join = round][line cap = round] (392.01,185.11) .. controls (392.2,185.11) and (392.4,185.11) .. (392.59,185.11) ;
%Straight Lines [id:da7581581540855162] 
\draw [color={rgb, 255:red, 208; green, 2; blue, 27 }  ,draw opacity=1 ][line width=1.5]    (439.8,193.01) -- (253.8,247.01) ;
%Shape: Free Drawing [id:dp11110061640278124] 
\draw  [line width=6] [line join = round][line cap = round] (380.01,209.11) .. controls (380.2,209.11) and (380.4,209.11) .. (380.59,209.11) ;
%Shape: Free Drawing [id:dp7467031482922981] 
\draw  [line width=6] [line join = round][line cap = round] (352.01,218.11) .. controls (352.2,218.11) and (352.4,218.11) .. (352.59,218.11) ;
%Shape: Free Drawing [id:dp7424231673302779] 
\draw  [line width=6] [line join = round][line cap = round] (329.01,224.11) .. controls (329.2,224.11) and (329.4,224.11) .. (329.59,224.11) ;
%Shape: Free Drawing [id:dp48302797041879875] 
\draw  [line width=6] [line join = round][line cap = round] (307.01,230.11) .. controls (307.2,230.11) and (307.4,230.11) .. (307.59,230.11) ;
%Shape: Free Drawing [id:dp29341047058441716] 
\draw  [line width=6] [line join = round][line cap = round] (289.01,217.11) .. controls (289.2,217.11) and (289.4,217.11) .. (289.59,217.11) ;
%Shape: Free Drawing [id:dp13541177497941526] 
\draw  [line width=6] [line join = round][line cap = round] (273.01,206.11) .. controls (273.2,206.11) and (273.4,206.11) .. (273.59,206.11) ;
%Shape: Free Drawing [id:dp8316183027844266] 
\draw  [line width=6] [line join = round][line cap = round] (253.01,187.11) .. controls (253.2,187.11) and (253.4,187.11) .. (253.59,187.11) ;
%Shape: Free Drawing [id:dp17475152276217587] 
\draw  [line width=6] [line join = round][line cap = round] (250.01,140.11) .. controls (250.2,140.11) and (250.4,140.11) .. (250.59,140.11) ;
%Shape: Free Drawing [id:dp013242577012332335] 
\draw  [line width=6] [line join = round][line cap = round] (282.01,132.11) .. controls (282.2,132.11) and (282.4,132.11) .. (282.59,132.11) ;
%Shape: Free Drawing [id:dp8874123110381877] 
\draw  [line width=6] [line join = round][line cap = round] (305.01,128.11) .. controls (305.2,128.11) and (305.4,128.11) .. (305.59,128.11) ;
%Shape: Free Drawing [id:dp4025623588094849] 
\draw  [line width=6] [line join = round][line cap = round] (334.01,121.11) .. controls (334.2,121.11) and (334.4,121.11) .. (334.59,121.11) ;
%Shape: Free Drawing [id:dp4860474360796607] 
\draw  [line width=6] [line join = round][line cap = round] (349.01,117.11) .. controls (349.2,117.11) and (349.4,117.11) .. (349.59,117.11) ;
%Shape: Free Drawing [id:dp35733340932774227] 
\draw  [line width=6] [line join = round][line cap = round] (251.01,156.11) .. controls (251.2,156.11) and (251.4,156.11) .. (251.59,156.11) ;
%Shape: Free Drawing [id:dp8125952596982753] 
\draw  [line width=6] [line join = round][line cap = round] (290.01,164.11) .. controls (290.2,164.11) and (290.4,164.11) .. (290.59,164.11) ;
%Shape: Free Drawing [id:dp46698442979862476] 
\draw  [line width=6] [line join = round][line cap = round] (301.01,167.11) .. controls (301.2,167.11) and (301.4,167.11) .. (301.59,167.11) ;
%Shape: Free Drawing [id:dp056835656806613066] 
\draw  [line width=6] [line join = round][line cap = round] (341.01,175.11) .. controls (341.2,175.11) and (341.4,175.11) .. (341.59,175.11) ;
%Shape: Free Drawing [id:dp8549644391891922] 
\draw  [line width=6] [line join = round][line cap = round] (351.01,177.11) .. controls (351.2,177.11) and (351.4,177.11) .. (351.59,177.11) ;
%Shape: Free Drawing [id:dp5218671377950745] 
\draw  [line width=6] [line join = round][line cap = round] (315.01,106.11) .. controls (315.2,106.11) and (315.4,106.11) .. (315.59,106.11) ;
%Shape: Free Drawing [id:dp6126819815914948] 
\draw  [line width=6] [line join = round][line cap = round] (375.01,147.11) .. controls (375.2,147.11) and (375.4,147.11) .. (375.59,147.11) ;
%Shape: Free Drawing [id:dp2501366975080235] 
\draw  [line width=6] [line join = round][line cap = round] (318.01,195.11) .. controls (318.2,195.11) and (318.4,195.11) .. (318.59,195.11) ;
%Shape: Free Drawing [id:dp3777540016944796] 
\draw  [line width=6] [line join = round][line cap = round] (349.01,131.11) .. controls (349.2,131.11) and (349.4,131.11) .. (349.59,131.11) ;
%Shape: Free Drawing [id:dp10862029576035548] 
\draw  [line width=6] [line join = round][line cap = round] (295.01,151.11) .. controls (295.2,151.11) and (295.4,151.11) .. (295.59,151.11) ;
%Shape: Free Drawing [id:dp3956233695953384] 
\draw  [line width=6] [line join = round][line cap = round] (351.01,166.11) .. controls (351.2,166.11) and (351.4,166.11) .. (351.59,166.11) ;

% Text Node
\draw (169,140) node [anchor=north west][inner sep=0.75pt]   [align=left] {\textbf{A}};
% Text Node
\draw (235,24) node [anchor=north west][inner sep=0.75pt]   [align=left] {\textbf{B}};
% Text Node
\draw (343,8) node [anchor=north west][inner sep=0.75pt]   [align=left] {\textbf{C}};
% Text Node
\draw (451,91) node [anchor=north west][inner sep=0.75pt]   [align=left] {\textbf{D}};
% Text Node
\draw (457,194) node [anchor=north west][inner sep=0.75pt]   [align=left] {\textbf{E}};
% Text Node
\draw (357,282) node [anchor=north west][inner sep=0.75pt]   [align=left] {\textbf{F}};
% Text Node
\draw (244,264) node [anchor=north west][inner sep=0.75pt]   [align=left] {\textbf{G}};

\end{tikzpicture}

\caption{Chord construction for $7$ atomic regions. \ Draw chords for the edges of the bulkless graph, then place a new vertex wherever two chords intersect, to produce a planar graph with $O(N^4)$ vertices and edges. \ Any given edge $e$ of the bulkless graph gets broken up into many (up to $O(N^2)$) new edges, each of which inherits the same weight $w(e)$ that the parent edge $e$ had.
\label{fig:chord}}
\end{figure}

\begin{lemma}[chord construction]
Given a bulkless graph $G$ with $N$ vertices which solves the DBRP for contiguous input data, there exists a planar graph $H$ with $O(N^4)$ vertices and edges which solves the DBRP for the same data.
\end{lemma}
\begin{proof}
The proof is essentially pictorial: see Figure \ref{fig:chord}. \ With the $N$ vertices of the bulkless graph $G$ equally spaced around a circle, we first draw all possible chords between pairs of vertices. \ We then place a new ``bulk'' vertex at each intersection of chords. \ There are $\binom{N}{2}$ chords, and each pair of chords can intersect at most once, so this produces a planar graph $H$ with $O(N^4)$ vertices as well as $O(N^4)$ edges.

All that remains is to choose edge weights for $H$ so that its min-cut structure is identical to that of $G$. \ To do this, we simply let the weight of each edge $e$ in $H$, equal the weight of the edge in $G$ (i.e., the chord) that $e$ came from. \ For example, all edges along the chord from $A$ to $C$ have weight $w_{AC}$.

Because all weights in $G$ are nonnegative, so too are all weights in $H$. \ Furthermore, given any contiguous boundary region $R$, any cut in $H$ separating $R$ from $[N]-R$ must intersect every chord of the form $(v,w)$, for vertices $v\in R$ and $w\not\in R$. \ Now, let $C$ be a cut that itself arises from a chord drawn between $R$ and $[N]-R$. \ Then $C$ intersects every such chord $(v,w)$ exactly once, and is therefore a min-cut between $R$ and $[N]-R$. \ Moreover, the weight of this $C$ is just the sum of the weights of all the chords $(v,w)$ that it intersects---the same as the weight of the corresponding min-cut in $G$. \ This shows that $H$ has the same min-cut structure as $G$, and solves the DBRP for the same input data that $G$ did.
\end{proof}

Note that, just as the bulkless graph for a given number of vertices was universal, so too is the chord construction: besides the number of atomic regions $N$, all of the solution information is contained in the weights rather than in the graph structure.

The chord construction solves the DBRP, at least in the case of contiguous entropy data that comes from a single 1D boundary. \ However, it still has the defect that the number of vertices and edges is $\sim N^4$: quadratically greater than the $\sim N^2$ parameters that we started with. \ So, in the next subsection, we give a different construction of a planar bulk graph, which solves the DBRP using only $\sim N^2$ vertices and edges: the information-theoretic minimum.

\subsection{The Diamondwork Construction}

\begin{figure}[!htb]
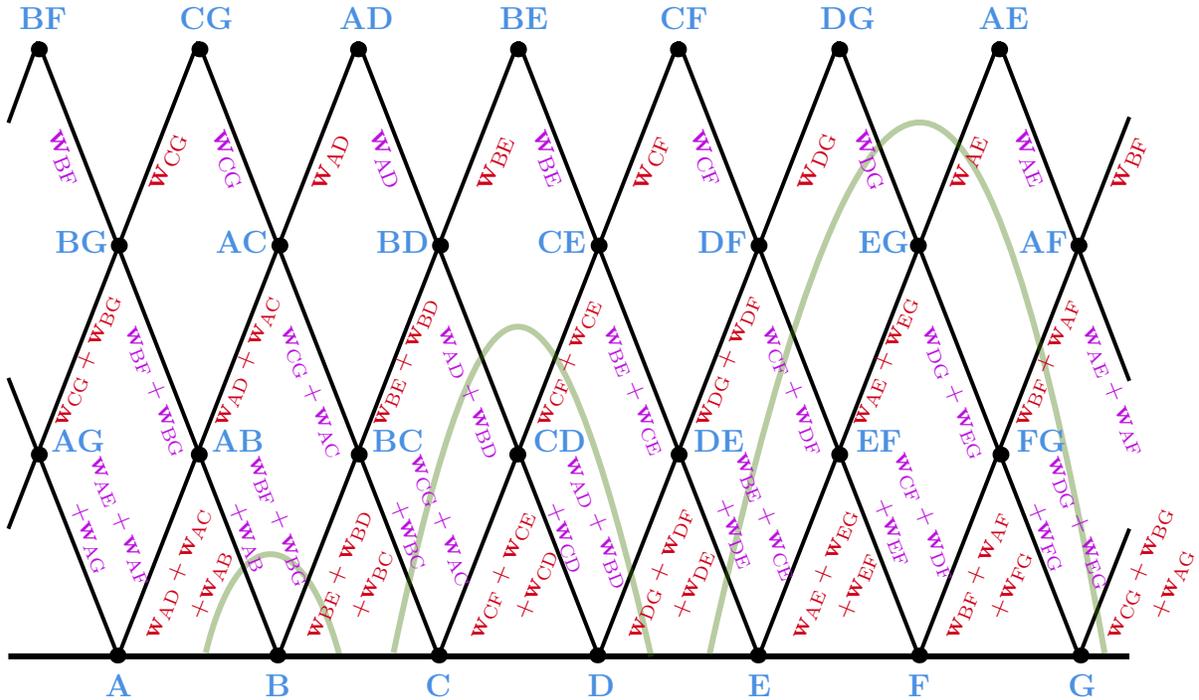

\centering

\tikzset{every picture/.style={line width=0.75pt}} %set default line width to 0.75pt        


\caption{The diamondwork construction for $N=7$ atomic regions. \ The boundary, at the bottom, is a circle, and so the left and right edges of the figure are identified. \ The $k^{th}$ layer inward from the boundary ($k^{th}$ layer upward in the figure) has one vertex for each contiguous size-$k$ region. \ Each vertex is connected inwards (up in the figure) to the vertices representing the two contiguous regions of size $k+1$ it is contained in, and outwards (down in the figure) to the two contiguous regions of size $k-1$ it contains. \ For example, the vertex $CE$, representing the size-3 contiguous region $CDE$, is connected inwards to $BE$ and $CF$, and outwards to $CD$ and $DE$. \ The graph for $N$ atomic regions has $N$ vertices per layer, and $(N+1)/2$ layers when $N$ is odd (the case where $N$ is even is slightly more complicated). \ The weight of an edge is obtained from the weights $w_{ij}$ of the bulkless graph by extending the edge inward and taking the sum of the weights corresponding to the vertices the extended edge hits. \ For example, the weight of the edge connecting $BC$ to $AC$, which extends through $AC$ to $CG$, is therefore $w_{AC}+w_{CG}$. \ As argued in the text, this means that min-cuts in the diamondwork graph have the same weights as min-cuts in the bulkless graph. \ Min-cuts for the boundary regions $B$, $CD$, and $EFG$ are shown in the Figure. \label{fig:diamondwork}}

\end{figure}

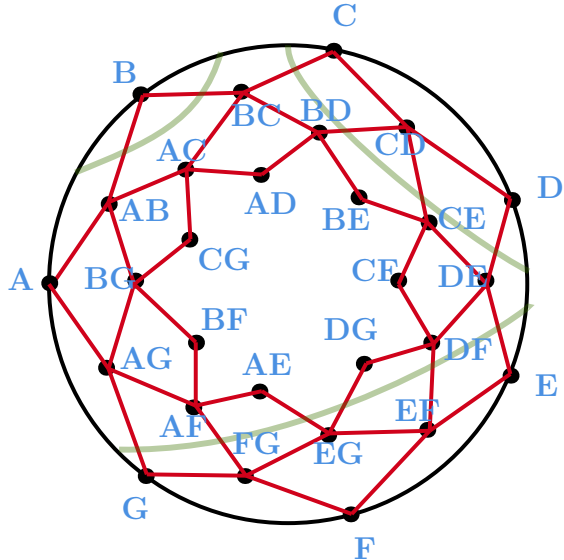
\begin{figure}[!htb]
\centering

\tikzset{every picture/.style={line width=0.75pt}} %set default line width to 0.75pt        

\begin{tikzpicture}[x=0.75pt,y=0.75pt,yscale=-1,xscale=1]
%uncomment if require: \path (0,303); %set diagram left start at 0, and has height of 303

%Shape: Circle [id:dp18095273953572089] 
\draw  [line width=1.5]  (204.59,150.61) .. controls (204.59,84) and (258.58,30) .. (325.19,30) .. controls (391.8,30) and (445.8,84) .. (445.8,150.61) .. controls (445.8,217.22) and (391.8,271.21) .. (325.19,271.21) .. controls (258.58,271.21) and (204.59,217.22) .. (204.59,150.61) -- cycle ;
%Shape: Free Drawing [id:dp881611277488342] 
\draw  [line width=6] [line join = round][line cap = round] (248.01,148.78) .. controls (248.2,148.78) and (248.4,148.78) .. (248.59,148.78) ;
%Shape: Free Drawing [id:dp21268996385411376] 
\draw  [line width=6] [line join = round][line cap = round] (273.34,92.78) .. controls (273.54,92.78) and (273.73,92.78) .. (273.92,92.78) ;
%Shape: Free Drawing [id:dp11292315419535237] 
\draw  [line width=6] [line join = round][line cap = round] (340.67,74.11) .. controls (340.87,74.11) and (341.06,74.11) .. (341.26,74.11) ;
%Shape: Free Drawing [id:dp5632031456358264] 
\draw  [line width=6] [line join = round][line cap = round] (396.01,119.45) .. controls (396.2,119.45) and (396.4,119.45) .. (396.59,119.45) ;
%Shape: Free Drawing [id:dp2688907443212407] 
\draw  [line width=6] [line join = round][line cap = round] (397.34,180.11) .. controls (397.54,180.11) and (397.73,180.11) .. (397.92,180.11) ;
%Shape: Free Drawing [id:dp06595174471002463] 
\draw  [line width=6] [line join = round][line cap = round] (345.34,226.78) .. controls (345.54,226.78) and (345.73,226.78) .. (345.92,226.78) ;
%Shape: Free Drawing [id:dp3150122036046157] 
\draw  [line width=6] [line join = round][line cap = round] (277.34,212.78) .. controls (277.54,212.78) and (277.73,212.78) .. (277.92,212.78) ;
%Shape: Free Drawing [id:dp6429939685969615] 
\draw  [line width=6] [line join = round][line cap = round] (234.67,110.11) .. controls (234.87,110.11) and (235.06,110.11) .. (235.26,110.11) ;
%Shape: Free Drawing [id:dp3232749196616653] 
\draw  [line width=6] [line join = round][line cap = round] (301.34,53.45) .. controls (301.54,53.45) and (301.73,53.45) .. (301.92,53.45) ;
%Shape: Free Drawing [id:dp42064304990896195] 
\draw  [line width=6] [line join = round][line cap = round] (384.67,71.45) .. controls (384.87,71.45) and (385.06,71.45) .. (385.26,71.45) ;
%Shape: Free Drawing [id:dp17755371971534895] 
\draw  [line width=6] [line join = round][line cap = round] (424.67,148.78) .. controls (424.87,148.78) and (425.06,148.78) .. (425.26,148.78) ;
%Shape: Free Drawing [id:dp8402732511594035] 
\draw  [line width=6] [line join = round][line cap = round] (395.34,224.11) .. controls (395.54,224.11) and (395.73,224.11) .. (395.92,224.11) ;
%Shape: Free Drawing [id:dp7912691268218033] 
\draw  [line width=6] [line join = round][line cap = round] (303.34,247.45) .. controls (303.54,247.45) and (303.73,247.45) .. (303.92,247.45) ;
%Shape: Free Drawing [id:dp5523209155622533] 
\draw  [line width=6] [line join = round][line cap = round] (233.34,192.78) .. controls (233.54,192.78) and (233.73,192.78) .. (233.92,192.78) ;
%Shape: Free Drawing [id:dp0723110390367998] 
\draw  [line width=6] [line join = round][line cap = round] (275.34,128.11) .. controls (275.54,128.11) and (275.73,128.11) .. (275.92,128.11) ;
%Shape: Free Drawing [id:dp3821570311005573] 
\draw  [line width=6] [line join = round][line cap = round] (311.34,95.45) .. controls (311.54,95.45) and (311.73,95.45) .. (311.92,95.45) ;
%Shape: Free Drawing [id:dp7870006258683737] 
\draw  [line width=6] [line join = round][line cap = round] (360.67,106.78) .. controls (360.87,106.78) and (361.06,106.78) .. (361.26,106.78) ;
%Shape: Free Drawing [id:dp9066845295433439] 
\draw  [line width=6] [line join = round][line cap = round] (380.67,148.78) .. controls (380.87,148.78) and (381.06,148.78) .. (381.26,148.78) ;
%Shape: Free Drawing [id:dp6339056128209413] 
\draw  [line width=6] [line join = round][line cap = round] (363.34,190.78) .. controls (363.54,190.78) and (363.73,190.78) .. (363.92,190.78) ;
%Shape: Free Drawing [id:dp7779821332273305] 
\draw  [line width=6] [line join = round][line cap = round] (310.67,204.78) .. controls (310.87,204.78) and (311.06,204.78) .. (311.26,204.78) ;
%Shape: Free Drawing [id:dp6793782904122072] 
\draw  [line width=6] [line join = round][line cap = round] (278.67,180.11) .. controls (278.87,180.11) and (279.06,180.11) .. (279.26,180.11) ;
%Straight Lines [id:da5177055666269743] 
\draw [color={rgb, 255:red, 208; green, 2; blue, 27 }  ,draw opacity=1 ][line width=1.5]    (207.56,148.87) -- (234.6,109.27) ;
%Shape: Free Drawing [id:dp6558731211883224] 
\draw  [line width=6] [line join = round][line cap = round] (204.67,150.11) .. controls (204.87,150.11) and (205.06,150.11) .. (205.26,150.11) ;
%Shape: Free Drawing [id:dp946593284556297] 
\draw  [line width=6] [line join = round][line cap = round] (250.67,54.78) .. controls (250.87,54.78) and (251.06,54.78) .. (251.26,54.78) ;
%Shape: Free Drawing [id:dp8654345362071698] 
\draw  [line width=6] [line join = round][line cap = round] (348.01,32.78) .. controls (348.2,32.78) and (348.4,32.78) .. (348.59,32.78) ;
%Shape: Free Drawing [id:dp670999665617013] 
\draw  [line width=6] [line join = round][line cap = round] (438.01,108.11) .. controls (438.2,108.11) and (438.4,108.11) .. (438.59,108.11) ;
%Shape: Free Drawing [id:dp17934325613451185] 
\draw  [line width=6] [line join = round][line cap = round] (437.34,196.78) .. controls (437.54,196.78) and (437.73,196.78) .. (437.92,196.78) ;
%Shape: Free Drawing [id:dp6071518392258288] 
\draw  [line width=6] [line join = round][line cap = round] (356.67,266.78) .. controls (356.87,266.78) and (357.06,266.78) .. (357.26,266.78) ;
%Shape: Free Drawing [id:dp19720017450373217] 
\draw  [line width=6] [line join = round][line cap = round] (253.34,247.45) .. controls (253.54,247.45) and (253.73,247.45) .. (253.92,247.45) ;
%Straight Lines [id:da5908800225462207] 
\draw [color={rgb, 255:red, 208; green, 2; blue, 27 }  ,draw opacity=1 ][line width=1.5]    (234.6,193.27) -- (204.59,150.61) ;
%Straight Lines [id:da32453778443515313] 
\draw [color={rgb, 255:red, 208; green, 2; blue, 27 }  ,draw opacity=1 ][line width=1.5]    (234.6,109.27) -- (251.93,55.27) ;
%Straight Lines [id:da21587042224925135] 
\draw [color={rgb, 255:red, 208; green, 2; blue, 27 }  ,draw opacity=1 ][line width=1.5]    (302.6,53.93) -- (251.93,55.27) ;
%Straight Lines [id:da5419044184901995] 
\draw [color={rgb, 255:red, 208; green, 2; blue, 27 }  ,draw opacity=1 ][line width=1.5]    (302.6,53.93) -- (347.93,33.27) ;
%Straight Lines [id:da30991710736536104] 
\draw [color={rgb, 255:red, 208; green, 2; blue, 27 }  ,draw opacity=1 ][line width=1.5]    (385.27,71.27) -- (347.93,33.27) ;
%Straight Lines [id:da4066221653819926] 
\draw [color={rgb, 255:red, 208; green, 2; blue, 27 }  ,draw opacity=1 ][line width=1.5]    (385.27,71.27) -- (437.27,107.27) ;
%Straight Lines [id:da18904677969873984] 
\draw [color={rgb, 255:red, 208; green, 2; blue, 27 }  ,draw opacity=1 ][line width=1.5]    (425.43,150.61) -- (437.27,107.27) ;
%Straight Lines [id:da9098527862686712] 
\draw [color={rgb, 255:red, 208; green, 2; blue, 27 }  ,draw opacity=1 ][line width=1.5]    (425.43,150.61) -- (437.27,195.93) ;
%Straight Lines [id:da030560372625337173] 
\draw [color={rgb, 255:red, 208; green, 2; blue, 27 }  ,draw opacity=1 ][line width=1.5]    (396.6,224.6) -- (437.27,195.93) ;
%Straight Lines [id:da9592374874779703] 
\draw [color={rgb, 255:red, 208; green, 2; blue, 27 }  ,draw opacity=1 ][line width=1.5]    (396.6,224.6) -- (357.27,266.6) ;
%Straight Lines [id:da7053660615747519] 
\draw [color={rgb, 255:red, 208; green, 2; blue, 27 }  ,draw opacity=1 ][line width=1.5]    (303.93,247.27) -- (357.27,266.6) ;
%Straight Lines [id:da6201373371820587] 
\draw [color={rgb, 255:red, 208; green, 2; blue, 27 }  ,draw opacity=1 ][line width=1.5]    (253.93,246.6) -- (303.93,247.27) ;
%Straight Lines [id:da3146733874612393] 
\draw [color={rgb, 255:red, 208; green, 2; blue, 27 }  ,draw opacity=1 ][line width=1.5]    (253.93,246.6) -- (234.6,193.27) ;
%Straight Lines [id:da04036788765544186] 
\draw [color={rgb, 255:red, 208; green, 2; blue, 27 }  ,draw opacity=1 ][line width=1.5]    (273.93,92.6) -- (302.6,53.93) ;
%Straight Lines [id:da8752674199256312] 
\draw [color={rgb, 255:red, 208; green, 2; blue, 27 }  ,draw opacity=1 ][line width=1.5]    (339.93,73.93) -- (302.6,53.93) ;
%Straight Lines [id:da6650753411368273] 
\draw [color={rgb, 255:red, 208; green, 2; blue, 27 }  ,draw opacity=1 ][line width=1.5]    (385.27,71.27) -- (339.93,73.93) ;
%Straight Lines [id:da17487501071181133] 
\draw [color={rgb, 255:red, 208; green, 2; blue, 27 }  ,draw opacity=1 ][line width=1.5]    (425.43,150.61) -- (395.27,119.27) ;
%Straight Lines [id:da4547849419927448] 
\draw [color={rgb, 255:red, 208; green, 2; blue, 27 }  ,draw opacity=1 ][line width=1.5]    (385.27,71.27) -- (395.27,119.27) ;
%Straight Lines [id:da9740392457996552] 
\draw [color={rgb, 255:red, 208; green, 2; blue, 27 }  ,draw opacity=1 ][line width=1.5]    (425.43,150.61) -- (398.6,180.6) ;
%Straight Lines [id:da5697815399778041] 
\draw [color={rgb, 255:red, 208; green, 2; blue, 27 }  ,draw opacity=1 ][line width=1.5]    (396.6,224.6) -- (398.6,180.6) ;
%Straight Lines [id:da07209286304507545] 
\draw [color={rgb, 255:red, 208; green, 2; blue, 27 }  ,draw opacity=1 ][line width=1.5]    (396.6,224.6) -- (345.93,225.93) ;
%Straight Lines [id:da17889635799471382] 
\draw [color={rgb, 255:red, 208; green, 2; blue, 27 }  ,draw opacity=1 ][line width=1.5]    (303.93,247.27) -- (345.93,225.93) ;
%Straight Lines [id:da0018405166170334386] 
\draw [color={rgb, 255:red, 208; green, 2; blue, 27 }  ,draw opacity=1 ][line width=1.5]    (303.93,247.27) -- (278.6,211.93) ;
%Straight Lines [id:da2840673267558613] 
\draw [color={rgb, 255:red, 208; green, 2; blue, 27 }  ,draw opacity=1 ][line width=1.5]    (234.6,193.27) -- (278.6,211.93) ;
%Straight Lines [id:da17848496054191632] 
\draw [color={rgb, 255:red, 208; green, 2; blue, 27 }  ,draw opacity=1 ][line width=1.5]    (247.52,150.61) -- (234.6,193.27) ;
%Straight Lines [id:da6456644403543281] 
\draw [color={rgb, 255:red, 208; green, 2; blue, 27 }  ,draw opacity=1 ][line width=1.5]    (247.52,150.61) -- (234.6,109.27) ;
%Straight Lines [id:da5704077107282621] 
\draw [color={rgb, 255:red, 208; green, 2; blue, 27 }  ,draw opacity=1 ][line width=1.5]    (273.93,92.6) -- (234.6,109.27) ;
%Straight Lines [id:da8494324266380102] 
\draw [color={rgb, 255:red, 208; green, 2; blue, 27 }  ,draw opacity=1 ][line width=1.5]    (311.93,95.27) -- (273.93,92.6) ;
%Straight Lines [id:da44080453217589044] 
\draw [color={rgb, 255:red, 208; green, 2; blue, 27 }  ,draw opacity=1 ][line width=1.5]    (275.93,127.93) -- (273.93,92.6) ;
%Straight Lines [id:da8882034633197378] 
\draw [color={rgb, 255:red, 208; green, 2; blue, 27 }  ,draw opacity=1 ][line width=1.5]    (275.93,127.93) -- (247.52,150.61) ;
%Straight Lines [id:da8837186880926211] 
\draw [color={rgb, 255:red, 208; green, 2; blue, 27 }  ,draw opacity=1 ][line width=1.5]    (278.6,179.93) -- (247.52,150.61) ;
%Straight Lines [id:da01783021182210276] 
\draw [color={rgb, 255:red, 208; green, 2; blue, 27 }  ,draw opacity=1 ][line width=1.5]    (278.6,179.93) -- (278.6,211.93) ;
%Straight Lines [id:da8663639107930778] 
\draw [color={rgb, 255:red, 208; green, 2; blue, 27 }  ,draw opacity=1 ][line width=1.5]    (311.93,95.27) -- (339.93,73.93) ;
%Straight Lines [id:da7821288436258658] 
\draw [color={rgb, 255:red, 208; green, 2; blue, 27 }  ,draw opacity=1 ][line width=1.5]    (339.93,73.93) -- (362.6,107.93) ;
%Straight Lines [id:da34066204106014575] 
\draw [color={rgb, 255:red, 208; green, 2; blue, 27 }  ,draw opacity=1 ][line width=1.5]    (362.6,107.93) -- (395.27,119.27) ;
%Straight Lines [id:da423856550853001] 
\draw [color={rgb, 255:red, 208; green, 2; blue, 27 }  ,draw opacity=1 ][line width=1.5]    (395.27,119.27) -- (380.86,150.61) ;
%Straight Lines [id:da6485781307790148] 
\draw [color={rgb, 255:red, 208; green, 2; blue, 27 }  ,draw opacity=1 ][line width=1.5]    (398.6,180.6) -- (380.86,150.61) ;
%Straight Lines [id:da10508146627769244] 
\draw [color={rgb, 255:red, 208; green, 2; blue, 27 }  ,draw opacity=1 ][line width=1.5]    (398.6,180.6) -- (364.6,190.6) ;
%Straight Lines [id:da1406416599120266] 
\draw [color={rgb, 255:red, 208; green, 2; blue, 27 }  ,draw opacity=1 ][line width=1.5]    (345.93,225.93) -- (364.6,190.6) ;
%Straight Lines [id:da9778167959171589] 
\draw [color={rgb, 255:red, 208; green, 2; blue, 27 }  ,draw opacity=1 ][line width=1.5]    (345.93,225.93) -- (311.93,204.6) ;
%Straight Lines [id:da5945151916454618] 
\draw [color={rgb, 255:red, 208; green, 2; blue, 27 }  ,draw opacity=1 ][line width=1.5]    (278.6,211.93) -- (311.93,204.6) ;
%Curve Lines [id:da12391881625823231] 
\draw [color={rgb, 255:red, 65; green, 117; blue, 5 }  ,draw opacity=0.37 ][line width=2.25]    (239.93,234.08) .. controls (349.93,236.08) and (454.6,157.41) .. (445.93,162.08) ;
%Curve Lines [id:da4651898862698931] 
\draw [color={rgb, 255:red, 65; green, 117; blue, 5 }  ,draw opacity=0.37 ][line width=2.25]    (325.19,30) .. controls (323.27,58.21) and (419.27,130.88) .. (446.6,144.08) ;
%Curve Lines [id:da8548901869660095] 
\draw [color={rgb, 255:red, 65; green, 117; blue, 5 }  ,draw opacity=0.37 ][line width=2.25]    (217.93,94.21) .. controls (253.93,79.54) and (281.27,68.88) .. (291.27,33.54) ;

% Text Node
\draw (182.23,141) node [anchor=north west][inner sep=0.75pt]  [color={rgb, 255:red, 74; green, 144; blue, 226 }  ,opacity=1 ] [align=left] {\textbf{A}};
% Text Node
\draw (235,36.31) node [anchor=north west][inner sep=0.75pt]  [color={rgb, 255:red, 74; green, 144; blue, 226 }  ,opacity=1 ] [align=left] {\textbf{B}};
% Text Node
\draw (345.77,7.67) node [anchor=north west][inner sep=0.75pt]  [color={rgb, 255:red, 74; green, 144; blue, 226 }  ,opacity=1 ] [align=left] {\textbf{C}};
% Text Node
\draw (449.21,94.67) node [anchor=north west][inner sep=0.75pt]  [color={rgb, 255:red, 74; green, 144; blue, 226 }  ,opacity=1 ] [align=left] {\textbf{D}};
% Text Node
\draw (448.08,194.67) node [anchor=north west][inner sep=0.75pt]  [color={rgb, 255:red, 74; green, 144; blue, 226 }  ,opacity=1 ] [align=left] {\textbf{E}};
% Text Node
\draw (356.77,276.67) node [anchor=north west][inner sep=0.75pt]  [color={rgb, 255:red, 74; green, 144; blue, 226 }  ,opacity=1 ] [align=left] {\textbf{F}};
% Text Node
\draw (239.6,256.67) node [anchor=north west][inner sep=0.75pt]  [color={rgb, 255:red, 74; green, 144; blue, 226 }  ,opacity=1 ] [align=left] {\textbf{G}};
% Text Node
\draw (238.03,106.38) node [anchor=north west][inner sep=0.75pt]  [color={rgb, 255:red, 74; green, 144; blue, 226 }  ,opacity=1 ] [align=left] {\textbf{AB}};
% Text Node
\draw (294.8,58.38) node [anchor=north west][inner sep=0.75pt]  [color={rgb, 255:red, 74; green, 144; blue, 226 }  ,opacity=1 ] [align=left] {\textbf{BC}};
% Text Node
\draw (366.9,73.04) node [anchor=north west][inner sep=0.75pt]  [color={rgb, 255:red, 74; green, 144; blue, 226 }  ,opacity=1 ] [align=left] {\textbf{CD}};
% Text Node
\draw (399.03,139.71) node [anchor=north west][inner sep=0.75pt]  [color={rgb, 255:red, 74; green, 144; blue, 226 }  ,opacity=1 ] [align=left] {\textbf{DE}};
% Text Node
\draw (377.39,207.04) node [anchor=north west][inner sep=0.75pt]  [color={rgb, 255:red, 74; green, 144; blue, 226 }  ,opacity=1 ] [align=left] {\textbf{EF}};
% Text Node
\draw (295.47,223.71) node [anchor=north west][inner sep=0.75pt]  [color={rgb, 255:red, 74; green, 144; blue, 226 }  ,opacity=1 ] [align=left] {\textbf{FG}};
% Text Node
\draw (238.67,181.01) node [anchor=north west][inner sep=0.75pt]  [color={rgb, 255:red, 74; green, 144; blue, 226 }  ,opacity=1 ] [align=left] {\textbf{AG}};
% Text Node
\draw (256.75,75.67) node [anchor=north west][inner sep=0.75pt]  [color={rgb, 255:red, 74; green, 144; blue, 226 }  ,opacity=1 ] [align=left] {\textbf{AC}};
% Text Node
\draw (329.52,55.67) node [anchor=north west][inner sep=0.75pt]  [color={rgb, 255:red, 74; green, 144; blue, 226 }  ,opacity=1 ] [align=left] {\textbf{BD}};
% Text Node
\draw (398.88,110.34) node [anchor=north west][inner sep=0.75pt]  [color={rgb, 255:red, 74; green, 144; blue, 226 }  ,opacity=1 ] [align=left] {\textbf{CE}};
% Text Node
\draw (402.08,176.34) node [anchor=north west][inner sep=0.75pt]  [color={rgb, 255:red, 74; green, 144; blue, 226 }  ,opacity=1 ] [align=left] {\textbf{DF}};
% Text Node
\draw (336.17,229.01) node [anchor=north west][inner sep=0.75pt]  [color={rgb, 255:red, 74; green, 144; blue, 226 }  ,opacity=1 ] [align=left] {\textbf{EG}};
% Text Node
\draw (258.29,215.01) node [anchor=north west][inner sep=0.75pt]  [color={rgb, 255:red, 74; green, 144; blue, 226 }  ,opacity=1 ] [align=left] {\textbf{AF}};
% Text Node
\draw (220.72,140.38) node [anchor=north west][inner sep=0.75pt]  [color={rgb, 255:red, 74; green, 144; blue, 226 }  ,opacity=1 ] [align=left] {\textbf{BG}};
% Text Node
\draw (277.93,130.93) node [anchor=north west][inner sep=0.75pt]  [color={rgb, 255:red, 74; green, 144; blue, 226 }  ,opacity=1 ] [align=left] {\textbf{CG}};
% Text Node
\draw (301.13,102.47) node [anchor=north west][inner sep=0.75pt]  [color={rgb, 255:red, 74; green, 144; blue, 226 }  ,opacity=1 ] [align=left] {\textbf{AD}};
% Text Node
\draw (279.61,161.17) node [anchor=north west][inner sep=0.75pt]  [color={rgb, 255:red, 74; green, 144; blue, 226 }  ,opacity=1 ] [align=left] {\textbf{BF}};
% Text Node
\draw (340.49,110.47) node [anchor=north west][inner sep=0.75pt]  [color={rgb, 255:red, 74; green, 144; blue, 226 }  ,opacity=1 ] [align=left] {\textbf{BE}};
% Text Node
\draw (355.36,136.47) node [anchor=north west][inner sep=0.75pt]  [color={rgb, 255:red, 74; green, 144; blue, 226 }  ,opacity=1 ] [align=left] {\textbf{CF}};
% Text Node
\draw (341.52,166.47) node [anchor=north west][inner sep=0.75pt]  [color={rgb, 255:red, 74; green, 144; blue, 226 }  ,opacity=1 ] [align=left] {\textbf{DG}};
% Text Node
\draw (300.14,183.8) node [anchor=north west][inner sep=0.75pt]  [color={rgb, 255:red, 74; green, 144; blue, 226 }  ,opacity=1 ] [align=left] {\textbf{AE}};

\end{tikzpicture}
\caption{The diamondwork construction for $N=7$ atomic regions, placed on a circular boundary. \ The same min-cuts as in Figure \ref{fig:diamondwork} are shown. \label{fig:diamondwork_circle}}
\end{figure}

Our final construction of a universal planar bulk graph yields the following:

\begin{theorem}[diamondwork graph]
Suppose we are given the $\binom{N}{2}$ entropies for the contiguous boundary regions of a single boundary with $N$ vertices, and suppose these entropies obey SSA. \ There exists a planar graph which solves the $DBRP$ for this input data with $N^2 / 2 +O(N)$ vertices and $N^2 +O(N)$ edges. \ Furthermore, for each $N$ this graph is universal: only the edge weights depend on the input data, not the graph itself or the min-cuts. \ The edge weights can be computed in $O(N^2)$ time, which is linear in the amount of input data. \label{thm:diamondwork}
\end{theorem}
\begin{proof}We will prove Theorem \ref{thm:diamondwork} in three steps. \ We first describe how to construct the diamondwork graph and assign weights to its edges in terms of the weights of the bulkless graph (Lemma \ref{lem:bulkless}) constructed from the same input data. \ We next show that for every contiguous region $L$, there exists a cut in the diamondwork graph with weight $S(L)$. \ Finally, we show that these cuts are min-cuts, and hence that the diamondwork graph solves the DBRP.

Throughout, we will refer extensively to Figures \ref{fig:diamondwork} and \ref{fig:diamondwork_circle}, which depict the diamondwork graph for $N=7$. \ We will therefore work directly with the case of $N$ odd, and only sketch the modifications required for the $N$ even case. \ It will sometimes be convenient to refer to atomic regions by numbers rather than Roman letters, e.g.\ $N$ denotes the last atomic region.

First we define the diamondwork graph. \ As shown in Figures \ref{fig:diamondwork} and \ref{fig:diamondwork_circle}, the graph contains one vertex for each contiguous boundary region of size at most $N/2$. \  The vertices are arranged in concentric layers: on the boundary, we have one vertex for each atomic boundary region. \ Then, one layer inward, there's a vertex for each contiguous boundary region of size $2$, then a vertex for each region of size $3$, and so on as we go deeper into the bulk. \ Each layer is connected to the layers before and after it in a diamondlike pattern (hence the name).

It remains to assign the edge weights. \ For convenience, we will write weights in terms of the weights $w_{ij}$ in the bulkless graph (Lemma \ref{lem:bulkless}), which in turn can be written in terms of the contiguous entropies via Eq.\ (\ref{eq:SL2}). \ The rule for assigning edge weights is shown pictorially in Figure \ref{fig:diamondwork}, and is this: the weight of any edge $e$, in the diamondwork graph, is the \emph{sum} of the bulkless weights $w_{ij}$, over all $(i,j)$ that correspond to the vertices that one reaches when one continues along $e$ toward the center of the diamondwork. \ For example, the weight of the edge connecting $A$ to $AB$ is $w_{AB} + w_{AC} + \cdots + w_{A,(N+1)/2}$. \ Since the $w_{ij}$'s are nonnegative, clearly these weights are nonnegative as well. 

Equivalently, one can think about the diamondwork graph as a sum of many superimposed triangles $T_{ij}$, one for each pair of boundary vertices $i$ and $j$. \ The three vertices of $T_{ij}$ are the boundary vertex $i$, the boundary vertex $j$, and the bulk vertex $ij$. \ The weight of any edge $e$ is then just the sum of $w_{ij}$, over all the triangles $T_{ij}$ to which $e$ belongs.

Next we show that, for every contiguous boundary region $L$, there exists a cut in the diamondwork graph of weight $S(L)$---the same as in the bulkless graph.

Recall (Eq.\ (\ref{eq:bulkless_mincut}) that in the bulkless graph, for every contiguous boundary region $L$, we had
\begin{equation}
S(L) =  \sum_{i\in L, j\in [N]-L} w_{ij}.\label{eq:bulkless_mincut_again}
\end{equation}
Assume for simplicity that $i<j$. \ Then it can be seen pictorially (Figure \ref{fig:diamondwork}) that the cut $C_{ij}$ that starts to the left of $i$, heads upwards and rightwards, cutting only left-directed edges until it cuts an edge connected to $ij$, then heads downwards and rightwards, cutting only right-directed edges, until it ends to the right of $j$, has precisely this total weight. 

We can prove this as follows. \ We've seen that, for every edge $e$, the weight of $e$ is the sum of $w_{ij}$, over all triangles $T_{ij}=(i,j,ij)$ that contain $e$. \ But this means that the weight of any cut $C$ can be thought of as the sum of $w_{ij}$ over all triangles $T_{ij}$ that $C$ intersects---provided that $C$ intersects each triangle at most once. \ But the cut $C_{ij}$ defined above does this, because both the cut and the triangles span at most $N/2$ boundary vertices each, so there can be no ``wraparound'' effects, where a cut intersects the same triangle twice.

Indeed, the cut $C_{ij}$ is \emph{precisely} the unique cut that passes exactly once through every triangle with one vertex in $L$ and one vertex in $[N]-L$. \ Hence, by Eq.\ (\ref{eq:bulkless_mincut_again}), its weight is exactly $S(L)$.

When $N$ is even, to ensure the above we need to modify the diamondwork construction slightly, as follows. \ We have $N/2$ ``normal'' layers, and then one innermost layer, with vertex labels of the form $k,k+N/2$. \ A triangle $T_{k,k+N/2}$ with a vertex in the innermost layer contributes a weight of $\frac{1}{2} w_{k,k+N/2}$ to each edge $e$ that it contains---so, half of a ``normal'' triangle's contribution. \ Observe that a cut $C_{ij}$, as defined above, will intersect a triangle of the form $T_{k,k+N/2}$ twice if it intersects it at all, and will pick up half of the $w_{k,k+N/2}$ contribution at each intersection. \ We thereby preserve the property that the weight of $C_{ij}$ equals the sum $S(L)$ of the appropriate weights from the bulkless graph.

Lastly, we show that these cuts $C_{ij}$, of weight $S(L)$, are in fact min-cuts of the diamondwork graph. \ We argued before that any cut separating $L$ from $[N]-L$ must pass at least once through every triangle with one vertex in $L$ and one vertex in $[N]-L$. \ The cut $C_{ij}$ passes \emph{exactly} once through every such triangle, and passes through no other triangles, so the result is immediate.

As a final observation, we can calculate the relevant sums of $w_{ij}$'s by starting with the innermost edges of the diamondwork graph and working our way toward the boundary, maintaining ``running totals.'' \ This yields an $O(N^2)$-time algorithm, which is linear in the amount of input data. \ Since we observed in Lemma \ref{lem:bulkless} that the bulkless weights $w_{ij}$ are themselves also computable in $O(N^2)$ time given the input data, the overall running time is linear.
\end{proof}

To build intuition, we can consider deforming the cut $C_{ij}$ either further into the bulk or closer to the boundary. \ We can see pictorially that both deformations increase the weight. \ For example, in Figure \ref{fig:diamondwork}, deforming the cut for $CD$ so it passes above $BD$ results in picking up additional weight $2w_{BD}$, while deforming the cut so it passes below $CD$ adds additional weight $2w_{CD}$. \ In either case, the cut has been forced to enter, and then leave again, a triangle that does not actually separate vertices in $L$ from vertices in $[N]-L$. \ The fact that the cut cannot be locally deformed without increasing its weight corresponds to the fact that a geodesic cannot be locally deformed without increasing its length.

\subsection{Diamondwork Examples}

It's instructive to consider a few examples, to see how varying the edge weights in the otherwise fixed diamondwork graph can encode different bulk geometries.

\begin{figure}[!htb]
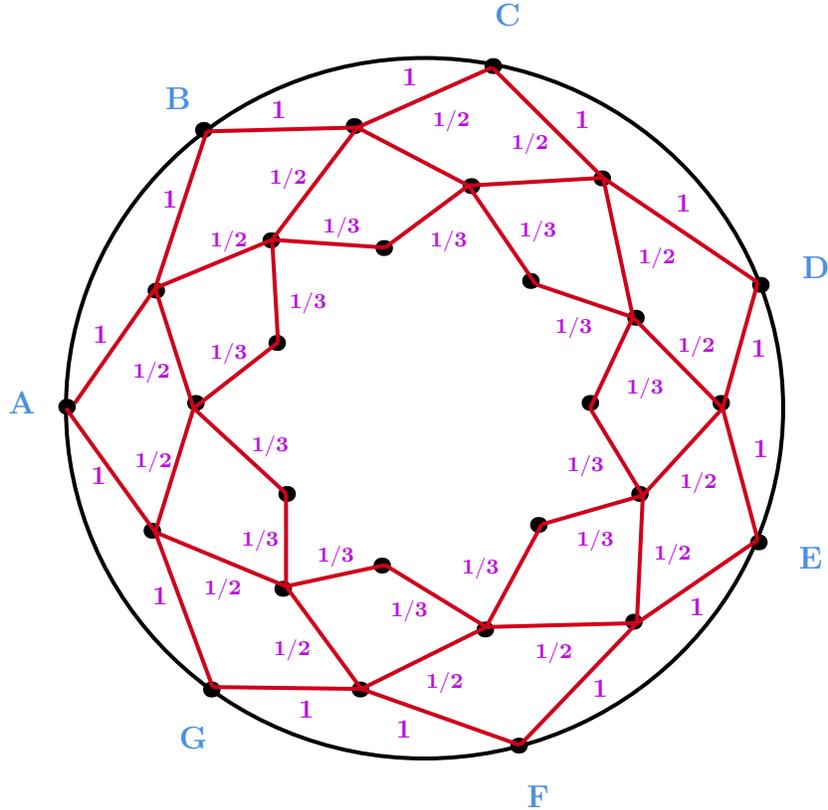

\centering

\tikzset{every picture/.style={line width=0.75pt}} %set default line width to 0.75pt        
\tikzset{every picture/.style={line width=0.75pt}} %set default line width to 0.75pt        



\caption{The $N=7$ diamondwork graph with harmonic edge weights.\label{fig:diamondwork_harmonic}}

\end{figure}

First consider the situation shown in Figure \ref{fig:diamondwork_harmonic}, where edges connecting vertices in layer $k$ and layer $k+1$ have weight $1/k$: in other words, the edge weights encode the harmonic series. \ As a direct consequence, the min-cut for a contiguous boundary region of size $L$ has a total weight of roughly $2\ln L$ (the factor of $2$ arising because the geodesic needs to go both inward and outward). \ Notably, this has the same dependence on $L$ as the length of a boundary-anchored geodesic in AdS geometry:
\begin{equation}
    \left|\gamma\right|_{\mathrm{AdS}} = 2L_{\mathrm{AdS}}\ln{L/a},
\end{equation}
with $a$ the UV cutoff, in the limit where $L\ll L_{\mathrm{AdS}}$, as well as the Cardy-Calabrese formula for the entanglement entropy of a small subregion $R$ of size $L$ in a $1+1$-dimensional conformal field theory \cite{Calabrese:2004eu},
\begin{equation}
    S(R)=\frac{c}{3}\ln{L/a}.
\end{equation}
Hence this example correctly reproduces, in our discrete setting, a central feature of AdS/CFT: namely that the larger a region on the boundary, the deeper the RT surface for that region penetrates into the bulk.

\begin{figure}[!htb]
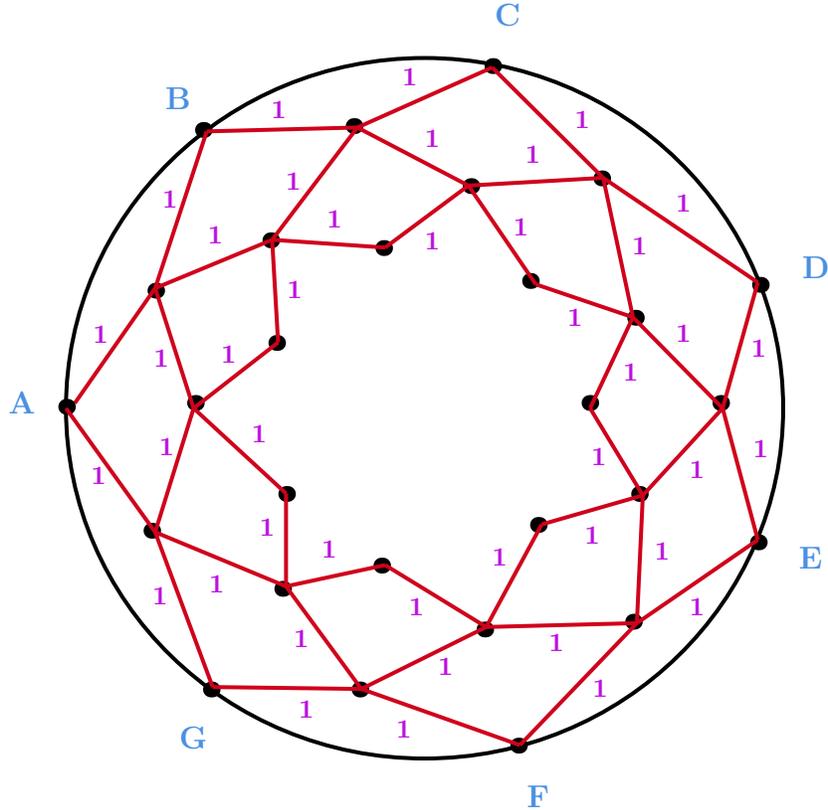

\centering

\tikzset{every picture/.style={line width=0.75pt}} %set default line width to 0.75pt        



\caption{The $7$-region diamondwork graph with constant edge weights.\label{fig:diamondwork_constant}}

\end{figure}

Second, consider another highly symmetric situation: a diamondwork graph with all edge weights equal to $1$, shown in Figure \ref{fig:diamondwork_constant}. \ Here the weight of the min-cut is simply equal to the number of edges it cuts, and hence the weight for a boundary region of size $L$ grows \emph{linearly} with $L$. \ This is what we would get if, instead of penetrating into the bulk, the geodesics simply hug the boundary. \ Imagine, for example, that the bulk is almost entirely filled with a gigantic black hole. \ In that case, geodesics that start and end at the boundary can never penetrate the event horizon; the best they can do is ``go around it the long way,'' traversing the thin ring between the event horizon and the boundary. \ One subtlety here is that, in Figure \ref{fig:diamondwork_constant}, there are \emph{also} min-cuts that penetrate arbitrarily deep into the bulk (indeed there must be, since all the edge weights are equal). \ To sustain the ``black hole'' interpretation, we have to say that those cuts are irrelevant; what is relevant is only whether there \emph{exist} cuts that hug the boundary (which there are). \ A second subtlety is that black holes are usually associated to a nonzero temperature, and to mixed states, while the diamondwork graph as we constructed it always describes a pure state---so if Figure \ref{fig:diamondwork_constant} represents a black hole, then it's a degenerate black hole.

\begin{figure}[!htb]
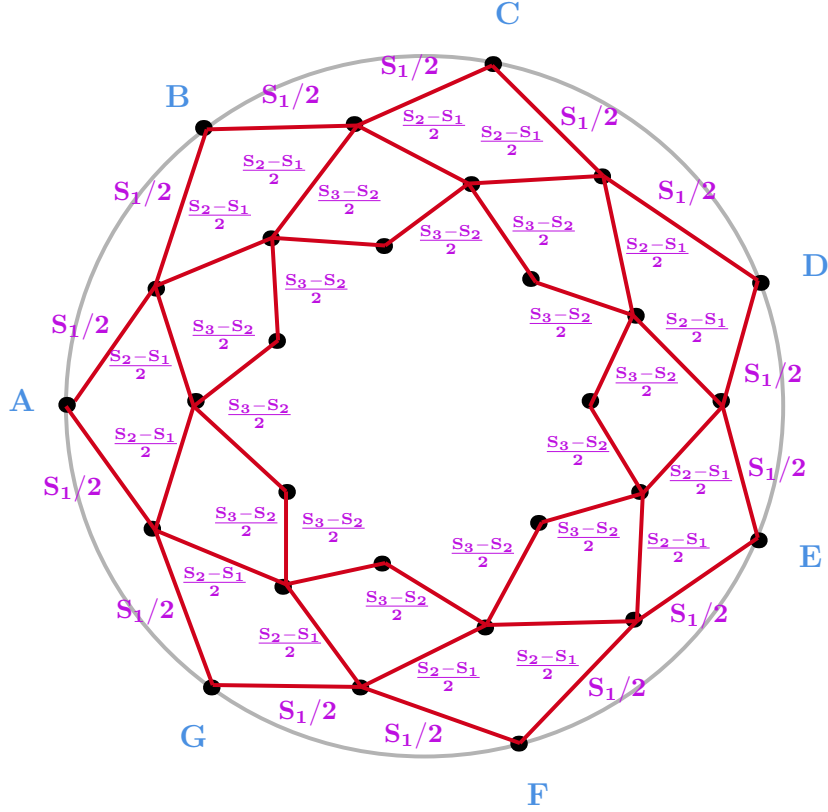

\centering

\tikzset{every picture/.style={line width=0.75pt}} %set default line width to 0.75pt        



\caption{The $7$-region diamondwork graph in the symmetric case where all one-party atomic regions have entropy $S_1$, all two-party regions have contiguous regions have entropy $S_2$, etc.\label{fig:diamondwork_symmetric}}

\end{figure}

In some particularly symmetric cases, we can write down the diamondwork weights directly, without needing to construct an auxiliary bulkless graph and pass through Eq.\ (\ref{eq:SL2}). \ For example, consider the case where all $k$-party contiguous unions of atomic regions have identical entropy $S_k$, i.e.\ the entropy of atomic regions is $S_1$, of unions of two atomic regions $S_2$, etc. \ Then the diamondwork graph, if it exists, can immediately be seen to have the form shown in Figure \ref{fig:diamondwork_symmetric}. \ Nonnegativity of the edge weights is enforced by Subadditivity and Strong Subadditivity: Subadditivity (\ref{eq:SA}) implies that
\begin{equation}
    S_1 \le S_2 \le  \cdots \le S_{k-1} \le S_k,
\end{equation}
while Strong Subadditivity (\ref{eq:SSA}) implies that
\begin{equation}
    S_k - S_{k-1}\le S_{k-1} - S_{k-2} \le \cdots \le S_3 - S_2\le S_2 - S_1.
\end{equation}

%%%%%%%%%%%%%%%%%%%%%%%%%%%%%%%%%%%
\section{The Multiple-Boundary Case}\label{sec:multiple}

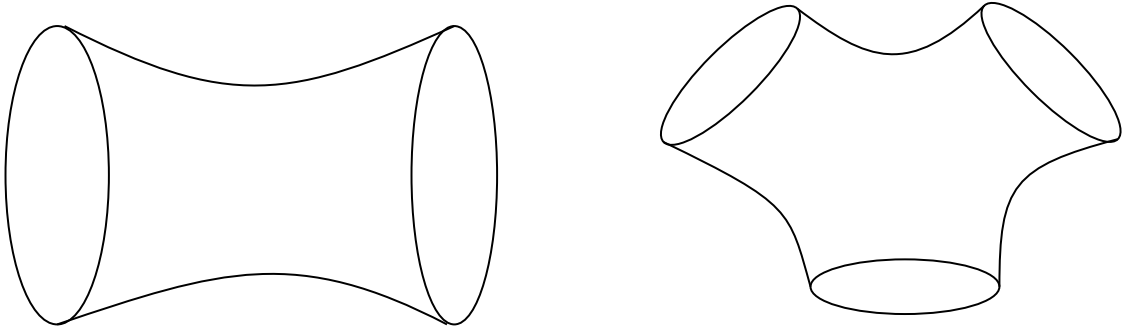
\begin{figure}[!htb]
\centering

\tikzset{every picture/.style={line width=0.75pt}} %set default line width to 0.75pt        

\begin{tikzpicture}[x=0.75pt,y=0.75pt,yscale=-1,xscale=1]

%Shape: Ellipse [id:dp946122083048057] 
\draw   (20,129.97) .. controls (20,88.38) and (31.67,54.67) .. (46.07,54.67) .. controls (60.47,54.67) and (72.14,88.38) .. (72.14,129.97) .. controls (72.14,171.55) and (60.47,205.27) .. (46.07,205.27) .. controls (31.67,205.27) and (20,171.55) .. (20,129.97) -- cycle ;
%Shape: Ellipse [id:dp9666288559036602] 
\draw   (224.76,129.97) .. controls (224.76,88.38) and (234.42,54.67) .. (246.35,54.67) .. controls (258.27,54.67) and (267.93,88.38) .. (267.93,129.97) .. controls (267.93,171.55) and (258.27,205.27) .. (246.35,205.27) .. controls (234.42,205.27) and (224.76,171.55) .. (224.76,129.97) -- cycle ;
%Curve Lines [id:da15347668160843453] 
\draw    (49.79,54.67) .. controls (132.85,97.05) and (163.48,92.27) .. (246.35,54.67) ;
%Curve Lines [id:da3620726847092015] 
\draw    (46.07,205.27) .. controls (128.88,177.17) and (166.89,165.81) .. (242.63,205.27) ;
%Shape: Ellipse [id:dp4156473585295717] 
\draw   (426,186.23) .. controls (426,193.85) and (447.33,200.03) .. (473.63,200.03) .. controls (499.94,200.03) and (521.27,193.85) .. (521.27,186.23) .. controls (521.27,178.61) and (499.94,172.43) .. (473.63,172.43) .. controls (447.33,172.43) and (426,178.61) .. (426,186.23) -- cycle ;
%Shape: Ellipse [id:dp7293305395462768] 
\draw   (351.95,113.25) .. controls (357.34,118.64) and (376.79,107.93) .. (395.39,89.32) .. controls (413.99,70.72) and (424.7,51.27) .. (419.32,45.88) .. controls (413.93,40.5) and (394.48,51.21) .. (375.88,69.81) .. controls (357.27,88.41) and (346.56,107.86) .. (351.95,113.25) -- cycle ;
%Shape: Ellipse [id:dp9102208964041327] 
\draw   (580.98,111.76) .. controls (586.37,106.37) and (575.66,86.92) .. (557.06,68.32) .. controls (538.46,49.72) and (519.01,39) .. (513.62,44.39) .. controls (508.23,49.78) and (518.94,69.23) .. (537.54,87.83) .. controls (556.14,106.44) and (575.59,117.15) .. (580.98,111.76) -- cycle ;
%Curve Lines [id:da8209882029391509] 
\draw    (351.95,113.25) .. controls (416.6,144.43) and (415.27,147.77) .. (426,186.23) ;
%Curve Lines [id:da8625374917142963] 
\draw    (521.27,186.23) .. controls (521.27,135.1) and (527.93,125.77) .. (580.98,111.76) ;
%Curve Lines [id:da5443452996619469] 
\draw    (419.32,45.88) .. controls (455.27,73.1) and (475.27,80.43) .. (513.62,44.39) ;

\end{tikzpicture}

\caption{Two- and three-boundary geometries with wormholes.\label{fig:multiboundary}}
\end{figure}

Having solved the case of a single 1D boundary with contiguous entropy data in Section \ref{sec:single}, it's natural to wonder how far we can generalize our bulk reconstruction methods. \ For starters, what about the case of \emph{two} 1D boundaries, which are connected via a wormhole? \ Or for that matter, $k$ 1D boundaries connected via a $k$-holed sphere (see Figure \ref{fig:multiboundary})?

A central difficulty, in this case, is simply that the quantum state associated with a single boundary $B$ is no longer necessarily pure---that is, we might have $S(B)>0$. \ And we have been unable to generalize the diamondwork construction from Section \ref{sec:single} from pure to mixed states.

One way to understand the problem with mixed states is that we no longer have the \textit{complementary recovery} identity $S(R)=S([N]-R)$. \ Because of this, given $N$ atomic regions on a single boundary, the contiguous entropies now comprise $N^2 - N + 1$ independent real parameters rather than merely $\binom{N}{2}$. \ Hence, just information-theoretically, any diamondwork graph would need to have $\sim N$ layers rather than only $\sim N/2$. \ But the proof of Theorem \ref{thm:diamondwork} breaks down once we go beyond $N/2$ layers.

Another way to understand the problem with mixed states is that we can no longer rely only on Strong Subadditivity (\ref{eq:SSA}) to ensure the existence of a bulk graph. \ At the least, we also need the Monogamy of Mutual Information (\ref{eq:MMI}), as shown by the following example of three atomic regions $A,B,C$ in an overall mixed state:
\begin{align}
    S(A)=S(B)=S(C)=4, \\
    S(AB)=S(BC)=S(AC)=5, \\
    S(ABC)=4.
\end{align}
This example satisfies SSA ($8<10$) but violates MMI ($16>15$). \ Since MMI is a cutting-and-pasting inequality (see Figure \ref{fig:MMI_proof}), it is satisfied by all graphs, and hence no graph with cut data matching these entropies exists.\footnote{We can purify the system by an additional region $D$, and think of this data as coming from a \emph{four}-party pure state. \ Then this example is a manifestation of our statement in Section \ref{entropineq} that when $N\ge 4$, the holographic entropy cone no longer coincides with the quantum entropy cone, but is a strict subset.}

\subsection{$2$-Boundary Counterexamples}\label{sub:counterexamples}

Even if the diamondwork construction could be extended to mixed states, we now point out yet a further problem with generalizing to multiple 1D boundaries. \ Namely, in the multi-boundary case, it seems natural to allow input data about regions that span more than one boundary. \ As soon as we do, however, we can get bulk graphs that are no longer embeddable onto the appropriate 2D surfaces.

In more detail, suppose we have $k$ circular boundaries $B_1,\ldots,B_k$, and the input data consists of $S(R_1 \cup \cdots \cup R_k)$ for all $k$-tuples of contiguous boundary regions $R_1\subseteq B_1,\ldots,R_k\subseteq B_k$. \ Then already when $k=2$, we claim that there \emph{cannot} be any solution as simple as our solution to the single-boundary case. \ We will prove this in two senses:

\begin{enumerate}
\item[(1)] Even when there exists a solution, there does not always exist a \emph{planar} solution (when $k=2$), or more generally, a solution that's embeddable onto a sphere with $k$ holes.
\item[(2)] The set of input vectors $v\in \mathbb{R}^{\binom{N}{2}^k}_{\ge 0}$ that \emph{do} admit planar solutions is not closed under nonnegative linear combinations.
\end{enumerate}

To start with (1), our counterexample will rely on the following lemma.

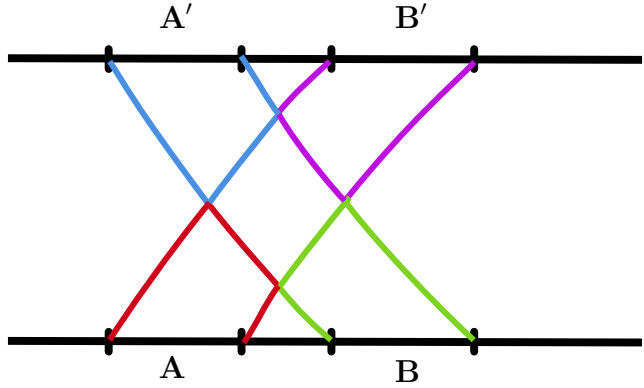
\begin{figure}[!htb]
    \centering

\tikzset{every picture/.style={line width=0.75pt}} %set default line width to 0.75pt        

\begin{tikzpicture}[x=0.75pt,y=0.75pt,yscale=-1,xscale=1]

%Straight Lines [id:da24810846500476424] 
\draw [line width=3]    (161,231) -- (318.61,231.4) -- (480.8,231.81) ;
%Shape: Free Drawing [id:dp775524204704314] 
\draw  [line width=3] [line join = round][line cap = round] (211.8,226.41) .. controls (211.8,229.75) and (211.8,233.08) .. (211.8,236.41) ;
%Shape: Free Drawing [id:dp5931429186976105] 
\draw  [line width=3] [line join = round][line cap = round] (278.8,226.4) .. controls (278.8,229.73) and (278.8,233.07) .. (278.8,236.4) ;
%Shape: Free Drawing [id:dp7215591566214712] 
\draw  [line width=3] [line join = round][line cap = round] (396.13,226.41) .. controls (396.13,229.75) and (396.13,233.08) .. (396.13,236.41) ;
%Shape: Free Drawing [id:dp12057354302056633] 
\draw  [line width=3] [line join = round][line cap = round] (324.13,226.4) .. controls (324.13,229.73) and (324.13,233.07) .. (324.13,236.4) ;
%Straight Lines [id:da5347743494328963] 
\draw [line width=3]    (161,88.33) -- (318.61,88.73) -- (480.8,89.15) ;
%Shape: Free Drawing [id:dp6325505359907975] 
\draw  [line width=3] [line join = round][line cap = round] (211.8,83.75) .. controls (211.8,87.08) and (211.8,90.41) .. (211.8,93.75) ;
%Shape: Free Drawing [id:dp7041553991361955] 
\draw  [line width=3] [line join = round][line cap = round] (278.8,83.7) .. controls (278.8,87.03) and (278.8,90.37) .. (278.8,93.7) ;
%Shape: Free Drawing [id:dp503031800231668] 
\draw  [line width=3] [line join = round][line cap = round] (396.13,83.75) .. controls (396.13,87.08) and (396.13,90.41) .. (396.13,93.75) ;
%Shape: Free Drawing [id:dp6249457204128044] 
\draw  [line width=3] [line join = round][line cap = round] (324.13,83.7) .. controls (324.13,87.03) and (324.13,90.37) .. (324.13,93.7) ;
%Curve Lines [id:da8274857138313505] 
\draw [color={rgb, 255:red, 74; green, 144; blue, 226 }  ,draw opacity=1 ][line width=2.25]    (212.27,89.93) .. controls (219.93,101.27) and (223.27,109.93) .. (261.6,161.27) ;
%Curve Lines [id:da9385079970240735] 
\draw [color={rgb, 255:red, 208; green, 2; blue, 27 }  ,draw opacity=1 ][line width=2.25]    (261.6,161.27) .. controls (280.6,184.6) and (269.27,171.27) .. (297.27,203.27) ;
%Curve Lines [id:da6512342810534493] 
\draw [color={rgb, 255:red, 126; green, 211; blue, 33 }  ,draw opacity=1 ][line width=2.25]    (297.27,203.27) .. controls (308.6,217.27) and (314.6,222.6) .. (323.93,230.6) ;
%Curve Lines [id:da13447647492593506] 
\draw [color={rgb, 255:red, 74; green, 144; blue, 226 }  ,draw opacity=1 ][line width=2.25]    (297.6,116.6) .. controls (288.93,127.27) and (274.27,145.27) .. (262.6,161.27) ;
%Curve Lines [id:da7490659110126918] 
\draw [color={rgb, 255:red, 189; green, 16; blue, 224 }  ,draw opacity=1 ][line width=2.25]    (323.27,89.93) .. controls (313.93,98.6) and (305.93,105.27) .. (297.27,115.93) ;
%Curve Lines [id:da7520779006893641] 
\draw [color={rgb, 255:red, 208; green, 2; blue, 27 }  ,draw opacity=1 ][line width=2.25]    (262.6,161.27) .. controls (242.93,185.27) and (219.6,218.6) .. (212.27,230.6) ;
%Curve Lines [id:da4767538447769808] 
\draw [color={rgb, 255:red, 208; green, 2; blue, 27 }  ,draw opacity=1 ][line width=2.25]    (297.27,203.27) .. controls (289.27,214.6) and (287.27,221.27) .. (279.93,231.93) ;
%Curve Lines [id:da8928524082162899] 
\draw [color={rgb, 255:red, 189; green, 16; blue, 224 }  ,draw opacity=1 ][line width=2.25]    (396.27,90.27) .. controls (383.93,103.27) and (365.27,117.93) .. (330.93,160.27) ;
%Curve Lines [id:da6366112388233915] 
\draw [color={rgb, 255:red, 126; green, 211; blue, 33 }  ,draw opacity=1 ][line width=2.25]    (333.27,158.6) .. controls (324.6,169.27) and (309.93,187.27) .. (298.27,203.27) ;
%Curve Lines [id:da16568160991516612] 
\draw [color={rgb, 255:red, 189; green, 16; blue, 224 }  ,draw opacity=1 ][line width=2.25]    (297.6,116.6) .. controls (307.93,134.6) and (329.27,159.27) .. (330.93,160.27) ;
%Curve Lines [id:da9898648300676942] 
\draw [color={rgb, 255:red, 74; green, 144; blue, 226 }  ,draw opacity=1 ][line width=2.25]    (278.6,87.27) .. controls (286.6,99.27) and (292.6,110.6) .. (297.6,116.6) ;
%Curve Lines [id:da9971097073519886] 
\draw [color={rgb, 255:red, 126; green, 211; blue, 33 }  ,draw opacity=1 ][line width=2.25]    (330.93,160.27) .. controls (338.6,171.6) and (369.93,206.6) .. (395.93,230.6) ;

% Text Node
\draw (235.67,237.33) node [anchor=north west][inner sep=0.75pt]   [align=left] {\textbf{A}};
% Text Node
\draw (354.67,239.33) node [anchor=north west][inner sep=0.75pt]   [align=left] {\textbf{B}};
% Text Node
\draw (235.7,59) node [anchor=north west][inner sep=0.75pt]   [align=left] {\textbf{A}$\displaystyle ^{\prime }$};
% Text Node
\draw (354.7,59) node [anchor=north west][inner sep=0.75pt]   [align=left] {\textbf{B}$\displaystyle ^{\prime }$};

\end{tikzpicture}
\caption{A cutting-and-pasting proof that, if two pairs of straddling geodesics cross each other, then they cannot be minimal surfaces. \ The four geodesics can be rearranged into four non-minimal surfaces that end on $A$, $A^\prime$, $B$ and $B^\prime$ respectively. \ Hence their combined length must exceed $S(A)+S(B)+S(A^\prime)+S(B^\prime)$.\label{fig:no_straddling}}

\end{figure}

\begin{lemma}
\label{straddlelem}Consider a wormhole with two 1D boundaries. \ Let $A$ and $B$ be disjoint contiguous regions on the left boundary, and let $A'$ and $B'$ be disjoint contiguous regions on the right boundary. \ Suppose the only RT surfaces for the regions $AB'$ and $BA'$ both involve `straddling' geodesics, i.e.\ geodesics that pass between two boundaries. \ Then those geodesics cannot cross each other.
\end{lemma}
\begin{proof}
We give a pictorial proof in Figure \ref{fig:no_straddling}. \ Briefly, we assume by contradiction that the only RT surfaces for $AB'$ and $BA'$ both contain straddling geodesics. \ We then cut-and-paste those geodesics, rearranging them into non-minimal surfaces for $A$, $B$, $A'$, and $B'$ separately. \ Taking unions, this gives us non-minimal surfaces for $AB'$ and $BA'$ separately. \ Since the total area of these surfaces equals the total area of the surfaces that we started with, and since neither new surface contains a straddling geodesic, this gives us our contradiction.
\end{proof}

Using Lemma \ref{straddlelem}, we can now show that, when there are two circular boundaries, it's no longer true that all input data can be explained by a planar graph if it can be explained by any graph at all.

\begin{theorem}
Consider a wormhole with two 1D boundaries. \ Let $A,B,C$ be disjoint contiguous regions on the left boundary, and let $A',B',C'$ be disjoint contiguous regions on the right boundary. \ Then there exist entropies for the $9$ regions $AA',AB',AC',BA',BB',BC',CA',CB',CC'$ that can be explained by \emph{some} bulk graph, but \emph{not} by any planar bulk graph (or equivalently, by any bulk graph embeddable onto the wormhole itself).
\label{nonplanarthm}
\end{theorem}
\begin{proof}
We set
\begin{equation}
S(A)=S(B)=S(C)=S(A')=S(B')=S(C')=3, 
\end{equation}
and
\begin{align} 
S(AA')&=S(AB')=S(AC')=S(BA')\nonumber\\
&=S(BB')=S(BC')=S(CA')=S(CB')=S(CC')=4. 
\end{align}
We then, crucially, have
\begin{equation} 
S(AA') < S(A)+S(A'),   \hspace{0.5 cm}  S(AB')<S(A)+S(B'),
\end{equation}
and so on for the other $7$ combinations. \ Now observe that, if there existed an RT surface for $AA'$ that did \emph{not} contain straddling geodesics, then we would necessarily have $S(A)=S(A)+S(A')$, and likewise for the other combinations. \ So we conclude that, for each of $9$ of the regions $AA'$, $AB'$, etc., any RT surface must indeed contain straddling geodesics.

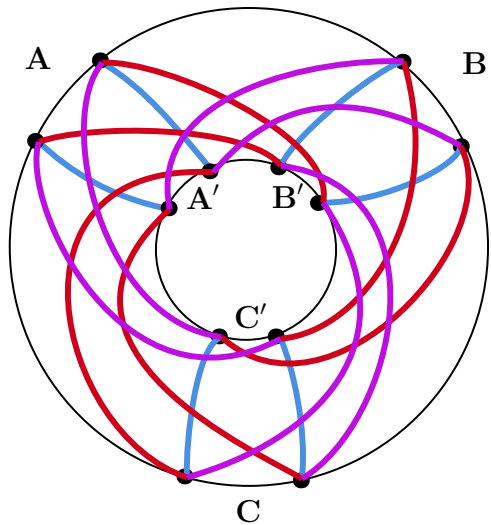
\begin{figure}[!htb]
    \centering

\tikzset{every picture/.style={line width=0.75pt}} %set default line width to 0.75pt        

\begin{tikzpicture}[x=0.75pt,y=0.75pt,yscale=-1,xscale=1]

%Shape: Circle [id:dp5183934017792586] 
\draw   (208.59,145.61) .. controls (208.59,79) and (262.58,25) .. (329.19,25) .. controls (395.8,25) and (449.8,79) .. (449.8,145.61) .. controls (449.8,212.22) and (395.8,266.21) .. (329.19,266.21) .. controls (262.58,266.21) and (208.59,212.22) .. (208.59,145.61) -- cycle ;
%Shape: Circle [id:dp3082700577466737] 
\draw   (282.31,147.06) .. controls (282.31,121.97) and (302.65,101.63) .. (327.74,101.63) .. controls (352.83,101.63) and (373.17,121.97) .. (373.17,147.06) .. controls (373.17,172.15) and (352.83,192.49) .. (327.74,192.49) .. controls (302.65,192.49) and (282.31,172.15) .. (282.31,147.06) -- cycle ;
%Curve Lines [id:da7306156263662904] 
\draw [color={rgb, 255:red, 74; green, 144; blue, 226 }  ,draw opacity=1 ][line width=2.25]    (223.27,92.6) .. controls (238.6,107.27) and (259.27,120.6) .. (289.27,126.6) ;
%Curve Lines [id:da24795279400636772] 
\draw [color={rgb, 255:red, 74; green, 144; blue, 226 }  ,draw opacity=1 ][line width=2.25]    (255.27,52.6) .. controls (277.27,65.93) and (292.27,84.93) .. (311.27,107.93) ;
%Shape: Free Drawing [id:dp15947381573369723] 
\draw  [line width=6] [line join = round][line cap = round] (221.34,92.1) .. controls (221.54,92.1) and (221.73,92.1) .. (221.92,92.1) ;
%Shape: Free Drawing [id:dp06024195510256347] 
\draw  [line width=6] [line join = round][line cap = round] (254.01,51.43) .. controls (254.2,51.43) and (254.4,51.43) .. (254.59,51.43) ;
%Shape: Free Drawing [id:dp6836398140708511] 
\draw  [line width=6] [line join = round][line cap = round] (406.67,52.1) .. controls (406.87,52.1) and (407.06,52.1) .. (407.26,52.1) ;
%Shape: Free Drawing [id:dp4006245162016937] 
\draw  [line width=6] [line join = round][line cap = round] (436.01,94.77) .. controls (436.2,94.77) and (436.4,94.77) .. (436.59,94.77) ;
%Shape: Free Drawing [id:dp35338625342877017] 
\draw  [line width=6] [line join = round][line cap = round] (309.34,107.43) .. controls (309.54,107.43) and (309.73,107.43) .. (309.92,107.43) ;
%Shape: Free Drawing [id:dp6795465608112434] 
\draw  [line width=6] [line join = round][line cap = round] (288.67,126.1) .. controls (288.87,126.1) and (289.06,126.1) .. (289.26,126.1) ;
%Shape: Free Drawing [id:dp19037993863775338] 
\draw  [line width=6] [line join = round][line cap = round] (296.67,261.43) .. controls (296.87,261.43) and (297.06,261.43) .. (297.26,261.43) ;
%Shape: Free Drawing [id:dp004665079397500271] 
\draw  [line width=6] [line join = round][line cap = round] (355.34,263.43) .. controls (355.54,263.43) and (355.73,263.43) .. (355.92,263.43) ;
%Shape: Free Drawing [id:dp9957075543182838] 
\draw  [line width=6] [line join = round][line cap = round] (344.01,105.43) .. controls (344.2,105.43) and (344.4,105.43) .. (344.59,105.43) ;
%Shape: Free Drawing [id:dp7320408482279621] 
\draw  [line width=6] [line join = round][line cap = round] (364.01,123.43) .. controls (364.2,123.43) and (364.4,123.43) .. (364.59,123.43) ;
%Shape: Free Drawing [id:dp04537495011309289] 
\draw  [line width=6] [line join = round][line cap = round] (314.01,190.77) .. controls (314.2,190.77) and (314.4,190.77) .. (314.59,190.77) ;
%Shape: Free Drawing [id:dp7728670510269777] 
\draw  [line width=6] [line join = round][line cap = round] (342.67,190.77) .. controls (342.87,190.77) and (343.06,190.77) .. (343.26,190.77) ;
%Curve Lines [id:da8072263820455463] 
\draw [color={rgb, 255:red, 74; green, 144; blue, 226 }  ,draw opacity=1 ][line width=2.25]    (435.93,94.6) .. controls (435.93,103.93) and (398.6,123.27) .. (366.6,123.27) ;
%Curve Lines [id:da7159701167368722] 
\draw [color={rgb, 255:red, 74; green, 144; blue, 226 }  ,draw opacity=1 ][line width=2.25]    (407.27,51.93) .. controls (396.6,53.93) and (353.27,87.27) .. (345.27,104.6) ;
%Curve Lines [id:da5428066200889776] 
\draw [color={rgb, 255:red, 74; green, 144; blue, 226 }  ,draw opacity=1 ][line width=2.25]    (345.27,191.93) .. controls (351.27,205.93) and (358.6,237.27) .. (355.93,262.6) ;
%Curve Lines [id:da7080045964384636] 
\draw [color={rgb, 255:red, 74; green, 144; blue, 226 }  ,draw opacity=1 ][line width=2.25]    (315.27,191) .. controls (302.6,195.67) and (297.27,243) .. (297.93,261.93) ;
%Curve Lines [id:da06085592595309963] 
\draw [color={rgb, 255:red, 208; green, 2; blue, 27 }  ,draw opacity=1 ][line width=2.25]    (255.27,52.6) .. controls (286.6,51) and (373.93,89.67) .. (366.6,123.27) ;
%Curve Lines [id:da5744363875840368] 
\draw [color={rgb, 255:red, 208; green, 2; blue, 27 }  ,draw opacity=1 ][line width=2.25]    (223.27,92.6) .. controls (246.6,83.67) and (329.27,83.67) .. (345.27,104.6) ;
%Curve Lines [id:da7375038233293147] 
\draw [color={rgb, 255:red, 208; green, 2; blue, 27 }  ,draw opacity=1 ][line width=2.25]    (435.93,94.6) .. controls (460.6,141.67) and (369.27,240.33) .. (315.27,191) ;
%Curve Lines [id:da659928790155659] 
\draw [color={rgb, 255:red, 208; green, 2; blue, 27 }  ,draw opacity=1 ][line width=2.25]    (289.27,126.6) .. controls (211.93,197) and (336.6,250.33) .. (355.93,262.6) ;
%Curve Lines [id:da6055768269407118] 
\draw [color={rgb, 255:red, 208; green, 2; blue, 27 }  ,draw opacity=1 ][line width=2.25]    (311.27,107.93) .. controls (199.93,101.67) and (231.93,239.67) .. (297.93,261.93) ;
%Curve Lines [id:da6471398494403213] 
\draw [color={rgb, 255:red, 208; green, 2; blue, 27 }  ,draw opacity=1 ][line width=2.25]    (407.27,51.93) .. controls (429.27,128.33) and (386.6,188.33) .. (345.27,191.93) ;
%Curve Lines [id:da9800901742206432] 
\draw [color={rgb, 255:red, 189; green, 16; blue, 224 }  ,draw opacity=1 ][line width=2.25]    (255.27,52.6) .. controls (231.27,81) and (254.6,183.67) .. (315.27,191) ;
%Curve Lines [id:da6684729649505488] 
\draw [color={rgb, 255:red, 189; green, 16; blue, 224 }  ,draw opacity=1 ][line width=2.25]    (223.27,92.6) .. controls (213.27,123.67) and (265.93,234.33) .. (345.27,191.93) ;
%Curve Lines [id:da8223540546046106] 
\draw [color={rgb, 255:red, 189; green, 16; blue, 224 }  ,draw opacity=1 ][line width=2.25]    (407.27,51.93) .. controls (365.27,50.33) and (277.27,66.33) .. (289.27,126.6) ;
%Curve Lines [id:da022083814060643858] 
\draw [color={rgb, 255:red, 189; green, 16; blue, 224 }  ,draw opacity=1 ][line width=2.25]    (435.93,94.6) .. controls (391.27,69) and (347.93,63.67) .. (311.27,107.93) ;
%Curve Lines [id:da3366204603213232] 
\draw [color={rgb, 255:red, 189; green, 16; blue, 224 }  ,draw opacity=1 ][line width=2.25]    (345.27,104.6) .. controls (430.6,115.67) and (403.27,234.33) .. (355.93,262.6) ;
%Curve Lines [id:da08761795349852708] 
\draw [color={rgb, 255:red, 189; green, 16; blue, 224 }  ,draw opacity=1 ][line width=2.25]    (366.6,123.27) .. controls (414.6,194.33) and (362.6,243.67) .. (297.93,261.93) ;

% Text Node
\draw (214.67,43) node [anchor=north west][inner sep=0.75pt]   [align=left] {\textbf{A}};
% Text Node
\draw (435.33,46) node [anchor=north west][inner sep=0.75pt]   [align=left] {\textbf{B}};
% Text Node
\draw (321,271.67) node [anchor=north west][inner sep=0.75pt]   [align=left] {\textbf{C}};
% Text Node
\draw (296,112.33) node [anchor=north west][inner sep=0.75pt]   [align=left] {\textbf{A}$\displaystyle ^{\prime }$};
% Text Node
\draw (339.33,112) node [anchor=north west][inner sep=0.75pt]   [align=left] {\textbf{B}$\displaystyle ^{\prime }$};
% Text Node
\draw (320.33,172.33) node [anchor=north west][inner sep=0.75pt]   [align=left] {\textbf{C}$\displaystyle ^{\prime }$};

\end{tikzpicture}

\caption{The non-planar graph $K_{3,3}$ embedded into a two-boundary geometry, via $9$ pairs of straddling geodesics.\label{fig:K33}}

\end{figure}

But this means that, whatever our bulk graph, it must give rise to pairs of straddling geodesics in a pattern isomorphic to the complete bipartite graph $K_{3,3}$ (see Figure \ref{fig:K33}). \ Since $K_{3,3}$ is non-planar, this implies that at least two pairs of straddling geodesics must cross each other in any planar graph model. \ But this contradicts Lemma \ref{straddlelem}. \ Hence there is no planar graph model.

On the other hand, the entropies above can easily be completed to a complete list of input entropies for which $K_{3,3}$ itself, with a weight of $1$ on each edge, provides a \emph{non}-planar graph model. \ We simply need to set
\begin{equation}
S(AB)=S(BC)=S(CA)=S(A'B')=S(B'C')=S(C'A')=6,
\end{equation}
\begin{equation}
S(ABC)=S(A'B'C')=9,
\end{equation}
\begin{equation}
 S(ABA')=S(ABB')=S(ABC')=S(BCA')=S(AA'B')=\cdots=5.
\end{equation}
\end{proof}

Contrast the above the case of a single boundary,
where it surprisingly turned out that every contiguous entropy vector that has any
realization at all has a planar realization.

We next prove statement (2).

\begin{theorem}
Consider a wormhole with two 1D boundaries. \ Let there be $3$ atomic regions per boundary, labeled $A,B,C$ on the left boundary and $A',B',C'$ on the right boundary. \ Let an input vector, $v\in \mathbb{R}_{\ge 0}^{63}$, specify $S(R)$ for every nonempty subset $R\subseteq \{A,B,C,A',B',C'\}$. \ Let $V$ be the subset of input vectors for which there exists a planar graph model. \ Then $V$ is not closed under convex combinations.
\label{nonconvexthm}
\end{theorem}
\begin{proof}
We'll define three entropy vectors, $v_1,v_2,v_3\in \mathbb{R}_{\ge 0}^{63}$. \ All three will have
\begin{equation} 
S(A)=S(B)=S(C)=S(A')=S(B')=S(C')=1.
\end{equation}
The values of $S(AB)$, $S(BC)$, $S(CA)$, $S(A'B')$, $S(B'C')$, $S(C'A')$, and
$S(ABC)=S(A'B'C')$ won't matter for this construction; we can simply choose any values for which planar graph models exist (which is not hard to arrange).

What \emph{does} matter is this:
\begin{itemize}
\item $v_1$ has $S(AA')=S(BB')=S(CC')=1$,
\item $v_2$ has $S(AB')=S(BC')=S(CA')=1$,
\item $v_3$ has $S(AC')=S(BA')=S(CB')=1$,
\end{itemize}
while in all three cases the ``other'' straddling entropies are all $2$: for example, $v_1$ has
\begin{equation}
    S(AB')=S(BC')=S(CA')=S(AC')=S(BA')=S(CB') = 2.
\end{equation}

We now consider the nonnegative linear combination $v_1+v_2+v_3$. \ This satisfies
\begin{equation} 
S(A)=S(B)=S(C)=S(A')=S(B')=S(C') = 3,
\end{equation}
while
\begin{align}
     S(AA')&=S(AB')=S(AC')=S(BA')=S(BB')=S(BC')\nonumber\\
     &=S(CA')=S(CB')=S(CC') = 2+2+1 = 5,
\end{align}
which is strictly less than
\begin{equation}
6 = S(A)+S(A') = S(A)+S(B') = \cdots.
\end{equation}
Just like in Theorem \ref{nonplanarthm}, this implies that the $9$ regions $AA', AB', AC'$, $BA', BB', BC'$,
$CA', CB', CC'$ \emph{all} require pairs of geodesics that straddle the wormhole. \ Hence these geodesic pairs have the pattern of the non-planar graph $K_{3,3}$ and there must be at least one crossing. \ So, again applying Lemma \ref{straddlelem}, we find that no graph model for $v_1+v_2+v_3$ can be planar. \ Hence the set $V$ of input data that admits a planar graph model is not closed under convex combination.
\end{proof}

 Note that, even with a \emph{single} boundary, if the input data can consist of entropies for non-contiguous regions, then by adapting the above counterexamples---and in particular, by embedding (say) the complete graph $K_5$ via straddling geodesics---we can create situations where there exists a bulk graph, but there does not exist any \emph{planar} bulk graph. \ In other words, the key to the above counterexamples is not multiple boundaries \emph{per se}, but rather the non-contiguous input data that multiple boundaries make natural.

\subsection{Size Upper Bound for Embeddable Graphs}

Stepping back, we saw in Subsection \ref{sub:counterexamples} that, when we try to generalize our methods to handle multiple boundaries, a new phenomenon arises: namely, we now encounter geodesics that are not, themselves, the RT surfaces of any boundary region. \ We called these ``straddling geodesics.'' \ These geodesics gave rise to graph models that were no longer embeddable on the relevant 2D surfaces---even by taking convex combinations of input vectors that \emph{do} give rise to embeddable graph models.

\begin{figure}[!htb]
\centering

\tikzset{every picture/.style={line width=0.75pt}} %set default line width to 0.75pt        

\begin{tikzpicture}[x=0.75pt,y=0.75pt,yscale=-1,xscale=1]

%Shape: Circle [id:dp5460041764088059] 
\draw   (42.59,150.61) .. controls (42.59,84) and (96.58,30) .. (163.19,30) .. controls (229.8,30) and (283.8,84) .. (283.8,150.61) .. controls (283.8,217.22) and (229.8,271.21) .. (163.19,271.21) .. controls (96.58,271.21) and (42.59,217.22) .. (42.59,150.61) -- cycle ;
%Shape: Free Drawing [id:dp18953977635475638] 
\draw  [line width=3] [line join = round][line cap = round] (273.8,149.41) .. controls (278.47,149.41) and (283.13,149.41) .. (287.8,149.41) ;
%Shape: Free Drawing [id:dp6984195753625391] 
\draw  [line width=3] [line join = round][line cap = round] (163.3,278.41) .. controls (163.3,275.01) and (163.3,271.61) .. (163.3,268.21) ;
%Shape: Free Drawing [id:dp8313597844056804] 
\draw  [line width=3] [line join = round][line cap = round] (38.8,148.41) .. controls (42.47,148.41) and (46.13,148.41) .. (49.8,148.41) ;
%Shape: Free Drawing [id:dp6846999867413079] 
\draw  [line width=3] [line join = round][line cap = round] (162.8,26.41) .. controls (162.8,29.75) and (162.8,33.08) .. (162.8,36.41) ;
%Curve Lines [id:da5086676195454811] 
\draw [color={rgb, 255:red, 208; green, 2; blue, 27 }  ,draw opacity=1 ][line width=2.25]    (172.8,162.43) .. controls (222.8,128.43) and (65.8,85.43) .. (42.59,150.61) ;
%Curve Lines [id:da36817513863492124] 
\draw [color={rgb, 255:red, 74; green, 144; blue, 226 }  ,draw opacity=1 ][line width=2.25]    (163.19,30) .. controls (223.8,84.43) and (104.8,108.43) .. (172.8,162.43) ;
%Curve Lines [id:da7292912800419995] 
\draw [color={rgb, 255:red, 208; green, 2; blue, 27 }  ,draw opacity=1 ][line width=2.25]    (283.8,150.61) .. controls (160.8,235.43) and (90.41,218.85) .. (172.8,162.43) ;
%Curve Lines [id:da5942066797545087] 
\draw [color={rgb, 255:red, 74; green, 144; blue, 226 }  ,draw opacity=1 ][line width=2.25]    (172.8,162.43) .. controls (179.8,220.43) and (59.19,216.61) .. (163.19,271.21) ;
%Shape: Circle [id:dp467663229984123] 
\draw   (384.59,146.57) .. controls (384.59,79.97) and (438.58,25.97) .. (505.19,25.97) .. controls (571.8,25.97) and (625.8,79.97) .. (625.8,146.57) .. controls (625.8,213.18) and (571.8,267.18) .. (505.19,267.18) .. controls (438.58,267.18) and (384.59,213.18) .. (384.59,146.57) -- cycle ;
%Shape: Free Drawing [id:dp38791482042196157] 
\draw  [line width=3] [line join = round][line cap = round] (615.8,145.38) .. controls (620.47,145.38) and (625.13,145.38) .. (629.8,145.38) ;
%Shape: Free Drawing [id:dp6896242669274142] 
\draw  [line width=3] [line join = round][line cap = round] (505.3,274.38) .. controls (505.3,270.98) and (505.3,267.58) .. (505.3,264.18) ;
%Shape: Free Drawing [id:dp17301607111582307] 
\draw  [line width=3] [line join = round][line cap = round] (380.8,144.38) .. controls (384.47,144.38) and (388.13,144.38) .. (391.8,144.38) ;
%Shape: Free Drawing [id:dp6337462956954356] 
\draw  [line width=3] [line join = round][line cap = round] (504.8,22.38) .. controls (504.8,25.71) and (504.8,29.05) .. (504.8,32.38) ;
%Curve Lines [id:da912717979199801] 
\draw [color={rgb, 255:red, 208; green, 2; blue, 27 }  ,draw opacity=1 ][line width=2.25]    (384.59,146.57) .. controls (395.8,112.82) and (448.8,105.82) .. (493.8,117.82) ;
%Curve Lines [id:da7189861897703547] 
\draw [color={rgb, 255:red, 208; green, 2; blue, 27 }  ,draw opacity=1 ][line width=2.25]    (490.8,205.01) .. controls (543.8,205.01) and (598.8,163.01) .. (625.8,146.57) ;
%Curve Lines [id:da22185564035373084] 
\draw [color={rgb, 255:red, 74; green, 144; blue, 226 }  ,draw opacity=1 ][line width=2.25]    (498.8,112.82) .. controls (510.8,69.82) and (546.8,68.82) .. (505.19,25.97) ;
%Curve Lines [id:da8003624999651533] 
\draw [color={rgb, 255:red, 74; green, 144; blue, 226 }  ,draw opacity=1 ][line width=2.25]    (505.19,267.18) .. controls (434.8,232.82) and (474.8,219.82) .. (490.8,204.82) ;
%Curve Lines [id:da6380509158107783] 
\draw [color={rgb, 255:red, 208; green, 2; blue, 27 }  ,draw opacity=1 ][line width=2.25]    (514.8,158.21) .. controls (497.8,147.82) and (492.8,133.82) .. (493.8,117.82) ;
%Curve Lines [id:da7079842431379122] 
\draw [color={rgb, 255:red, 208; green, 2; blue, 27 }  ,draw opacity=1 ][line width=2.25]    (490.8,204.82) .. controls (498.8,195.82) and (517.8,177.82) .. (514.8,158.21) ;
%Curve Lines [id:da7976220174175797] 
\draw [color={rgb, 255:red, 74; green, 144; blue, 226 }  ,draw opacity=1 ][line width=2.25]    (495.8,199.82) .. controls (503.8,190.82) and (530.8,180.82) .. (519.8,153.21) ;
%Curve Lines [id:da7018949911270476] 
\draw [color={rgb, 255:red, 74; green, 144; blue, 226 }  ,draw opacity=1 ][line width=2.25]    (519.8,153.21) .. controls (502.8,142.82) and (497.8,128.82) .. (498.8,112.82) ;
%Shape: Free Drawing [id:dp032436005792494615] 
\draw  [line width=6] [line join = round][line cap = round] (152.01,122.11) .. controls (152.2,122.11) and (152.4,122.11) .. (152.59,122.11) ;
%Shape: Free Drawing [id:dp30420256254807154] 
\draw  [line width=6] [line join = round][line cap = round] (171.01,162.11) .. controls (171.2,162.11) and (171.4,162.11) .. (171.59,162.11) ;
%Shape: Free Drawing [id:dp1049830412965147] 
\draw  [line width=6] [line join = round][line cap = round] (146.01,208.11) .. controls (146.2,208.11) and (146.4,208.11) .. (146.59,208.11) ;
%Shape: Free Drawing [id:dp41292479307389507] 
\draw  [line width=6] [line join = round][line cap = round] (491.01,204.11) .. controls (491.2,204.11) and (491.4,204.11) .. (491.59,204.11) ;

% Text Node
\draw (44,46) node [anchor=north west][inner sep=0.75pt]   [align=left] {\textbf{A}};
% Text Node
\draw (270,44) node [anchor=north west][inner sep=0.75pt]   [align=left] {\textbf{B}};
% Text Node
\draw (275,238) node [anchor=north west][inner sep=0.75pt]   [align=left] {\textbf{C}};
% Text Node
\draw (43,229) node [anchor=north west][inner sep=0.75pt]   [align=left] {\textbf{D}};
% Text Node
\draw (386,41.97) node [anchor=north west][inner sep=0.75pt]   [align=left] {\textbf{A}};
% Text Node
\draw (612,39.97) node [anchor=north west][inner sep=0.75pt]   [align=left] {\textbf{B}};
% Text Node
\draw (617,233.97) node [anchor=north west][inner sep=0.75pt]   [align=left] {\textbf{C}};
% Text Node
\draw (385,224.97) node [anchor=north west][inner sep=0.75pt]   [align=left] {\textbf{D}};

\end{tikzpicture}

\caption{Ensuring that two geodesics cross at most once. \ In the left figure, the red and blue curve cross each other multiple times. \ However, between each crossing, we can pick whichever curve has smaller length between the crossings, and use that for both curves, offsetting them by a slight amount to prevent intersection. \ This process can be continued until at most one crossing remains. \ The right figure shows the use of this procedure to eliminate the top two crossings in the left figure. \label{fig:pulling_apart_geodesics}}
\end{figure}
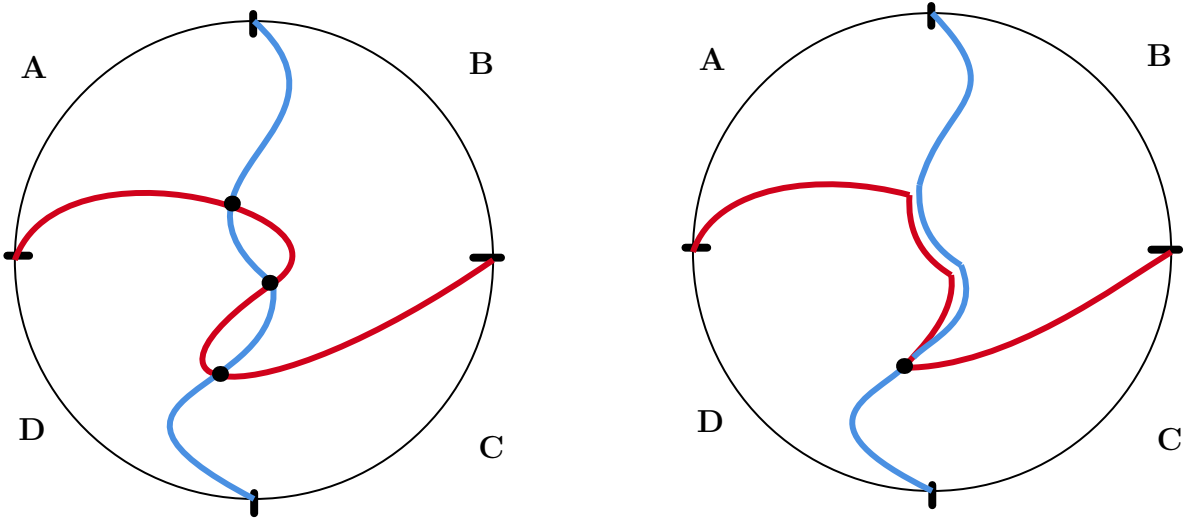

Nevertheless, straddling geodesics are still minimal geodesics. \ As such, they have the crucial property that any two such geodesics intersect in at most one point (see Figure \ref{fig:pulling_apart_geodesics}). \ By using this property, we can prove an upper bound of $O(N^4)$ on the number of vertices that could ever be needed in a bulk graph, \emph{assuming} that the bulk graph is embeddable onto a 2D surface with one hole for each boundary. \ This, of course, is dramatically better than the generic $2^{2^N}$ upper bound on the number of vertices from Proposition \ref{thm:double_exponential}. \ It's not as good as the $O(N^2)$ from the diamondwork construction, nor does it lead to a polynomial-time (let alone linear-time) algorithm for \emph{finding} the graph. \ These we leave as open problems.
 
\begin{theorem}[embeddable graphs need only $O(N^4)$ vertices]
Let $\mathcal{S}$ be a $2D$ surface with $k$ circular boundaries, which are divided into $N$ atomic regions in total. \ Given a list of entropies for various unions of those atomic regions (which need not be contiguous), suppose there exists a weighted, undirected graph $G$ solving the DBRP which can be embedded onto $\mathcal{S}$. \ Then there exists another graph $H$ with at most $O(N^4)$ vertices, which is also embeddable onto $\mathcal{S}$ and which has the same min-cuts as $G$ for all boundary regions.\label{thm:embeddable}
\end{theorem}

\begin{proof}
Because $G$ is embeddable onto a 2D surface $\mathcal{S}$, every min-cut can be represented as (possibly a finite union of) curves on $\mathcal{S}$, which we call \emph{geodesics}. \ It suffices to consider the $\binom{N}{2}$ geodesics that start and end at the points between atomic boundary regions. \ Just as in the proof of Proposition \ref{thm:double_exponential}, we can reproduce the min-cut structure of $G$ using at most one vertex in each region formed from the intersections of these geodesics. \ So it suffices to upper-bound the number of such intersection regions.

Recall that we can always choose the geodesics so that every pair crosses each other at most once (see Figure \ref{fig:pulling_apart_geodesics}). \ Let $M\le \binom{N}{2}$ be the number of minimally-intersecting geodesics, and imagine drawing them one at a time according to some ordering. \ Then the $t^{th}$ geodesic can create a new region by intersecting one of the previous $t-1$ geodesics, or by intersecting the boundary at the end of its route. \ Hence, letting $f(t)$ be the number of regions formed by the first $t$ geodesics, we have the recurrence relation $f(t) \le f(t-1)+t$, with base case $f(0)=1$. \ The solution is of course
\begin{equation}
f(M) = \frac{M(M+1)}{2} + 1 = O(N^4).
\end{equation}
\end{proof}

We note that, not by coincidence, the sequence of integers $f(1),f(2),\ldots$ from Theorem \ref{thm:embeddable} is the so-called ``Lazy Caterer's Sequence,'' giving the maximum number of pieces into which a circular cake can be divided using $M$ straight cuts.

From an algorithmic standpoint, the obvious drawback of Theorem \ref{thm:embeddable} is that it requires an embeddable bulk graph as its starting point. \ Nevertheless, Theorem \ref{thm:embeddable} is already enough to imply the following:

\begin{corollary}
Let $\mathcal{S}$ be a $2D$ surface with $k$ circular boundaries. \ Then given as input the entropies $S(R_1),\ldots,S(R_L)\ge 0$ for \emph{any} list of boundary regions $R_1,\ldots,R_L$ (which need not be contiguous), the decision problem of whether the $S(R_i)$'s admit a graph model that's embeddable onto $\mathcal{S}$ is in $\mathsf{NP}$ (Nondeterministic Polynomial-Time).
\end{corollary}
\begin{proof}
Note that we can express each $R_i$ as the union of at most $N\le 2L$ atomic boundary regions. \ So by Theorem \ref{thm:embeddable}, if a graph model $G$ embeddable onto $\mathcal{S}$ exists, then such a $G$ exists with at most $O(L^4)$ vertices. \ We are almost done, but to get an $\mathsf{NP}$ witness, we also need the technical fact that whenever $G$ exists, we can choose weights for its edges that are expressible with only polynomially many bits each. \ This follows from the fact that, once we know the set of edges in the min-cut for each $R_i$, we can express the vector of edge weights as the solution to a linear program, albeit one with $\exp(L)$ constraints. \ We can then construct a basic feasible solution using Gaussian elimination.
\end{proof}

Note that, whenever we can find disjoint unions of boundary regions that all require straddling geodesics, such that those geodesics have the pattern of $K_{3,3}$ or some other non-planar graph, that constitutes a certificate that there can be \emph{no} planar bulk graph. \ And likewise for any higher-genus surface $\mathcal{S}$: to construct a no-certificate, it suffices to find straddling geodesics forming a graph that cannot be embedded onto $\mathcal{S}$. \ However, it is unclear whether such certificates exist whenever there is no embeddable graph, \emph{and} whether they can be found in polynomial time whenever they exist. \ Note that, if the first statement is true, then the embeddable DBRP is in the class $\mathsf{NP}\cap\mathsf{coNP}$; if both are true, then the problem is in $\mathsf{P}$.

\section{Discussion and Open Problems}\label{sec:discussion}

Having presented our main results, we conclude with some possible generalizations and open research directions. 

\paragraph{Higher Dimensions}

The immediate obstacle we face when the boundary has more than one spatial dimension is that there is now an \emph{exponential}, rather than polynomial, number of contiguous regions. \ It is an interesting question whether there is a different, polynomially-sized set of entropic data which suffices, in some cases, to reproduce the rest of the entropy---that is, is there a higher-dimensional analogue of Eq.\ (\ref{eq:noncontiguous})? \ One obvious candidate on a $2$-dimensional boundary would be the set of discs of a fixed radius. \ However, it is unclear what the appropriate generalizations of our bulkless, chord, or diamondwork constructions would be. \ Perhaps we no longer want a graph at all, but a \emph{simplicial complex}.

\paragraph{Beyond RT: Dynamics and Corrections}

Note that (\ref{eq:noncontiguous}) requires the RT formula to hold \emph{exactly}; if we create, for example, an EPR pair shared between $A$ and $C$, it is clear that the left hand side will decrease while the right hand side will not change. \ (If we create \emph{many} EPR pairs, the ER=EPR conjecture \cite{Maldacena:2013xja} suggests that we will form a wormhole, and thus be back in the purview of Section \ref{sec:multiple}.) \ All of our graph constructions should therefore be understood to be discretizations of fully classical geometries. \ Furthermore, we have tacitly assumed that all of the extremal surfaces lie on a single timeslice of the bulk. \ In general, this need not be the case: there exist dynamical spacetimes, such as collapsing shockwave geometries, where extremal surfaces are not spacelike. \ In cases like this, we should use not the RT formula but the HRT formula \cite{Hubeny:2007xt,Wall:2012uf}, which correctly prescribes the appropriate way to extremize over such surfaces. \ We did not consider such dynamical spacetimes further in the body of the paper, where we were really trying to reconstruct a discrete bulk geometry at one point in time rather than a full spacetime. \ There doesn't seem to be any fundamental obstacle to trying to discretize a spacetime rather than a graph, but it would have to be a discretization of at least a three-dimensional manifold, with all the difficulties discussed before. \ For some existing work on entropy relations and discretizations for dynamical spacetimes, see \cite{May:2016dgv,Rota:2017ubr}.

\paragraph{Beyond Min-Cuts} We commented at the outset that in some sense the DBRP is the \emph{inverse} of the classical min-cut problem: instead of being given a fixed graph and asked to compute the min-cut separating two sets of vertices, we are given cut-values and asked to construct a graph. \ It would be interesting to understand the relation of the DBRP to other min-cut (and max-flow) problems. \ For example, in the multiflow literature, a flow is said to \emph{lock} a region if it saturates the bound given by the max-flow/min-cut theorem, and various ``locking theorems'' guarantee the existence of flows that lock all contiguous boundary regions simultaneously \cite{doi:10.1137/S0895480195287723,Headrick:2020gyq}. \ Is there any relation to our results, which also involve a special role for contiguous boundary regions? \ Also, once we're considering max-flows in addition to min-cuts, do the so-called \emph{quantum max-flows} (e.g., \cite{Cui:2015pla}) play any interesting role in AdS/CFT, and if so can we construct bulk graphs that account for them?

\paragraph{More about Multiple Boundaries} We showed in Section \ref{sec:multiple} that, when straddling geodesics exist, in general there is no planar graph that reproduces the boundary entropies. \ Nevertheless, Theorem \ref{thm:embeddable} tells us that there exists a polynomial set of input data---the lengths of \emph{all} $N^2$ geodesics, both RT surfaces and those passing between boundaries---which suffices to reproduce the entropy vector. \ One obstruction to passing from this observation to a graph is that it is unclear how to derive the lengths of geodesics that are not themselves RT surfaces from the entropy vector. \ Can this be done? \ Along similar lines, in Section \ref{sec:multiple} we worked with the $O(N^{2k})$ entropies of unions of contiguous regions. \ Yet Theorem \ref{thm:embeddable} again shows that embeddable graphs, which include those with the topology of a $k$-punctured sphere, need at most $O(N^4)$ vertices, independent of the value of $k$. \ Does this mean there is a smaller set of input data---for example, the entropies of all \emph{pairs} of contiguous boundary regions---that already suffices to determine the entropies of \emph{all} boundary regions? \ If so, is there some way to use that input data to construct versions of bulkless or diamondwork graphs that work for multi-boundary geometries, or even just a single boundary in a mixed state?

\paragraph{From Weights to Geometry} As shown in Section \ref{sec:single}, our various graph constructions (bulkless, chord, diamondwork) for a single boundary encode all of the geometric information in the edge weights and none in the graph structure: that is what it means for them to be ``universal.'' \ To think of the graph as an actual discrete geometry, we might instead want to move the geometric information from the weights to the graph structure. \ Forthcoming work by one of us and collaborators \cite{CPW} presents an approximate numerical algorithm to accomplish this task.

%\paragraph{Characterizing the Holographic Entropy Cone} In Section \ref{entropineq}, we raised the question of whether all inequalities that characterize the holographic entropy cone can be proved by cutting and pasting. \ Either answer would have interesting foundational implications for holography. \ If the answer is no, and yet holographic entropies are dual to areas, this would suggest that the areas of minimal surfaces differ from the areas of other surfaces in some not-yet-known way. \ More generally, in some special holographic states it is known \cite{brown2015non} that entropies can obey nonlinear inequalities. \ Does this have any implications for graph structure?

\paragraph{Graphs beyond Holography} Because our graph constructions take only the contiguous entropies as input, they can be applied even to entropy vectors \emph{outside} the holographic entropy cone. \ In that case, however, our constructions would produce graphs that generate ``wrong'' predictions for the entropies of some non-contiguous regions. \ Can we understand the failure of those predictions in any quantitative detail? \ What about for entropy vectors that are outside the holographic entropy cone, but still in the stabilizer entropy cone (see, e.g., \cite{Keeler:2022ajf})? \ More generally, what insight can we glean from our graph constructions about the holographic entropy cone itself?

\paragraph{Recovering the Quantum State} In general, an entropy vector represents an equivalence class of quantum states living on $N$-partite Hilbert spaces, all of which share the same von Neumann entropies of their reduced density matrices. \ Given an entropy vector, can we efficiently prepare a state $\ket{\psi}$ with the appropriate entropies? \ In other words, can we go in the opposite direction from this paper: from entropy vector to boundary, rather than from entropy vector to bulk? \ Can we at least do this for matching vectors, in which all entropies can be obtained from the contiguous ones? \ There are existing constructions in the literature to provide CFT states \cite{Bao:2015bfa} and stabilizer tensor networks \cite{Hayden:2016cfa} which automatically satisfy the RT formula for a given set of boundary subregions, but both require an infinitely large Hilbert space. \ Can we do better than this? \ We note that it's not hard to construct a state $\ket{\psi}$ that reproduces the entropies of the \emph{contiguous} regions, by starting with the bulkless graph, with edge weights $\{ w_{ij} \}$, and then inserting $w_{ij}$ EPR pairs (or fractional EPR pairs) between the atomic regions $i$ and $j$, for all $i,j$. \ However, this state will correspond to a geometry with numerous wormholes, and will yield wildly wrong predictions for the entropies of \emph{non}-contiguous regions.

\paragraph{Unordered Regions} In this paper, we studied how to produce a bulk graph given entropies of contiguous 1D boundary regions---but always under the assumption that we knew the \emph{order} of atomic regions around the boundary, as well as which entropies correspond to which boundary regions. \ What if we no longer know one or both of those things, and are just handed a ``bag of entropies'' to make sense of? \ Do we then get an $\mathsf{NP}$-hard reconstruction problem? \ Or possibly something equivalent to the so-called \emph{turnpike problem} in computer science \cite{dakic2000turnpike}?

\acknowledgments

The authors thank Ning Bao, Charles Cao, Netta Engelhardt, Matt Headrick, Sergio Hernandez-Cuenca, Alexander Jahn, Greg Kuperberg, Sarah Racz, Vincent Steffan, Lenny Susskind, and Benson Way for helpful discussions and comments.

% \begin{appendices}
% \addtocontents{toc}{\protect\setcounter{tocdepth}{1}}

% \end{appendices}

\bibliographystyle{hapalike}
\bibliography{paper}

\end{document}